\documentstyle[psfig]{mn}
\newcommand\beq{\begin{equation}}
\newcommand\eeq{\end{equation}}
\newcommand\QALKP{the {\em HST} QSO Absorption Line Key Project}
\newcommand\Lya{{Ly$\alpha$ }}
\newcommand\kpc{{\rm \,kpc}}
\newcommand\erg{{\rm \, erg}}
\newcommand\km{{\rm \,km}}
\newcommand\cm{{\rm \,cm}}
\newcommand\second{{\rm \,s}}
\newcommand\Mpc{{\rm \,Mpc}}
\newcommand\yr{{\rm \,yr}}

\title[Low Redshift \Lya Systems Associated with Galaxies]{Low Redshift QSO \Lya Absorption Line Systems Associated with Galaxies}

\author[W.P. Lin, G. B\"orner and H.J. Mo]
  {W.P.~Lin,$^1$$^2$$^3$\thanks{{\it E-mail} address: linwp@mpa-garching.mpg.de}
   G.~B\"orner$^1$, H.J.~Mo$^1$ \\
   $^1$ Max-Planck-Institut f\"ur Astrophysik, 85741 Garching, Germany\\
   $^2$ Beijing Astronomical Observatory and National Astronomical Observatories, Beijing 100012, P. R. China\\
   $^3$ Beijing Astrophysics Centre\thanks{BAC is jointly sponsored by the Chinese Academy of Sciences and Peking University.}, Peking University, Beijing 100871, P. R. China}

\date{
in original form March 2000}
\pagerange{\pageref{firstpage}--\pageref{lastpage}}
\pubyear{2000}

\begin{document}

\maketitle
\label{firstpage}

\begin{abstract}
In this paper we present semi-analytic models and Monte-Carlo simulations of QSO \Lya absorption line
systems which originate in gaseous galactic haloes, galaxy discs 
and dark matter (DM) satellites around big central haloes. 
The aim is to estimate the number density per unit redshift of \Lya absorption lines related with galaxies
and to investigate the properties of the predicted galaxy/absorber systems, such as equivalent widths $W_r$, projected distances $\rho$, galaxy luminosities $L_B$, as well as absorber redshifts $z$.
It is found that for strong \Lya absorption lines ($W_r\geq 0.3${\AA}) galactic haloes and satellites can explain $\sim$ 20 per cent and 40 per cent of the line number density of {\QALKP} respectively. 
The population of DM satellites is adopted from numerical simulations by Klypin et al. (1999). 
If big galaxies indeed possess such large numbers of DM satellites and they possess gas, these satellites may play an important role for strong \Lya lines. 
However the predicted number density of Lyman-limit systems by satellites is $\sim 0.1$ (per unit redshift), which is four times smaller than that by halo clouds.
Including galactic haloes, satellites and HI discs of spirals, the predicted number density of strong lines can be as much as 60 per cent of the {\em HST} result. The models can also predict all of the observed Lyman-limit systems.
For strong lines the average covering factor within $250 h^{-1} \kpc$ is estimated to be $\sim 0.36$, which is in good agreement with observations.
And the effective absorption radius of a galaxy (with unit covering factor) is estimated to be $\sim 150 h^{-1} \kpc$.
There exist correlations of $W_r$ versus $\rho$, $L_B$ and $z$. 
The models predict $W_r \propto \rho^{-\alpha} L_B^{\beta} (1+z)^{-\gamma}$ with $\alpha \sim 0.5, \beta \sim 0.15, \gamma \sim 0.5$.

To compare with results of imaging and spectroscopic surveys, we study the selection effects of selection criteria similar to the surveys.
We simulate mock observations through known QSO lines-of-sight and find that selection effects can statistically tighten the dependence of line width on projected distance. 
This result confirms previous suggestions in the literature.
After applying selection criteria, the models can predict 
similar distributions of $W_r$, $\rho$, $L_B$, absolute magnitudes and absorber redshifts to those of imaging and spectroscopic surveys. 
Finally we find that the total redshift interval of present observations ($\sim 5$) is not large enough for the models to reveal the real relationships if adopting the selection criteria. An adequate total redshift interval might be $\sim$ 10.
This may conciliate contraditious conclusions about the anti-correlation of equivalent widths versus projected distances by different authors.
\end{abstract}


\section{Introduction}
\label{sec1}

The origin and nature of low redshift \Lya line absorbers are still a matter of debate (cf. Chen et al. 1998, hereafter CLWB; Tripp, Lu, \& Savage 1998).
These absorbers are thought to arise either from gaseous galactic haloes/discs or from the underdense web-like regions of filaments and sheets in the large scale structure of the cosmic matter. 
The first suggestion seems reasonable because galaxies in general seem to possess extended gaseous haloes (Bahcall \& Spitzer 1969) or huge gaseous discs (Maloney 1992; Hoffman et al. 1993).
The second suggestion is drawn from the studies of high redshift \Lya absorption lines.
At high redshift, it is widely believed that most \Lya absorption lines are tracers of intergalactic hydrogen as suggested by Sargent et al. (1980). A diffuse intergalactic medium (IGM) model for the \Lya forest was investigated by Bi, B\"orner, \& Chu (1992).
Several authors investigated the scenario of the \Lya forest produced by the IGM in the context of cosmological simulations (e.g., Cen et al. 1994; Petitjean, M\"ucket, \& Kates 1995; Miralda-Escud\'e et al. 1996; Hernquist et al. 1996; Haehnelt, Steinmetz, \& Rauch 1996; Cen \& Simcoe 1997; Zhang et al. 1997, 1998; Theuns et al. 1998; Machacek et al. 1999).
It is natural to extend this model to low redshift. Some insights have been provided by hydrodynamic simulations predicting absorber properties down to $z=0$. 
For instance, Theuns, Leonard, \& Efstathiou (1998) find that the observed decrease in the rate of evolution of \Lya absorption lines at $z \leq 2$ can be explained by the steep decline in the photoionizing background resulting from the rapid decline in the quasar numbers at low redshift (see also Riediger, Petitjean, \& M\"ucket 1998).
Dav\'e et al. (1999) find that shocked or radiatively cooled gas of higher overdensity can give rise to the majority of strong \Lya lines at every redshift.
However on the observational side, Lanzetta et al. (1995, hereafter LBTW) \& CLWB claim that luminous galaxies can account for at least about $50$ per cent (and even more) of the strong \Lya absorption lines ($W_r \geq 0.3$ {\AA}) observed by the {\em Hubble Space Telescope (HST) Quasar Absorption Line Key Project} (Bahcall et al. 1993; Bahcall et al. 1996; Weymann et al. 1998).
Although the fraction is still quite uncertain because of the unknown galaxy space density (or luminosity function), the unknown gas absorption cross section of galaxies and the uncertainties of the observed number density of \Lya absorption systems, 
there is no doubt in principle that some absorbers are physically associated with galaxies especially for those with column densities above $10^{14} \cm^{-2}$.
Thus the origin of \Lya lines at low redshift is still an unsolved problem.

In order to clarify this question, it is particularly important to identify galaxies giving rise to QSO absorption and to analyse their physical properties.
So far, tremendous efforts have been made to locate responsible galaxies in QSO fields (e.g., Morris et al. 1993; Lanzetta et al. 1995; Bowen et al. 1996; Le Brun, Bergeron, \& Boiss\'e 1996; van Gorkom et al. 1996; CLWB; Bowen, Pettini, \& Boyle 1998; Tripp et al. 1998; Impey, Petry, \& Flint 1999). 
In general, these investigations of the physical link between an individual galaxy and an individual absorption system aim at answering
(1) whether absorbers are physically associated with galaxies and what the percentage of absorption lines is arising from galaxies,
(2) whether there is an anti-correlation between projected distance (i.e., impact parameter) from the line-of-sight (hereafter LOS) to the galaxy centre and the absorption rest frame equivalent width (hereafter REW). 
For example, Morris et al. (1993) carried out a redshift survey of galaxies in the field of 3C273 and found no galaxies within a projected distance of $230 h_{80}^{-1} \kpc$ of any of the 12 lines toward 3C273 with REW exceeding 50 m{\AA}. 
One of their conclusions is that the absorbers are not randomly distributed with respect to the galaxies, though the absorber-galaxy correlation is not as strong as the galaxy-galaxy correlation.
In a similar program but with different LOS, Stocke et al. (1995) and Shull et al. (1996) suggested that most of the \Lya absorbers are located within large-scale galaxy structures. 
In contrast, using an imaging and spectroscopic survey (in the field of HST spectroscopic target QSOs), 
LBTW claimed that the fraction of absorbers arising from galaxies is quite high and there is a distinct anti-correlation of REW versus projected distance. 
CLWB confirmed these results with more LOSs and concluded that most galaxies are surrounded by extended gaseous envelopes of $\approx 170 h^{-1} \kpc$ in radius and that many or most \Lya absorption systems arise in galaxies. 
Tripp et al. (1998) reached similar conclusions, but cautioned that selection effects could artificially tighten the anti-correlation, and also that the galaxy survey may be incomplete. 
They pointed out that there could be fainter galaxies at smaller projected distance (also suggested by Linder 1998; see also Impey et al. 1999) which could be revealed in a deeper survey. Moreover, they found some missing lines from the CLWB samples and from their LOSs. These missing lines would weaken the anti-correlation, if included. 
Recently, Impey, Petry \& Flint (1999) studied \Lya QSO absorbers in the nearby universe ($0<z<0.22$) based on the spectroscopy of ten quasars obtained with the Goddard High Resolution Spectrograph (GHRS) of the HST. 
At odds with the results of LBTW \& CLWB, they concluded that nothing in their data would specifically lead to associate absorbers preferentially with haloes of luminous galaxies.

Another very useful tool for the analysis of physical properties of absorbers is provided by observations of the intervening absorption in multiple LOSs either from close quasar pairs or from multiple, gravitational lensed quasar images (see Rauch 1998 for the part of review on this subject).
For example, double LOSs observations (e.g., Dinshaw et al. 1995; Dinshaw et al. 1994; Fang et al. 1996; Petitjean et al. 1998) have shown that the absorber size is about hundreds of kiloparsecs,
Rauch et al. (1999) analysed spectra of images of a lensed quasar at redshift 3.628 and found some low-ionization lines arising from the ISM.

Given these observational results, it is important to build theoretical models to understand the origin of the \Lya absorbers at low redshift.
Some theoretical efforts have been made to relate absorption systems with galaxies (e.g., Mo \& Morris 1994; Mo 1994; Morris \& van den Bergh 1994; Mo \& Miralda-Escud\'e 1996; Linder 1998; Linder 1999).
Unfortunately, even for those \Lya absorption lines genuinely arising from galaxy haloes, 
we do not know {\em a priori} which part of the galaxy gives rise to the absorption.
In other words, it is unclear whether the absorbing clouds are located in the outer regions of the halo as infalling clouds or in the rotating disc as interstellar medium clouds or in the satellites.
There are some competing models. 
Morris and van den Bergh (1994) estimated that tidal tails can explain $\sim 20$ per cent of the low redshift \Lya absorbers, but so far there is a lack of detailed models.
Mo \& Miralda-Escud\'e (1996) concluded that gaseous galaxy haloes can account for all absorbers with HI column density $N_{\rm HI} \geq 10^{17} \cm^{-2}$ at redshift $z\leq 2$. 
Recently a model in which absorption is due to gas in an extended disc was proposed by Linder (1998, 1999), who argued that high surface brightness galaxies together with low surface brightness galaxies can account for the majority of \Lya absorption line systems. 
However this picture only incorporates spiral galaxies while there are some absorbers associated with E/S0 galaxies (cf. CLWB). This model requires a large number of low surface brightness galaxies and it is unknown if a spiral galaxy can possess a disc extending beyond $100 h^{-1} \kpc$.

In current models of galaxy formation, galaxies are considered to possess haloes, discs (for spiral galaxies) and satellites which can give rise to absorption. 
Motivated by these considerations, we perform Monte-Carlo simulations using semi-analytic models with plausible assumptions, given our current knowledge about the properties of these components.
Our aim is to study: (1) which component is most important. 
(2) whether current observational results can be explained by the models. 
(3) whether selection effects (which should be applied when pairing an absorber with a luminous galaxy) can tighten the correlations between REW and projected distance. 
(4) what kind of future observations are needed to discriminate models and to examine the correlations.

The paper is organized as follows. In \S~\ref{secSS}, we describe our Monte-Carlo simulation methods and give results. 
We construct simulations with allowed parameters and compare the predicted line number density $(\frac{dN}{dz})$ with observational results for $W_r \geq 0.3${\AA} \Lya lines, Lyman-limit systems, and damped \Lya systems. 
Detailed properties of absorbers, correlations of equivalent width versus projected distance, galaxy luminosity and redshift, are studied. 
The average covering factor is estimated.
In \S~\ref{secMO}, we study selection effects. After applying selection criteria, we compare our results with imaging and spectroscopic surveys. At last we make some predictions. 
A discussion is presented in \S~\ref{secDis}. A summary of the results is given in \S~\ref{secCl}.
Throughout the paper, we adopt a dimensionless Hubble constant $h=H_0/(100\,\km\,{\second}^{-1} \Mpc^{-1})$.
Our presentation is mainly based on the $\Lambda$CDM cosmogony (with $\Omega_0=0.3,\Omega_{\Lambda,0}=0.7, h=0.7$), results based on the SCDM cosmogony (with $\Omega_0=1.0,\Omega_{\Lambda,0}=0.0, h=0.5$) are also discussed for comparison.

\section{The Monte-Carlo Simulations}
\label{secSS}
We start our Monte-Carlo simulations with galaxy samples (whose luminosity distribution is consistent with observational luminosity functions) along a QSO LOS.
These galaxies are placed along the LOS randomly within a cylinder volume in co-moving coordinate (then we assign a redshift and a projected distance for each galaxy).
For one particular galaxy whose halo is characterized by the circular velocity (derived from the Tully-Fisher relation or the Faber-Jackson law), 
we model its absorbing components (discs, halo gas clouds and satellites) in detail
so as to determine its gas cross section and cloud properties, such as HI column density, temperature and LOS velocity.
The total equivalent line width is then calculated assuming a Voigt profile for each cloud.
This procedure produces a catalogue of absorber-galaxy pairs 
with information about the absorbing galaxies for further analysis. 

\subsection{Galaxy samples}
\label{sec.gsample}
There will be a number of galaxies intersecting a particular LOS with random projected 
distances to a given QSO at redshift $z_{\rm q}$. We consider those galaxies in a 
cylinder within a radius of $R_{\rm cy}=400 h^{-1} \kpc$ to the QSO LOS, since in our models there 
is no absorption cloud outside of $400 h^{-1} \kpc$ and the upper limit of the projected distance in imaging surveys of absorbers is less than this radius. 
The number of galaxies at redshift $z$ with interval $\Delta z$ is 
\beq
\Delta N_{\rm g} = n_c (1+z)^3 \pi R_{\rm cy}^2 \frac {c dt}{dz}\Delta z, 
~~~~0 < z < z_{\rm q},
\eeq
where
\beq
\frac{dt}{dz} = \frac{1}{(1+z) H(z)},
\eeq
\beq
H(z) = H_0 [\Omega_{\Lambda,0}+(1-\Omega_{\Lambda,0}-\Omega_0)(1+z)^2+
       \Omega_0 (1+z)^3]^{1/2}.
\eeq
The co-moving galaxy density $n_c$ is obtained by integration over the B-band Schechter luminosity function
\[
n_c = \int^{\infty}_{L_{B\rm min}} \phi(L_{B}) dL_{B}
\]
\beq
~~~ = \int^{\infty}_{L_{B\rm min}} \phi^* (L_{B}/L_{B*})^{-\alpha} e^{-L_{B}/L_{B*}}dL_{B}/L_{B*},
\eeq
where $L_{B\rm min}$ is the minimum B-band luminosity.
The luminosities of these galaxies are selected in such a way that their distribution is consistent with the luminosity function $\phi(L_B)$. 

The luminosity functions for different morphological types over the range $-22 \leq M_B+5\log h \leq -14$ are as follows
(Marzke et al. 1998):
(1) For late-type galaxies (Spiral), $\phi^*=8.0\pm 1.4 \times 10^{-3} h^3 \Mpc^{-3}$, 
$\alpha=1.11^{+0.07}_{-0.06}$, $M_{B*}=-19.43^{+0.08}_{-0.08}+5\log h$, and $0.0067 L_{B*} \leq L_{B} \leq 10.7 L_{B*}$.
(2) For early-type galaxies (E/S0), $\phi^*=4.4\pm 0.8 \times 10^{-3} h^3 \Mpc^{-3}$, $\alpha=1.00^{+0.09}_{-0.09}$, $M_{B*}=-19.37^{+0.10}_{-0.11}+5\log h$, 
and $0.007  L_{B*} \leq L_{B} \leq 11.3 L_{B*}$.
We do not consider Irr/Pec galaxies because they are rare ($\phi^*=0.2\pm 0.08\times 10^{-3} h^3 \Mpc^{-3}$)
\setcounter{footnote}{0}
\footnote{Only normal galaxies are considered in the models. In reality however, galaxies may contain HI tidal tails and could give rise to absorption. We will discuss the problem in \S~\ref{secDis}.}.

The luminosity functions above are valid only at very low redshift ($z<0.05$). 
They are  derived from the recently enlarged Second Southern Sky Redshift Survey (SSRS2).
Some other determinations give higher normalizations. For example, the galaxy luminosity functions from the ESO Slice Project (ESP) galaxy redshift survey (Zucca et al. 1997) is characterized by $\phi^*=2.0\pm0.4 \times 10^{-2}h^3 \Mpc^{-3}$,
$M_*=-19.61^{+0.06}_{-0.08}$ and $\alpha=1.22^{+0.06}_{-0.07}$ over the redshift interval $z<0.3$. 
The ESP luminosity functions are in good agreement with those of the AUTOFIB redshift
survey (Ellis et al. 1996) which are characterized by $\phi^*=2.45^{+0.37}_{-0.31}\times 10^{-2} h^3 \Mpc^{-3}$, $\alpha=-1.16^{+0.15}_{-0.12}$, $M_*=-19.30^{+0.15}_{-0.12}$ over the redshift interval $0.02<z<0.15$,
$\phi^*=1.48^{+0.30}_{-0.19}\times 10^{-2} h^3 \Mpc^{-3}$, $\alpha=-1.41^{+0.12}_{-0.07}$, $M_*=-19.65^{+0.12}_{-0.10}$ over the redshift interval $0.15<z<0.35$, 
and $\phi^*=3.55^{+2.91}_{-2.00}\times 10^{-2} h^3 \Mpc^{-3}$, $\alpha=-1.45^{+0.16}_{-0.18}$, $M_*=-19.38^{+0.27}_{-0.25}$ over the redshift interval $0.35<z<0.75$.
The luminosity function of Lilly et al. (1995)
is characterized by $\phi^*=2.72\pm0.4\times 10^{-2} h^3 \Mpc^{-3}$, 
which is about twice that of SSRS2 (Marzke et al. 1998), 
and $\alpha=1.03$ over the redshift interval $0.2<z<0.5$. 

Since the faint-end slope $\alpha$ and the characteristic magnitude $M_*$ of various luminosity functions are not much different, 
we apply the AUTOFIB luminosity function normalization over $0.02 < z <0.15$
in our simulations and assign it to the two morphological types with the same ratio for spirals 
and E/S0 galaxies as in the SSRS2 luminosity function. 
Namely, the characteristic luminosity is 
$\phi^* \sim 1.58 \times 10^{-2} h^3 \Mpc^{-3}$ for spiral galaxies 
and $\phi^* \sim 0.87 \times 10^{-2} h^3 \Mpc^{-3}$ for E/S0 galaxies.
We will discuss results for other luminosity functions.


As the luminosity of a galaxy in the cone along the LOS and its redshift $z_{\rm g}$ are 
known, we calculate the apparent magnitude of the galaxy applying the k-correction and cosmological dimming,
\beq
m_B = M_B+5\log (D_L)+25+k_B(z_{\rm g}),
\label{mB}
\eeq
where $D_L=(1+z)D_M$ is luminosity distance of the galaxy in $\Mpc$, and $D_M$, the proper motion distance is
\beq
D_M=c \int^{z_{\rm g}}_0 H(z)^{-1} dz
\eeq
for a flat universe ($\Omega_{k} = 1-\Omega_0-\Omega_{\Lambda}=0$), which is the case in this paper.
We adopt B-band $k$ corrections for galaxies of different morphological types as in Pence (1976). 

At any given epoch, haloes can be parameterized by their circular velocity $V_{\rm cir}$,
which is simply related with galaxy morphological type and luminosity. 
Empirical relations are known between B-band magnitude and LOS velocity dispersion, $\sigma$, of the matter in galaxies both for ellipticals and for spirals (Faber and Jackson 1976; Tully and Fisher 1977). 

The Faber-Jackson relation is
\beq
-M_B+5\log h = (19.39\pm0.07)+10(\log \sigma -2.3)
\eeq
for ellipticals, and 
\beq
-M_B+5\log h = (19.75\pm0.07)+10(\log \sigma -2.3)
\eeq
for S0's (Fukugita \& Turner 1991).
The circular velocity of the halo for elliptical and S0 galaxies is $V_{\rm cir}=\sqrt{2}\sigma$. 
Using the E/S0 type luminosity function of Marzke et al. (1998), we get $81.6\,\km\, {\second}^{-1} \leq V_{\rm cir} \leq 514.6\, \km\, {\second}^{-1}$ ($V^{*}_{\rm cir}\simeq 280.1\, \km\, {\second}^{-1}$) for Elliptical galaxies 
and $75.1\, \km\, {\second}^{-1} \leq V_{\rm cir} \leq 473.6\, \km\, \second^{-1}$ ($V^{*}_{\rm cir}\simeq 258.5\, \km\, \second^{-1}$) for S0 galaxies. 

For spirals the Tully-Fisher relation is used to derive the LOS
velocity width $\Delta v$. We take (Fukugita \& Turner 1991)
\beq
-M_B+5\log h = (19.18\pm0.10)+(6.56\pm0.48)(\log \Delta v -2.5).
\eeq
The halo circular velocity for a spiral is $V_{\rm cir}=\Delta v /2$.
We get $25.7\, \km\, {\second}^{-1} \leq V_{\rm cir} \leq 425.4\, \km\, {\second}^{-1}$ ($V^{*}_{\rm cir}\simeq 172.6\, \km\, {\second}^{-1}$) using the spiral-type luminosity function of Marzke et al. (1998).
Note that the upper limits of the circular velocity are very large, there are, however, no such large galaxies in the sample due to the exponential cutoff in the Schechter luminosity function.

\subsection{Gaseous galactic haloes}

We model the gaseous galactic haloes following the work by Mo \& Miralda-Escud\'e (1996). 
In such semi-analytic models, it is assumed that the gas in a halo has a two-phase
(a hot phase and a cold phase) structure which, 
in principle is described by the density profiles and the temperature profile. 
The density profiles are characterized by the so-called cooling radius and virial radius. And the temperature profile of the hot gas is characterized by the virial temperature.
Our modeling is summarized as follows (see Mo \& Miralda-Escud\'e 1996 for more details):

In cooling flow models, when the gravitational potential is important, 
the core radius of the hot gas profile is similar to the cooling radius (Waxman \& Miralda-Escud\'e 1995). 
Thus, a self-similar density profile for hot gas is assumed as,
\beq
\rho_h(r) = \rho_h(r_c) \frac {2 r_c^2}{r(r+r_c)},
\eeq
where
\beq
\rho_h(r_c) = \frac{5 \mu k T_v}{2 \Lambda(T_v) t_{\rm \scriptscriptstyle M}}.
\eeq 
We assume
\beq
\rho_h(r_c)=\frac{f_g V_{\rm cir}^2}{4 \pi G r_c^2},
\eeq
so that the density of the hot gas at this radius is a fraction $f_g$
of the total density of the halo. 
$\Lambda(T_v)$ is the cooling rate of the gas at the virial temperature.
The gas mass fraction $f_g$ is assumed $\sim 0.03 - 0.05$.
$\mu$ is the average mass per particle, 
which is $\sim 0.6 m_{\rm H}$ with $m_{\rm H}$ being the mass of hydrogen nucleus.
The cooling radius $r_c$ is determined by eq. (11)
and eq. (12),
\beq
r_c = \sqrt{\frac{f_g \Lambda(T_v) t_{\rm \scriptscriptstyle M}}{5 \pi G \mu^2}}\simeq 124.6\,\kpc \sqrt {\Lambda_{-23} t_{\rm \scriptscriptstyle M}/10 {\rm G\,yrs}},
\eeq
where $\Lambda_{-23}$ is the cooling rate in units of $10^{-23} {\rm erg}\,\second^{-1} \cm^{3}$.
The hot gas is taken to be isothermal, so that $T_h(r) \equiv T_v=\mu V_{\rm cir}^2/2k$.
$t_{\rm \scriptscriptstyle M} = t/(1+\Omega_0)$, is the time interval between major mergers, since the gas
is then shock heated to a stage from which it starts cooling.
The cooling function $\Lambda(T_v)$ is adopted from Sutherland \& Dopita (1993).
The age of a halo at redshift $z_1$ is an integration of eq. (2)
\beq
t=\int^\infty_{z_1} \frac{1}{(1+z)\,H(z)}dz.
\eeq
The virial radius is
\beq
r_v=V_{\rm cir}/[10 H(z)].
\eeq

When the hot gas is shock heated and starts to cool, it will sink to
the galaxy centre with velocity $\vec{u}=-\hat{r}u(r)$. The cooling flow can be described as,
\beq
\frac{\partial{\rho_c}}{\partial{t}}+\nabla \cdot (\rho_c \vec{u})=\frac{\Lambda(T_h)}{\frac{5}{2}\mu k T_h}\rho^2_h(r).
\eeq

\noindent We assume $u(r)$ to be a constant:
\beq
u(r)=v_c
\eeq
where $v_c$ is the infall velocity which must be of the order of $V_{\rm cir}$. In eq. (13), $r_c$ is a function of $t$ (here $t$ stands for $t_{\rm \scriptscriptstyle M}$, and we omit the subscript
hereafter), and we have $\dot{r_c}=\frac{d r_c}{dt}=r_c/2t$ if we assume a constant $T_v$.
Let us set 
\[
\begin{array}{lll}
x & \equiv & \frac{r}{r_c} \\
\bar{\rho} & \equiv & \frac{\rho} {\rho_h(r_c)} \\
\bar{u} & \equiv & \frac{u}{\dot{r_c}}.
\end{array}
\]

\noindent Equation (16) has only one variable $r$ and can be simplified to
\beq
(x+\bar{u})\frac{d \bar{\rho_c}}{d x} +\bar{\rho_c} \frac{1}{x^2}\frac{d}{dx}(x^2\bar{u}) = \frac{-8}{x^2(1+x)^2}.
\eeq
This equation can be solved analytically (see Appendix A for details).

When $r_c >r_v$, $r_c$ is simply a parameter rather than a physical cooling radius. 
Thus we assume that the residual hot gas at $r_v$ is still at the virial temperature and has a density such that its cooling time is equal to t,
so that $T_h(r_v)=T_v$ and $\rho_h(r_v)=(5\mu k T_v)/[2\Lambda(T_v) t]$.
In such a case, we replace $r_c$ in eq. (10), (11) by $r_v$.
And we can not use eq. (16) to describe the cooling flow. 
In this case, most of the accreted gas will have cooled. 
Since the total gas mass accreted in a halo with circular velocity $V_{\rm cir}$ is $f_g V^2_{\rm cir} r_v/G$, the total mass of gas that has been in the cold phase can be written as
\beq
M = \frac{f_g V^2_{\rm cir} r_v}{G} - \int^{r_v}_{0} 4 \pi x^2 \rho_h(x) dx,
\eeq
where $\rho_h(x)$ is the density of gas in the hot phase as discussed above. The cold gas should form clouds that will fall through the halo. Assume a mass flow rate $\dot{M}=M/t$, and assume that the clouds move to the halo centre with a constant velocity $v_c$ (which is of the same order as the halo circular velocity). Assuming also spherical symmetry for the gas distribution, we can write the density of the cold gas as (see Mo \& Miralda-Escud\'e 1996 for more details),
\beq
\rho_c(r)=\frac{f_g V_{\rm cir}^2 r_{\rm min}}{4 \pi G r^2 v_c t}\left[1-\frac{r_{\rm min}^2}{r_c^2}\int^1_0 x^2 \bar{\rho_h}(x) dx \right],
\label{Denrv}
\eeq
where $r_{\rm min}=\min(r_v,r_c)$.


The cold gas is assumed to be, for simplicity, in spherical clouds with masses constrainted by various physical processes, 
such as gravitational instability, evaporation by hot gas, hydrodynamic 
instability, etc. Too large clouds will eventually collapse to form stars and 
small clouds will evaporate by heat conduction and also be disrupted by 
hydrodynamic instabilities. 
The net effects of these processes is to  preferentially destroy low mass clouds, so it is possible to end-up with a log-normal mass function, like the mass function of star clusters observed in the galaxy (e.g., Gnedin \& Ostriker 1997), if we begin with a power-law mass function.
For this reason and for lack of knowledge of the mass distribution of cold clouds,
we assume a log-normal distribution of cloud masses $M_c$,
\beq
p(M_c)dM_c=\frac{1}{\sqrt{2\pi}\sigma_{\rm \scriptscriptstyle M}} exp\left[-\frac{ln^2(M_c/\bar{M})}{2\sigma_{\rm \scriptscriptstyle M}^2}\right]\frac{dM_c}{M_c}.
\eeq
Here $\bar{M}$ is the cloud mean mass. As discussed by Mo \& Miralda-Escud\'e (1996), the mean mass of the clouds is approximately $10^5 - 10^6 M_{\odot}$.
In this paper, we choose $\bar{M}=$ several $\times 10^5 M_{\odot}$ and $\sigma_{\rm \scriptscriptstyle M} = 0.1 \sim 0.3$. 
We also use a constant cloud mass in simulations, but the results do not change much.

We model the clouds as spheres of uniform, isothermal photoionized 
gas confined by the pressure of the hot medium, so that 
$\rho_{\rm cloud} T_c=\rho_h T_h$, where $T_c$, the temperature of the clouds,
is about $2\times 10^4 K$.
 The cloud radius is $R_c=\left(3 M_c /4\pi \rho_{\rm cloud}\right)^{1/3}$ (typically $R_c \sim 1-10 \kpc$ at a radius within $100 \kpc$, cf. Mo 1994).
The total hydrogen column density through the cloud centre will be $N_{0}({\rm H})=R_c \rho_{\rm cloud} /2.3\mu$, and the H number density is $n({\rm H})=\rho_{\rm cloud}/2.3\mu$.

We assume that the cloud is almost completely photoionized by a constant UV background, and in ionization equilibrium.
The fraction of hydrogen in the neutral state (HI atom), is determined by
the flux of the UV background ionization field $J(\nu)$ and $n_{\rm H}$. We take
\beq
J(\nu)=J_{-21}(z)\times 10^{-21}\left(\frac{\nu}{\nu_{\rm \scriptscriptstyle HI}}\right)^{-\alpha} \Theta(\nu){\rm erg}\cm^{-2}{\rm sr}^{-1}{\rm Hz}^{-1}{\second}^{-1},
\eeq
where $\nu_{\rm \scriptscriptstyle HI}$ is the hydrogen Lyman limit frequency, $J_{-21}(z)=0.5$ for 
$z > 2$, and $J_{-21}=0.5 \times [(1+z)/3]^2$ for $z<2$. 
A break in the spectrum at $\nu_4 \equiv 4 Ry$ (due to continuum absorption by He II),
with $\Theta(\nu <\nu_4)=1$ and $\Theta_4 \equiv \Theta(\nu \geq \nu_4)=0.1$, is included (cf. Miralda-Escud\'e \& Ostriker 1990; Madau 1992).
We take $\alpha=0.5$.

For $N({\rm H}) \leq 10^{19}\,\cm^{-2}$, which is the case for most clouds, 
the cloud is optically thin to the ionizing field with ionization parameter $U=\frac{\Phi({\rm H})}{n({\rm H})c}$, 
where the ionizing photon flux $\Phi({\rm H})=\int ^\infty _{\nu_{\rm HI}} \frac{4 \pi J(\nu)}{{\rm h} \nu} d\nu$, h is Planck's constant.
The neutral hydrogen column density $N_{\rm HI}$ can be derived from the code CLOUDY 90 (Ferland 1996).
Then we obtain the HI column density of a cloud at a distance to the LOS $l$,\beq
N_{\rm HI}=N_{\rm HI}(0) \sqrt{1-l^2/R_c^2},
\eeq
where $N_{\rm HI}(0)$ is the HI column density through the cloud centre.

There might be one or more absorbing clouds in the LOS with different velocities with respect to the galaxy centre. The velocity structure follows eq. (17) and the LOS velocity can be calculated easily.

\subsection{Dark matter satellites}

According to the hierarchical clustering scenario, galaxies are assembled by 
merging and accretion of numerous dark matter satellites of different sizes and masses. 
As pointed out by Klypin et al. (1999), this ongoing process 
does not destroy all the accreted satellites.
Their paper gives results of satellite population around a big galaxy-size halo by high-resolution cosmological 
simulations. The VDF (velocity distribution function) of satellites within 
$200 h^{-1} \kpc$ and $400 h^{-1} \kpc$ is
\beq
n(>V_{\rm cir,sat})\approx 5000\,(V_{\rm cir,sat}/10\,\km\,\second^{-1})^{-2.75} h^3\,\Mpc^{-3}
\label{VDF1}
\eeq 
and
\beq
n(>V_{\rm cir,sat})\approx 1200\,(V_{\rm cir,sat}/10\,\km\,{\second}^{-1})^{-2.75} h^3\,\Mpc^{-3},
\label{VDF2}
\eeq 
respectively, where $V_{\rm cir,sat}=(10 - 70) \km\, {\second}^{-1}$. 
This number of satellites is roughly proportional to $(V_{\rm cir}/220\,\km\,{\second}^{-1})^3$.
The velocity dispersion of the satellites is of the order of the circular velocity of the central halo. 
The number of satellites in the models and in the Local Group
agrees well for massive satellites with $V_{\rm cir}>50\, \km\, {\second}^{-1}$, but disagrees by a factor of ten for low mass satellites with $V_{\rm cir}$ about $10\, \km\, {\second}^{-1} - 30\,\km\, {\second}^{-1}$ (see Klypin et al. 1999 for discusion). 
Possibly, most of these low mass satellites are dark matter mini-haloes or analogy of high-velocity clouds (HVCs) at distance $>100 \kpc$ in the halo of Milky Way galaxy. In addition, the distant HVCs are interpreted as gas contained within DM `minihalos'. (e.g., Blitz et al. 1999). It is possible that these satellites possess gas but have little or no star formation, so that they are faint and can not be found in optical surveys. 
The simulation of the survival of DM satellites has included dynamics friction and tidal stripping. 
Gas in DM satellite also suffers from ram-pressure stripping by hot gas in the central halo.
If the surviving DM satellites can accrete gas and the gas is not stripped away,
they can contribute to QSO absorption
because there could be some fraction of gas in the neutral state with detectable 
HI column density (Mo \& Morris 1994), 
\[
N_{\rm HI}\approx 6\times 10^{14}\cm^{-2}\left(\frac{V_{\rm cir,sat}}{30\km{\second}^{-1}} \right)^{4}
\]
\beq
~~~~~~~ \times \left(\frac{10\kpc}{R}\right)^{3}\left(\frac{f_g}{0.05}\right)^{2}\frac{T_{4.5}^{-3/4}}{J_{-21}},
\label{NHI}
\eeq
where $R$ is the galaxy projected distance (distance from satellite centre to LOS) and $T_{4.5}=T/10^{4.5}K$.
We choose $T_{4.5}=1.0$ here.
The spatial distribution of these satellites in the vicinity of the central galaxy is assumed to follow an inverse square law of distance to the centre.

If the population of DM satellites around big central haloes is not 
predicted correctly by N-body simulations or these satellites do not possess much gas 
(for example, due to tidal striping, ram-pressure striping, photoevaporation, or supernova-driven ejection), the absorption by satellites will be overestimated.

\subsection{Galaxy discs}

Observations of low redshift damped \Lya systems show that
some of these systems are possibly not in normal disc galaxies and their host galaxies are ambiguous (e.g., they could be low surface brightness galaxies or faint dwarf galaxies, or 
failed galaxies which are not detected, see Steidel et al. 1994, Le Brun et al. 1997, Rao \& Turnshek 1998). 
But in general, some damped \Lya systems must arise from galaxy discs (e.g, Prochaska \& Wolfe 1997, 1998 and references therein). 
In this paper we only simulate damped systems arising from galaxy discs.

Unfortunately the HI extent of discs is uncertain so far. 
Several studies involving $21 \cm$ mapping of galaxy discs have found `sharp edges' where the HI column density falls off dramatically from a few times $10^{19} \cm^{-2}$ to an undetectable level ($\sim 4 \times 10^{18} \cm^{-2}$). 
Such edges have been explained by models where the ionizing level increases rapidly from the inner optically thick to the outer optically thin regime (Maloney 1993; Corbelli \& Salpeter 1993; Dove \& Shull 1994a). 
Maloney (1993) assumed an exponential hydrogen disc and used a transition region model to calculate the HI column density of NGC 3198. His results are in good agreement with observations. 
Other authors have suggested an extended power-law disc in the highly ionized regime (Hoffman et al. 1993; Linder 1998; Linder 1999). 
However we will take the plausible assumption of an exponential disc extending from the centre to
the outer part of a galaxy. 
For an exponential disc, we adopt the model of Mo, Mao, \& White (1998, hereafter MMW).
The galaxy disc is assumed to be thin, to be in centrifugal balance, and to have an exponential surface density profile,
\beq
\Sigma(R)=\Sigma_0 exp(-R/R_d).
\eeq
Here $R_d$, $\Sigma_0$ and $R$ are the disc scalelength, central surface 
density and distance to the centre respectively.
Following MMW, we have
\beq
R_d \approx 8.8~ h^{-1}\kpc\left(\frac{\lambda}{0.05}\right)\left(\frac{V_c}{250  \km\,{\second}^{-1}}\right)\left[\frac{H(z)}{H_0}\right]^{-1}\left(\frac{j_d}{m_d}\right),
\eeq
and
\[
\Sigma_0 \approx 4.8\times 10^{22}~h~\cm^{-2} m_{\rm H} \left(\frac{m_d}{0.05}\right) \left(\frac{\lambda}{0.05}\right)^{-2} 
\]
\beq
~~~~~~ \times \left(\frac{V_c}{250~\km{\second}^{-1}}\right)
\left[\frac{H(z)}{H_0}\right] \left(\frac{m_d}{j_d}\right)^{2},
\eeq
where $m_d$ and $j_d$ are the fixed ratios of disc mass to halo total mass and disk angular momentum of halo total angular 
momentum respectively. Generally, we choose $m_d \approx j_d \approx 0.05$ 
throughout this work without considering the instability of galaxy discs and 
evolution of these two parameters.
$\lambda$ is defined as the halo spin parameter, whose distribution is 
\beq
p(\lambda)d\lambda=\frac{1}{\sqrt{2\pi}\sigma_{\lambda}} exp\left[-\frac{ln^2(\lambda/\bar{\lambda})}{2\sigma_{\lambda}^
2}\right]\frac{d\lambda}{\lambda},
\eeq
where $\bar{\lambda}=0.05$ and $\sigma_{\lambda}=0.5$ (MMW).

A galaxy disc is thought to have a vertical structure. 
It is a good assumption (expect for very flattened haloes) to ignore the change in the halo density with a height $Z$ above the midplane (cf. Maloney 1993). 
Then in the limit of negligible self-gravity the vertical profile of the gas will be a Gaussian,
\beq
n_{\rm H}(R,Z) = n_{\rm H}(R,0)e^{-Z^2/2\sigma_h^2}
\eeq
where the scale height is given by
\beq
\sigma_h(R) \simeq R \frac{\sigma_{zz}}{V_A}.
\eeq
Here we assume the core radius of the halo to be much smaller than $R$ (cf. Maloney 1992). The asymptotic velocity $V_A$ is assumed to be of the same order as the halo circular velocity $V_{\rm cir}$. 
And we take the typical velocity dispersion $\sigma_{zz} \simeq 6 \,\km\, {\second}^{-1}$.
The midplane density is
\beq
n_{\rm H}(R,0) = \frac{N_{\rm H}^{\rm tot}(R)}{(2\pi)^{1/2}\sigma_h},
\eeq
where the total hydrogen column density $N_{\rm H}^{\rm tot}(R)=\Sigma(R)/m_{\rm H}$.
The incident ionizing photons come from the top of the gas disc with a flux $\phi_{i,ex}$ photons $\cm^{-2} {\second}^{-1}$. 
This photon flux will ionize the gas to a depth $Z_i$ at which the column recombination rate equals the ionizing photon flux, i.e.,
\beq
\int^{\infty}_{Z_i}\alpha_{\rm rec}n_{\rm H}^2(R,0) e^{-Z^2/\sigma_h^2} dZ = \frac{1}{2} \phi_{i,ex}.
\label{eq-ionization}
\eeq
Here $\alpha_{\rm rec}=4.18\times 10^{13}\, \cm^3\, {\second}^{-1}\, T_{e,4}^{-0.72}$ is the recombination coefficient at a temperature of $T_{e,4}=T_e/10,000 K$. We assume $T_{e}=20,000 K$.
In the optically thick regime, the UV ionizing field is incident from one side so that the ionizing photon flux is $\frac{1}{2} \phi_{i,ex}$, while $\phi_{i,ex}= 5.4 \times 10^4 I_{\rm ly}\, {\rm photons}\, \cm^{-2}\, s^{-1}$ and $I_{\rm ly}=J_{-21}/(0.04\, {\rm erg}\, \cm^{-2}\, sr^{-1}\, {\rm Hz}^{-1}\, \second^{-1})$. 
We define 
\[
b \equiv \frac{\Phi_{i,ex}}{\sqrt{\pi}\alpha_{\rm rec}n_{\rm H}^2(R,0)\sigma_h} \sim 1.36\times 10^3 \left(\frac{\sigma_h}{\kpc}\right) T_{e,4}^{-0.72} I_{\rm ly}^{-1} /N_{18}^2,
\]
where $N_{18}=N^{tot}_{\rm H}/(10^{18} \cm^{-2})$. Then the equation (\ref{eq-ionization}) becomes
\beq
1- {\rm Erf}(Z_i/\sigma_h) = b,
\label{eqn-zi}
\eeq
where Erf is the error function.

When $b=1$, $Z_i=0$, one can get a critical column density below which the hydrogen will be highly ionized. This critical column density is $\sim$ a few $\times 10^{19}\, \cm^{-2}$ (see Maloney 1993). 

When $b < 1$, the disc is optically thick and 
one can derive a $Z_i$ from equation (\ref{eqn-zi}).
The HI in disc has a sandwich structure.
If $Z < Z_i$, all the hydrogen is assumed to be neutral (the central HI layer) so that the HI fraction is
\beq
\chi_{\rm \scriptscriptstyle HI} \simeq 1. 
\eeq
Above the central HI layer, which is the case for $Z \geq Z_i$, the hydrogen is highly ionized. Thus assuming ionization equilibrium, the fraction of HI is
\beq
\chi_{\rm \scriptscriptstyle HI} \simeq 2 n_{\rm H} (\alpha +3 ) T_e^{-0.72} I_{\rm ly}^{-1},
\eeq
given the ionizing field is incident from one side of the disc. 

When $b >1$, the whole disc becomes optically thin and thus the fraction of HI can be estimated as (given the ionizing field is incident from both sides of the disk)
\beq
\chi_{\rm \scriptscriptstyle HI} \simeq n_{\rm H} (\alpha +3 ) T_e^{-0.72} I_{\rm ly}^{-1}
\eeq
assuming ionization equilibrium (cf. Maloney 1992). 
Here $\alpha$ is the spectral index of the UV background ionizing field, which is $\sim 0.5 - 1.5$.

Once $\chi_{\rm \scriptscriptstyle HI}$ is known, we can calculate the total HI column density along a LOS by
\beq
N_{\rm HI} = \int_{LOS}\chi_{\rm \scriptscriptstyle HI} n_{\rm H}(R,Z') dZ',
\eeq
where the LOS has mid-plane distance $R_0$, inclination angle $\gamma$ and orientation angle $\alpha$. For a point ($R$, $Z'$) along the LOS, we have
\beq
R(Z') = \sqrt{R^2_0 + (Z' \tan \gamma)^2 + 2 R_0 Z' \tan\gamma \cos \alpha}.
\eeq
Thus we can calculate $N_{\rm HI}$ numerically.
We also do calculations using Cloudy 90 (Ferland 1996) and find the result is consistent with our calculation of the above sandwich structure of the disc using the analytic methods.

In the optically thick regime, our predicted HI column density is in good agreement with the calculation by Maloney (1993) and the fraction of hydrogen in the HI phase is about 2/3.
However as we shall see in the next section, the gas cross section of this regime could be too large and the models predict too many damped \Lya systems (DLAs) at low redshift.
Thus we assume the fraction of the hydrogen column density in the HI phase to be $\kappa$ in this regime and simulate for different values of $\kappa$ to compare with the observational number density of DLAs.

\subsection{Rest frame equivalent width}

We `observe' haloes, discs, and satellites by random LOS and
obtain HI clouds  which contribute to \Lya absorption. Each absorbing 
cloud (halo cloud, disc, satellite) is assumed to have the Voigt profile.
The optical depth of a line is
\beq
\tau_{\nu} \approx 2.65\times 10^{-2} f_{jk} N_j \phi (\nu;\nu_{jk})
\eeq
for $h\nu_{jk} \ll kT$, where $f_{jk}$ is the oscillator strength of the line and 
$N_j$ is column density, $\phi$ is the Voigt profile, $j, k$ are the lower and upper energy level indexs respectively (for \Lya line, $j=1, k=2$).
The REW, defined in frequency units is,
\beq
W=\int (1-e^{-\tau_{\nu}})d\nu.
\eeq
In accordance with observational usage, $W$ is defined in wavelength units, so the value must be multiplied by $\lambda /\nu$.
For the HI \Lya line, $\lambda=1215.670${\AA}, and $f_{jk}=0.4162$.

Because there may be $n$ components in a LOS (a direct sum or a blend of two or more lines), we compute ${\displaystyle \tau(\nu)
=\sum_{i=1}^{n}\tau_{\nu,i}}$ and then calculate $W$ numerically. 
For $N_{HI}>10^{19} \cm^{-2}$, which is the case when LOS intersects an optically thick disc, the REW is determined from the column density accurately as
(Petitjean 1998)
\beq 
W_r=\sqrt{N_{\rm HI}/(1.88\times 10^{18}\cm^{-2})}\, {\rm \AA}.
\eeq

\subsection{The predicted line number densities}
\label{secSR}
There are two kinds of basic observational facts for \Lya absorbers at low redshift. 
From spectroscopic surveys in QSO spectra one can derive the line number densities per unit redshift interval, $(\frac{dN}{dz})$ (for absorbers with $W_r \geq 0.3${\AA}, for metal absorption systems, for LLSs, DLAs). 
Another kind of observation is the imaging and spectroscopic survey in the QSO fields from which one can relate absorbers with luminous galaxies. 
In this part of the paper, we compare the predicted $(\frac{dN}{dz})$ with that observed to set constraints on model parameters and absorbing components.
The only selection criterion for $(\frac{dN}{dz})$ is the lower limit of the line width (or HI column densities).
However, to compare our results with results of imaging and spectroscopic surveys, it is necessary to study selection effects (such as limitations on apparent magnitude, velocity separation, angular separation) when relating absorbers with luminous galaxies and studying the properties of absorber-galaxy pairs. This study will be presented in \S~\ref{secMO}.

Several models are simulated with different absorbing components and different parameters in plausible ranges.
The parameters used to describe the absorbing components are as follows: 
$f_g$ (the baryon fraction), $Z$ (the metallicity, in unit of $Z_{\odot}$), 
$M_5$ (the mean mass of halo clouds, in units of $10^5 M_{\odot}$), 
$v_{\rm inf}$ (the infall velocity of halo clouds, scaled by the halo circular velocity, $V_{\rm cir}$), 
$\kappa$ (the HI fraction of total hydrogen in the optically thick part of the galaxy disc), 
the velocity distribution function (VDF) chosen for satellites [we denote the VDF of satellites following eq.(\ref{VDF1}) as VDF1 and VDF following eq.(\ref{VDF2}) as VDF2], 
and the flux of the UV background ionizing field $J_{-21}(z)=J^0_{-21} (1+z)^2$ ($J^0_{-21}$ is the flux at $z=0$).

Five types of models are considered: (1) Model A: discs only, (2) Model B: halo clouds only, (3) Model C: satellites only, (4) Model D: disks and halo clouds, (5) Model F: disks, halo clouds and satellites.
Model A, B, C are used to predict the fraction of absorption by galaxy discs, halo clouds and satellites respectively.
Model D and F are used to see what fraction the plausible combination of absorbing components can explain observational line number densities. 
The standard model is Model F3 which includes all absorption components and has standard parameters:\\
$f_g=0.05, Z=0.3 Z_{\odot}, M_5=5, v_{\rm inf}=1.0, J^0_{-21}=0.056,
\kappa=0.1$, and VDF2
(the reasons of choosing these values are given below).

Model A has only one parameter, $\kappa$ (see below for its meaning) which is listed in Table~\ref{A}. 
The detailed parameters of the other models are listed in the second column of Table~\ref{C}. In the table, we only list those non-standard parameters and assume a $\Lambda$CDM cosmogony. But we also discuss the SCDM cosmogony in some cases (see notation in the table).

To get statistically stable results, we `observe' through sufficient numbers of LOSs (typically several hundreds).
The redshift interval for each LOSs is from 0 to 1, and contains approximately 300 galaxies in the simulations with $V_{\rm cir}\geq 30 \km{\second}^{-1}$ within a column with a radius of $400\,h^{-1}\,\kpc$ along a LOS. 
We simulate 500 LOSs for mode A and 200 LOSs for model B to F (i.e., with $\sim 6 \times 10^4$ galaxies in total) to get stable results. 
The results are listed in Table~\ref{A} and Table~\ref{C}. The details of the models are as follows:

\begin{enumerate}
 \item Model A: 
As mentioned above, the HI fraction of the disc model in the inner part of the disk is $\sim$ 2/3. This predicts too many low redshift DLAs (see below and Table~\ref{A}).
However some gas in the disc may form stars or may be in molecular phase, only a fraction of the gas in the inner part of disk is in HI phase. 
So we define this fraction as a free parameter $\kappa$,
and simulate for a set of $\kappa$ to get constraints on it by comparing the predicted DLA number densities with the observational results.
Note that we adopt the traditional definition of DLA with $N_{\rm HI}\geq 2\times 10^{20} \cm^{-2}$.
The observational number density of DLAs derived by Rao et al. (1995) is $n_{\rm \scriptscriptstyle DLA}(z)=dN_{\rm \scriptscriptstyle DLA}/dz=(0.015\pm0.004)(1+z)^{2.27\pm 0.25}$.
For a mean redshift of $z=0.5$, $n_{\rm \scriptscriptstyle DLA}(0.5)\simeq 0.038\pm0.014$.
The result of {\QALKP} (Jannuzi et al. 1998) at $z=0.58$ is about 0.020. 
Our simulation results for 500 LOSs with total redshift interval of 500 are listed in Table~\ref{A}. 
Our predicted number densities of DLAs are $\sim 0.04-0.06$, comparable to the observed counterparts, if $\kappa=0.1 - 0.2$.
Models with larger $\kappa$ will predicted too many low redshift DLAs.
Thus a reasonable value of $\kappa$ is 0.1 and we use this as the standard parameter.

\begin{table}
\caption{Results of disc-absorbers with different $\kappa$. See text for discusion.}
\label{A}
\begin{tabular}{ccccc}
\hline \hline
{$\kappa$} & $(\frac{dN}{dz})_{\rm \scriptscriptstyle DLA}$ & $(\frac{dN}{dz})_{\rm \scriptscriptstyle LLS}$ & $(\frac{dN}{dz})(W_r\geq 0.3{\rm \AA})$ &$(\frac{dN}{dz})_{\rm disc}$$^*$ \\
0.1 & 0.03 & 0.17 & 0.27 & 0.59 \\ 
0.2 & 0.06 & 0.13 & 0.29 & 0.60 \\ 
0.3 & 0.07 & 0.12 & 0.29 & 0.57 \\ 
0.4 & 0.09 & 0.09 & 0.26 & 0.59 \\ \hline
\end{tabular}
$^*$$(\frac{dN}{dz})_{\rm disc}$ is the line number density for all disk absorbers with $N_{\rm HI}$ down to $10^{12} \cm^{-2}$.
\end{table}

 \item Model B: Using model B, we perform simulations for different values of 
$f_g$, metallicity $Z$, mean cloud mass as well as cloud infall velocity. 
We only take into account cold gas inside $r_{\rm min}$ of haloes since we assume that there is no cold gas outside this radius.
The changing pattern of total absorber number with different parameters is conceivable. 
For comparison (see Table~\ref{C}), the results of models B1, B2 and B3 show,  
that the predicted number density will decline for larger
mean cloud masses because there are fewer clouds and thus the covering factor is smaller.
The mean cloud mass will be several $\times 10^5 M_{\odot}$. Cold clouds of such mass can survive from evaporation by heat conduction and gravitational instability, as well as hydrodynamic instability (Mo \& Miralda-Escud\'e 1996).
The results of models B3 and B4 show that the predicted number densities do not change significantly for $Z=0.3 Z_{\odot}$ and $Z=0.1 Z_{\odot}$. We use $Z \simeq 0.3 Z_{\odot}$ hereafter.
The results of models B2 and B5 show, that the predicted number density will increase when the gas fraction $f_g$ increases (thus increase the total gas cross section). 
Comparing B2 and B6, if the infall velocities of the clouds is only some fraction of the halo circular velocity $V_{\rm cir}$, the predicted number density increases because there should be more cold gas nowadays in the haloes, which is apparent from eq.~(\ref{Denrv}). 
The results of model B7 shows, if the UV background ionizing field were lower at $z<1$ due to the declining quasar density, more absorbers can be expected.

 \item Model C: Using model C, we compare results for different $f_g$ and VDF of satellites.
The simulation results are as follows (see Table~\ref{C}): model C1 (C3) predicts more absorbers than model C2 (C4) because of the larger gas fraction [see eq.~(\ref{NHI})]; 
model C3 (C4) can produce more absorbers than model C1 (C2) because more satellites are included.
Model C3 can predict $\sim 40$ per cent of the line number density of the observed strong \Lya lines.
Our population of satellites is chosen from Klypin et al. (1999). If it is true that a big galaxy possesses a large number of satellites and they possess gas, theses satellites may play an important role for strong \Lya absorption lines (with $W_r \ge 0.3${\AA}).

 \item Model D: The predicted number density of $W_r \ge 0.3${\AA} lines in model D is similar to those of model B since the contribution by galaxy discs is only a small fraction (see Table~\ref{A} \& Table~\ref{C}).
Galaxy halo clouds together with galaxy discs cannot explain the majority of observational number densities for the strong lines.

 \item Model F: All possible components are considered in this type of model.
The predicted absorber number densities are larger than those in halo-only models and satellite-only models (see Table~\ref{C}).
For example, model F3 predicts that the number density of absorbers can be as high as $55\pm 22$ per cent of the observed number density of {\QALKP}. 
Model F5 can predict $59\pm23$ per cent of the observed strong absorption lines. This is in good agreement with the results of LBTW and CLWB.
It is easy to understand that the predicted number density should be higher if the UV background is weaker: For example, in model F6, whose UV background is assumed to be five times lower, the predicted fraction can reach 92 per cent, because more gas in the galactic haloes is in the HI phase and the total absorption cross section of satellite-halos becomes larger. This model, however, predicts too many Lyman-limit systems (see below).

 \item If we choose a SCDM cosmogony, for example in model B8, D4, F7, the total absorber number density will decrease for two reasons.
One reason is,  that for a fixed baryon fraction, the cosmic time becomes shorter than in a $\Lambda$CDM model so that less gas can cool 
(the cooling radius decreases, as does absorbing gas cross section). Another reason is that in a SCDM universe there are fewer galaxies (about 221) along a LOS than in a $\Lambda$CDM universe.
In summary adopting a SCDM cosmogony reduces the fraction of absorber number density to only about 28 per cent in model F7, which is too small compared with the
{\em HST} result.
We conclude that models with a SCDM cosmogony can not predict sufficient absorbers at low redshift under the same assumption.
\end{enumerate}

The observed $(\frac{dN}{dz})$ of {\QALKP} (Bahcall et al. 1996) for strong \Lya lines ($W_r \geq 0.3${\AA}) is
\[
(18.2\pm5.0) \, (1+z)^\gamma, \gamma=0.58\pm0.50.
\]
The observed $(\frac{dN}{dz})$ for Lyman-limit systems is
\[
\sim 0.5\pm0.3 \,\,{\rm at}\, z=0.5
\]
or
\[
\sim 0.7\pm0.2 \,\,{\rm with}\, <z_{\rm \scriptscriptstyle LLS}>=0.7
\]
(Stengler-Larrea et al. 1995; see also Mo \& Miralda-Escud\'e 1996 for model predictions). 
The observed $(\frac{dN}{dz})$ for damped \Lya systems has been given above.
The results of LBTW \& CLWB show that at least 30 per cent of the absorbers with $W_r \geq 0.3${\AA} are related with luminous galaxies.
With comparison to these results, we summarize:
(1) model B can be ruled out because the predicted $(\frac{dN}{dz})$ of strong \Lya lines are small and no damped systems can be predicted in such models.
(2) model C can be ruled out for lack of damped systems and for insufficient $(\frac{dN}{dz})$ of Lyman-limits systems.
(3) models D, F2 and F7 can be excluded because they give insufficient $(\frac{dN}{dz})$ of strong \Lya lines.
(4) models B7 and F6 predict too many Lyman-limit systems and should be ruled out also.
Thus, we come to conclusion that models F1, F3, F4, F5 are reasonable models, because they predict plausible number densities for strong \Lya lines ($W_r \geq 0.3${\AA}), Lyman-limit systems, and DLAs.

\begin{table*}
\begin{minipage}{155mm}
\caption{Parameters and results of models}
\label{C}
\begin{tabular}{@{}ccccccccc}\hline\hline
 &  & N & f1 & f2 & f3& $(\frac{dN}{dz})_{\rm mean}$& $(\frac{dN}{dz})_{\rm mean}$ & \\
{Model}& parameters &(total)&(S)&(S0)&(E)& (LLS$^a$)& ($>$0.3{\AA}) & fraction$^b$ \\
{B1}& $M_5= 1$          &{1357}&{46.4}&{24.1}&{32.2}&{0.40}& {~4.6}&{$21.1\pm~8.3\%$} \\ 
{B2}& standard          &{1078}&{45.9}&{22.3}&{31.8}&{0.45}& {~3.7}&{$17.1\pm~6.7\%$} \\
{B3}& $M_5=10$          &{ 925}&{46.4}&{21.6}&{32.0}&{0.35}& {~2.9}&{$13.4\pm~5.2\%$} \\
{B4}& $Z=0.1 Z_{\odot}$ &{ 844}&{51.1}&{19.3}&{29.6}&{0.42}& {~2.7}&{$12.4\pm~4.9\%$} \\
{B5}& $f_g=0.03$        &{ 807}&{48.7}&{21.2}&{30.1}&{0.20}& {~2.2}&{$10.1\pm~4.0\%$} \\
{B6}& $v_{\rm inf}=0.5$     &{1339}&{47.2}&{20.7}&{32.1}&{0.61}& {~4.6}&{$21.2\pm~8.3\%$} \\
{B7}& $J^0_{-21}=0.006$&{1259}&{51.3}&{20.3}&{28.4}&{1.22}& {~4.9}&{$22.6\pm~8.9\%$} \\
{B8}& SCDM              &{ 472}&{51.1}&{16.5}&{32.4}&{0.16}& {~1.5}&{$~6.9\pm~2.7\%$} \\
{C1}& VDF1              &{1984}&{33.9}&{23.3}&{42.7}&{0.07}& {~4.9}&{$22.6\pm~8.9\%$} \\
{C2}& $f_g=0.03$,VDF1   &{1897}&{32.1}&{26.3}&{41.5}&{0.06}& {~3.4}&{$15.7\pm~6.1\%$} \\
{C3}& standard          &{4454}&{32.4}&{24.3}&{43.4}&{0.15}& {~9.8}&{$45.2\pm17.7\%$} \\
{C4}& $f_g=0.03$        &{4190}&{32.0}&{24.2}&{43.8}&{0.10}& {~6.1}&{$28.1\pm11.0\%$} \\
{D1}& standard          &{1041}&{47.6}&{20.4}&{32.1}&{0.48}& {~3.3}&{$15.2\pm~6.0\%$} \\
{D2}& $f_g=0.03$        &{ 741}&{50.2}&{20.2}&{29.6}&{0.36}& {~2.0}&{$~9.2\pm~3.6\%$} \\
{D3}& $v_{\rm inf}=0.5$     &{1370}&{44.0}&{21.3}&{34.7}&{0.69}& {~4.9}&{$22.6\pm~8.9\%$} \\
{D4}& SCDM              &{ 571}&{50.4}&{20.7}&{28.9}&{0.28}& {~2.1}&{$~9.7\pm~3.8\%$} \\
{F1}& VDF1              &{2591}&{37.9}&{23.1}&{39.0}&{0.70}& {~7.3}&{$33.6\pm13.2\%$} \\
{F2}& $f_g=0.03$, VDF1  &{2335}&{38.2}&{21.9}&{39.9}&{0.44}& {~4.9}&{$22.6\pm~8.9\%$} \\
{F3}& standard          &{4885}&{34.7}&{23.5}&{41.8}&{0.69}& {11.9}&{$54.8\pm21.5\%$} \\
{F4}& $f_g=0.03$        &{4545}&{34.3}&{24.0}&{41.7}&{0.42}& {~7.7}&{$35.5\pm13.9\%$} \\
{F5}& $v_{\rm inf}=0.5$     &{5076}&{33.8}&{24.5}&{41.6}&{0.81}& {12.9}&{$59.4\pm23.2\%$} \\
{F6}& $J^0_{-21}=0.01$ &{5017}&{35.4}&{22.9}&{41.7}&{1.55}& {20.1}&{$92.6\pm36.3\%$} \\
{F7}& SCDM              &{3281}&{33.3}&{23.6}&{43.2}&{0.33}& {~6.1}&{$28.1\pm11.0\%$} \\ \hline
\end{tabular}

$^a$ The Lyman-limit systems (LLS) are defined as absorbers with HI column density $10^{17} \cm^{-2} \leq N_{\rm HI} \leq 2 \times 10^{20} \cm^{-2}$.\\
$^b$ The fraction of simulated lines related to galaxies to the number of lines per unit redshift at redshift $z=0.35$, $(dN/dz)(W_r \geq 0.3{\rm \AA})\simeq 21.7\pm 8.5$ predicted by {\QALKP} (Bahcall et al. 1996, see also CLWB).
\end{minipage}
\end{table*}

\subsubsection{Velocity spread of absorption systems}
\begin{figure}
{\psfig{figure=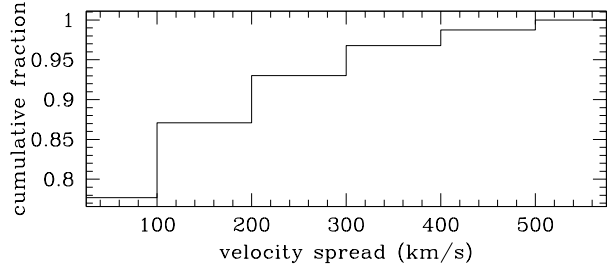,width=8.4cm}}
\caption{The cumulative distribution of velocity spreads of the simulated galaxy/absorber systems for model F3.}
\label{Figdv}
\end{figure}

The line number densities above are predicted by measuring the overall REW of all the subcomponents of a galaxy/absorber.
There are some cases where more than one subcomponent in a galaxy is crossed by the LOS. 
Sometimes it happens that the velocity spread of a line system is very large.
When the velocity spread of a line is less than $100 - 200 \km \second^{-1}$, the {\em HST} key project would possibly have detected just one line, but otherwise this would have been counted as 2 lines. 
To study this effect, we plot the cumulative distribution of velocity spread of the simulated galaxy/absorber systems for model F3 in Fig.\ref{Figdv}. 
As we can see, about 78 (87) per cent of the systems have a velocity spread less than $100 (200) \km \second^{-1}$.
So, if we treat those lines with velocity spread larger than $200 \km \second^{-1}$ as 2 lines, the predicted line number density will increase by about 15 per cent.

\subsubsection{Photoionization flux contributed by galaxies}

Hot stars (O, B stars) in a galaxy can contribute some ionizing photons (at $\lambda < 912 {\AA}$, namely Lyman continuum, hereafter Lyc), which may reduce the neutral fraction in the galactic halo.
Here we make a simple estimation about the effect of extra photoionization by star formation in the galaxy itself.

Stellar synthesis models suggest that the number of ionizing photons emitted from a galaxy in a unit time ($Q(H^0)$) is related to the star formation rate (SFR) by
\[{\rm SFR}=1.08 \left(\frac{Q(H^0)}{10^{53}\second^{-1}}\right) M_{\odot} \yr^{-1}
\]
(see Kennicutt 1998 and references therein).
Of the Lyc photons, only $\sim 14$ per cent can escape the OB association
and enter the diffuse ionized medium (``Reynolds layer'') above the galaxy disk (Dove \& Shull 1994b). 
About 65 per cent of the escaping photons are not absorbed in the H{\small II} layer and either escape from the galaxy or are absorbed by additional gas at high latitude (Dove \& Shull 1994b).
Thus approximately 9 per cent of the Lyc photons can reach the outer halo. 
However, Leitherer et al. (1995) observed a sample of 4 starburst galaxies and concluded that less than 3 per cent of the ionizing photons can escape. 
As an approximation, we use an escaping fraction $f_{esc}=0.05$.
The rate of the escaping Lyc photons is then $f_{esc} Q(H^0)$. 
The ionizing photon flux at galactic distance $R$ is
\[
\Phi_{Lyc}\simeq 10^{8} f_{esc} Q_{51} R_{\kpc}^{-2} \cm^{-2} \second^{-1},
\]
where $R_{\kpc}=R/{\kpc}$, $Q_{51}=Q(H^0)/(10^{51} \second^{-1})$ and the energy flux
\footnote{If we take the shape of $J_v$ into account, the result should be multiplied by some factor. However because all the estimations here are quite uncertain, we omit this factor.}
is
\[
J_\nu=\frac{\rm h}{4 \pi} \Phi_{Lyc} \simeq \left(\frac{\Phi_{Lyc}}{10^7 \cm^{-2} \second^{-1}}\right) \erg \cm^{-2} {\rm \,sr}^{-1} \second^{-1} {\rm Hz}^{-1},
\]
where h is Planck's constant. In dimensionless form we have
\[
J_{-21} \simeq 10 f_{esc} Q_{51} R_{\kpc}^{-2}.
\]

For a normal spiral galaxy like the Milky Way Galaxy, with a SFR of $1 \sim 2 M_{\odot} \yr^{-1}$, $Q_{51}$ is about 100, 
and $J_{-21}=50 R_{\kpc}^{-2}$. At a distance of 10 kpc, $J_{-21}=0.5$, which is larger than what we have used ($J_{-21}=0.05$ at $z=0$) by an order of magnitude. At a distance of 30 kpc, $J_{-21}=0.05$, which is comparable to what we have used.
At a distance of 100 kpc $J_{-21}=0.005$.
For most of the halo clouds in our models, the typical distances to the galaxy center are larger than 30 kpc, so the additional photoionization flux by the galaxy itself may be negligible. 
This is also true for the satellites, because they are located at even larger distances.
The typical absorption radius is about 100 kpc. If we make the extreme assumption that all clouds within 30 kpc are fully ionized, the number of lines would be reduced by about 10 per cent.
For early type galaxies, the SFR is lower and so the local photoionization can be neglected.
The situation may be different for starburst galaxies with SFR bigger than $\sim 10 M_{\odot} \yr^{-1}$, but the number density of starburst galaxies is small and furthermore, the rate of emission of ionizing photons drops quickly in a short period of time (see Nulsen, Barcons, \& Fabian 1998 for discussion).
To estimate the effect we have made simulations of the model F3 but with an ionizing flux increased by a factor of 2.
The value of $\frac{dN}{dz}$ for strong lines is reduced to $\sim 8.8$ from 11.9.
We therefore expect that our results will not be affected significantly by the local ionizing sources.


\subsection{Properties of absorbing galaxies}

\begin{figure*}
\begin{minipage}{178mm}
\centerline{\psfig{figure=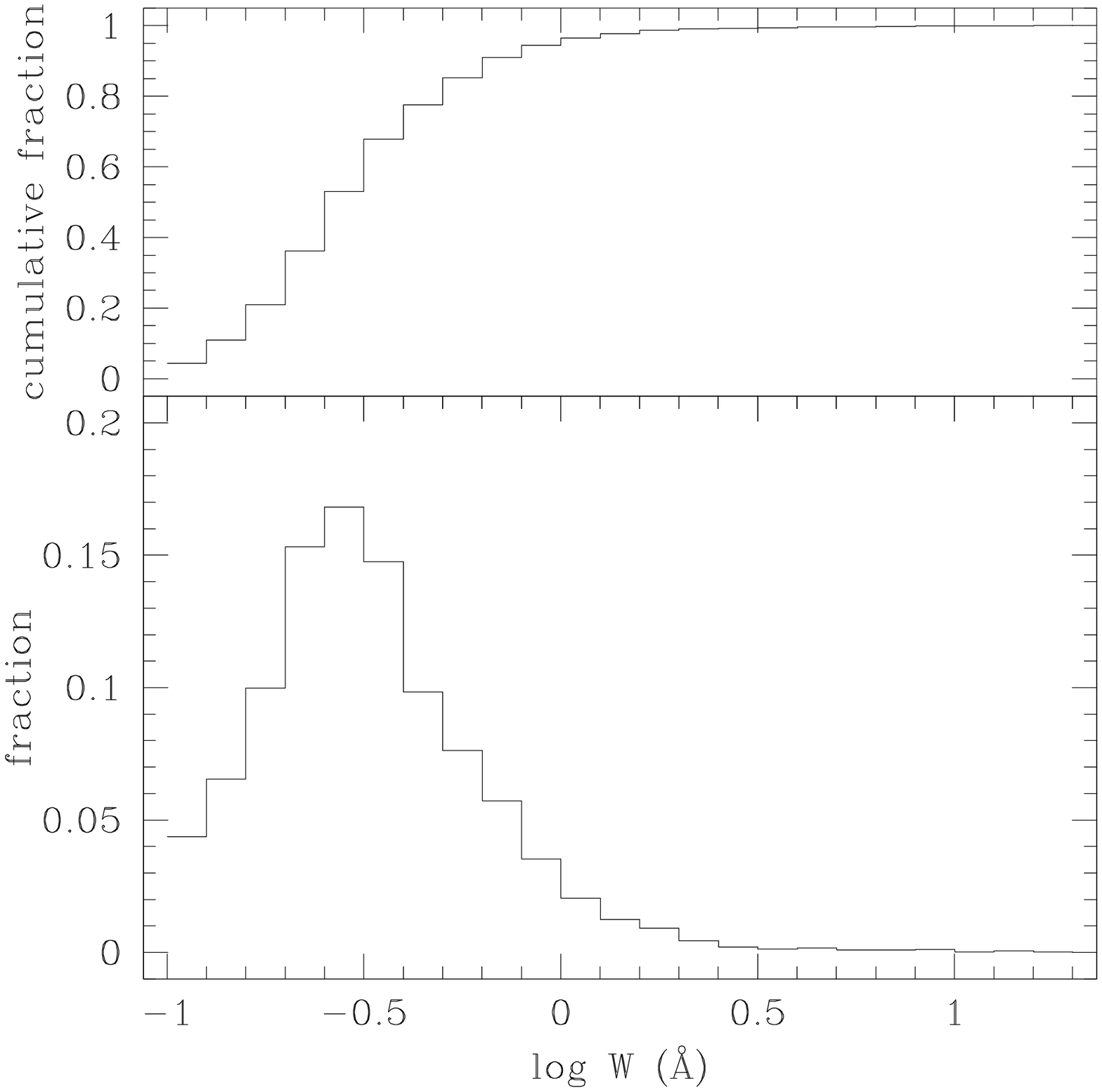,height=5.5cm,width=8.5cm}\psfig{figure=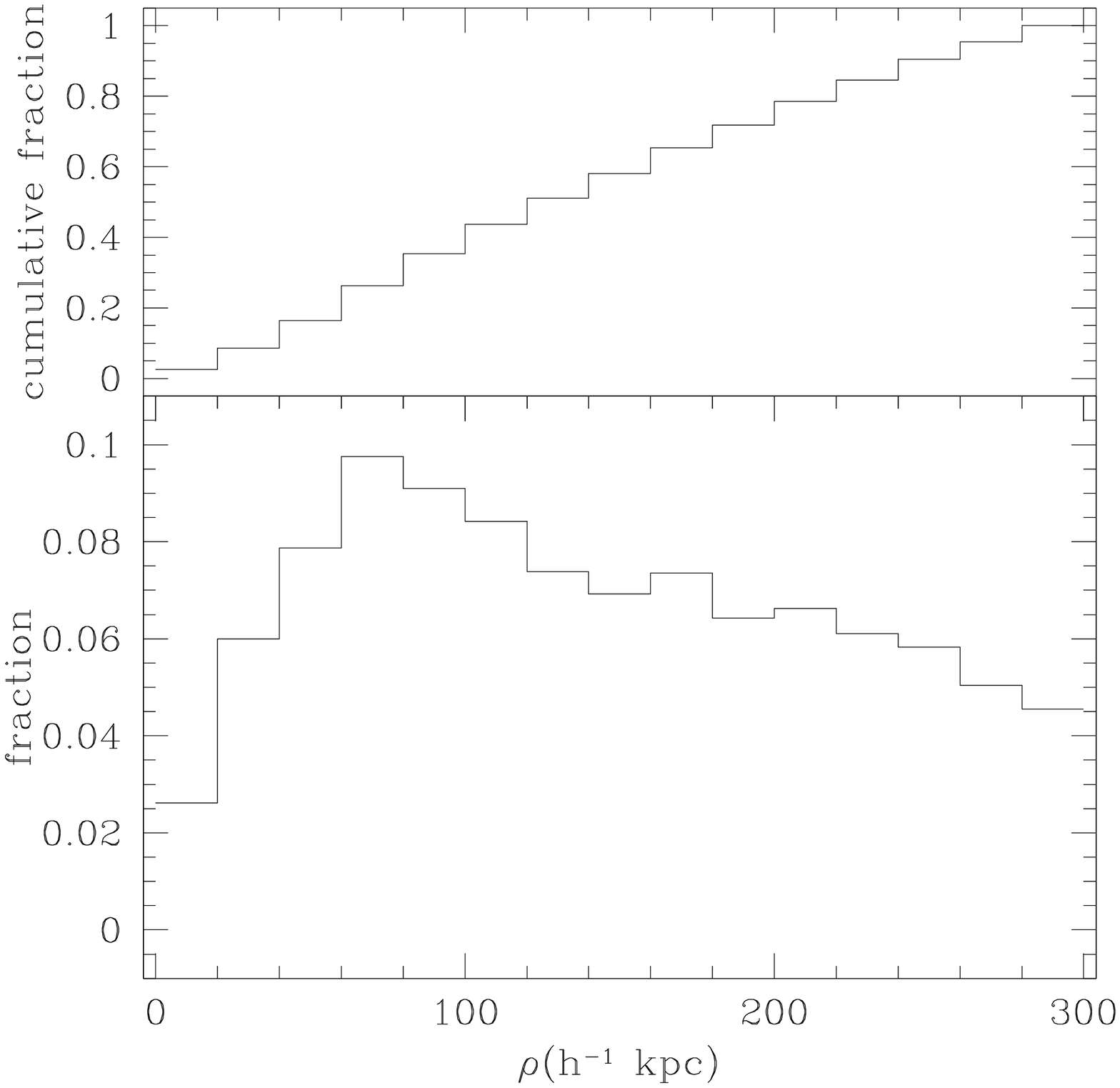,height=5.5cm,width=8.5cm}}
\centerline{\psfig{figure=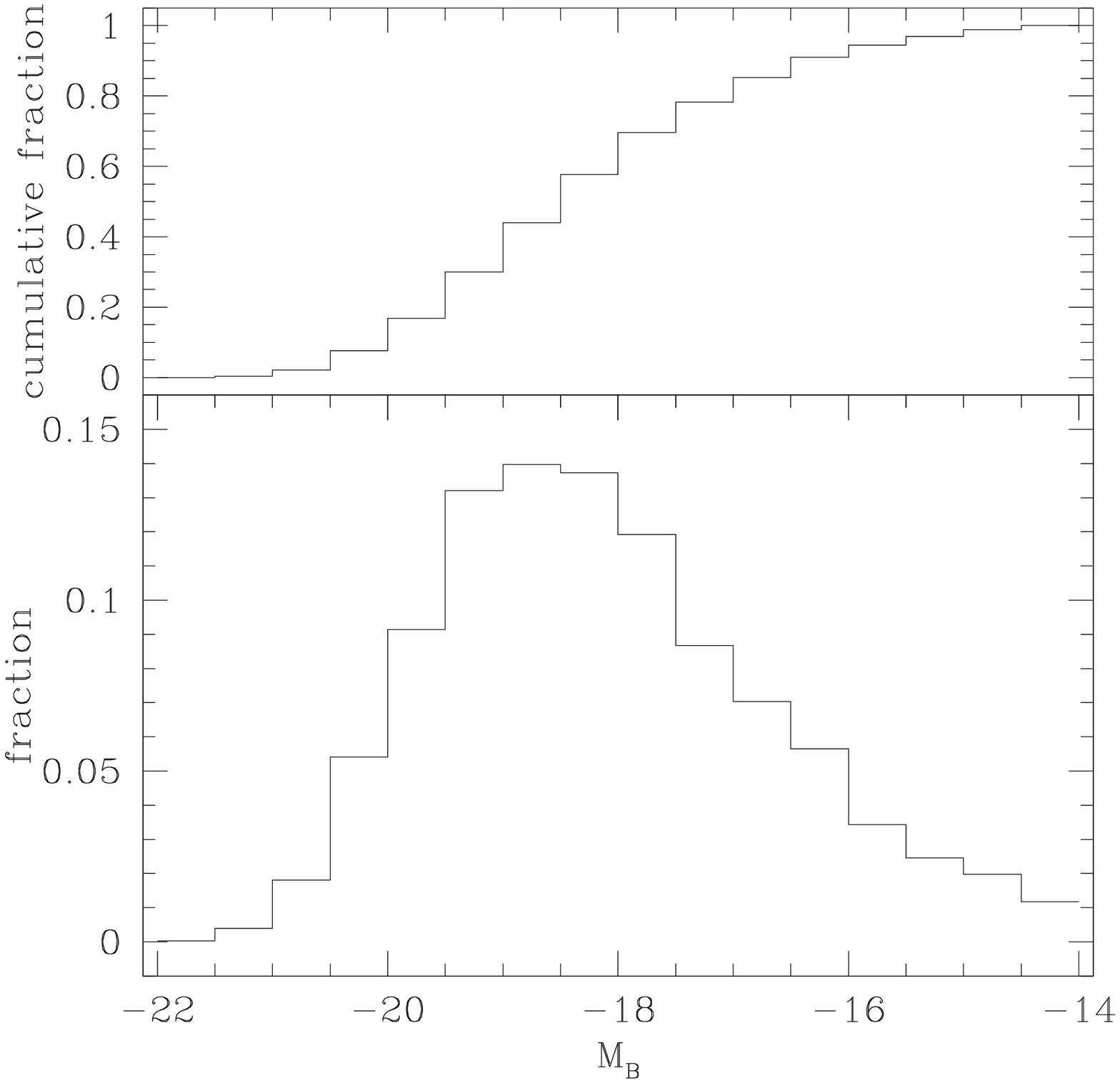,height=5.5cm,width=8.5cm}\psfig{figure=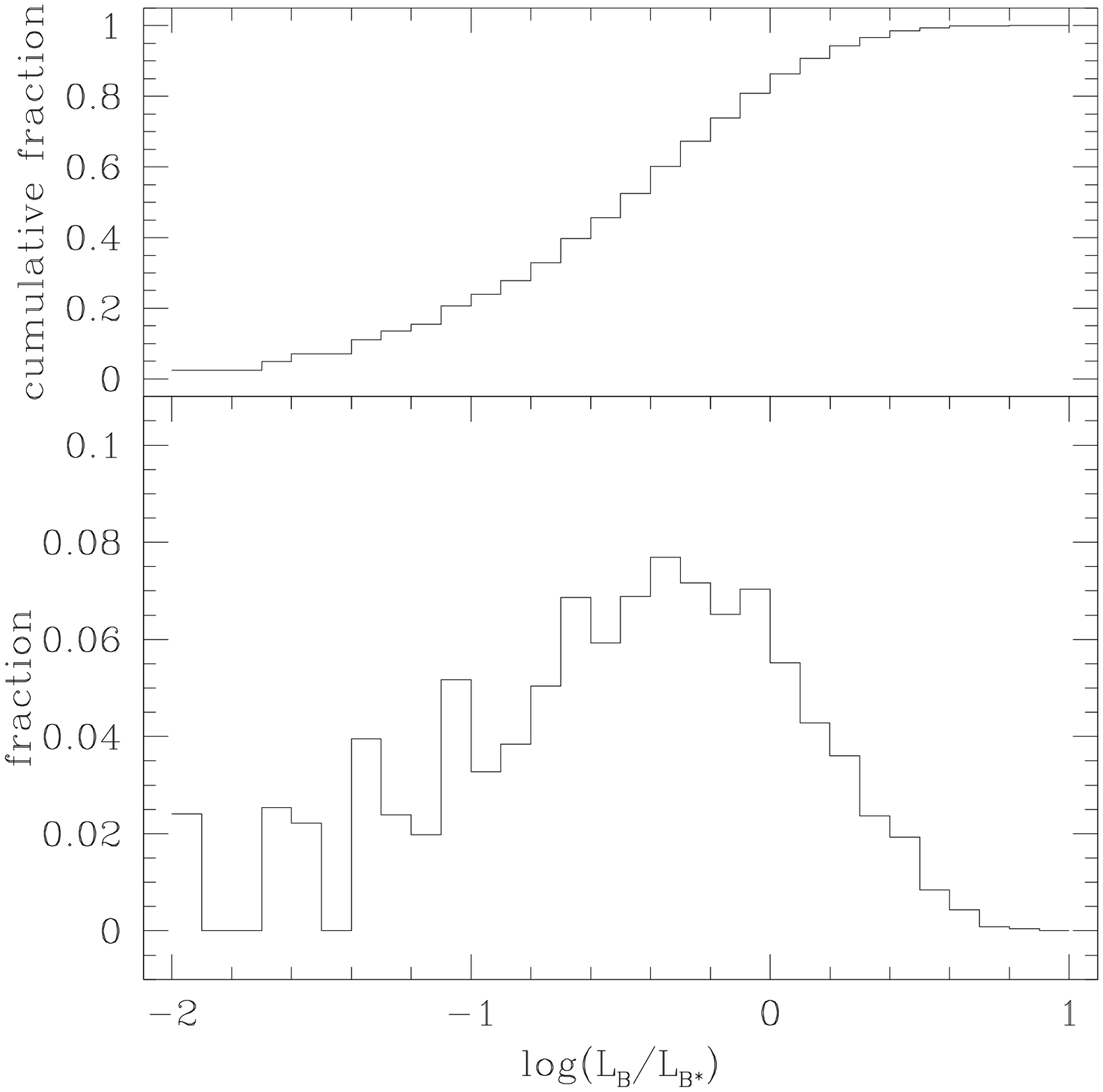,height=5.5cm,width=8.5cm}}
\centerline{\psfig{figure=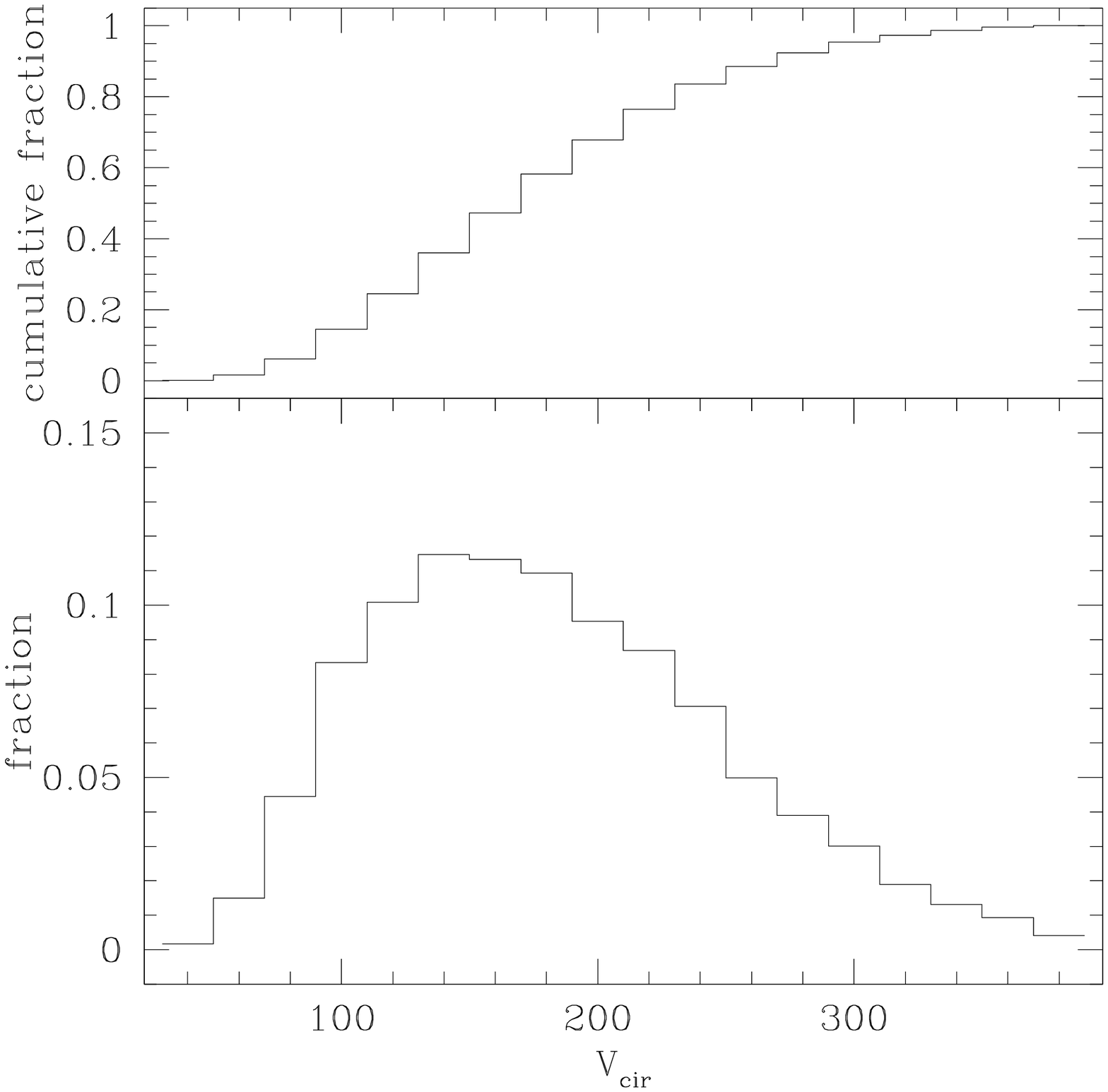,height=5.5cm,width=8.5cm}\psfig{figure=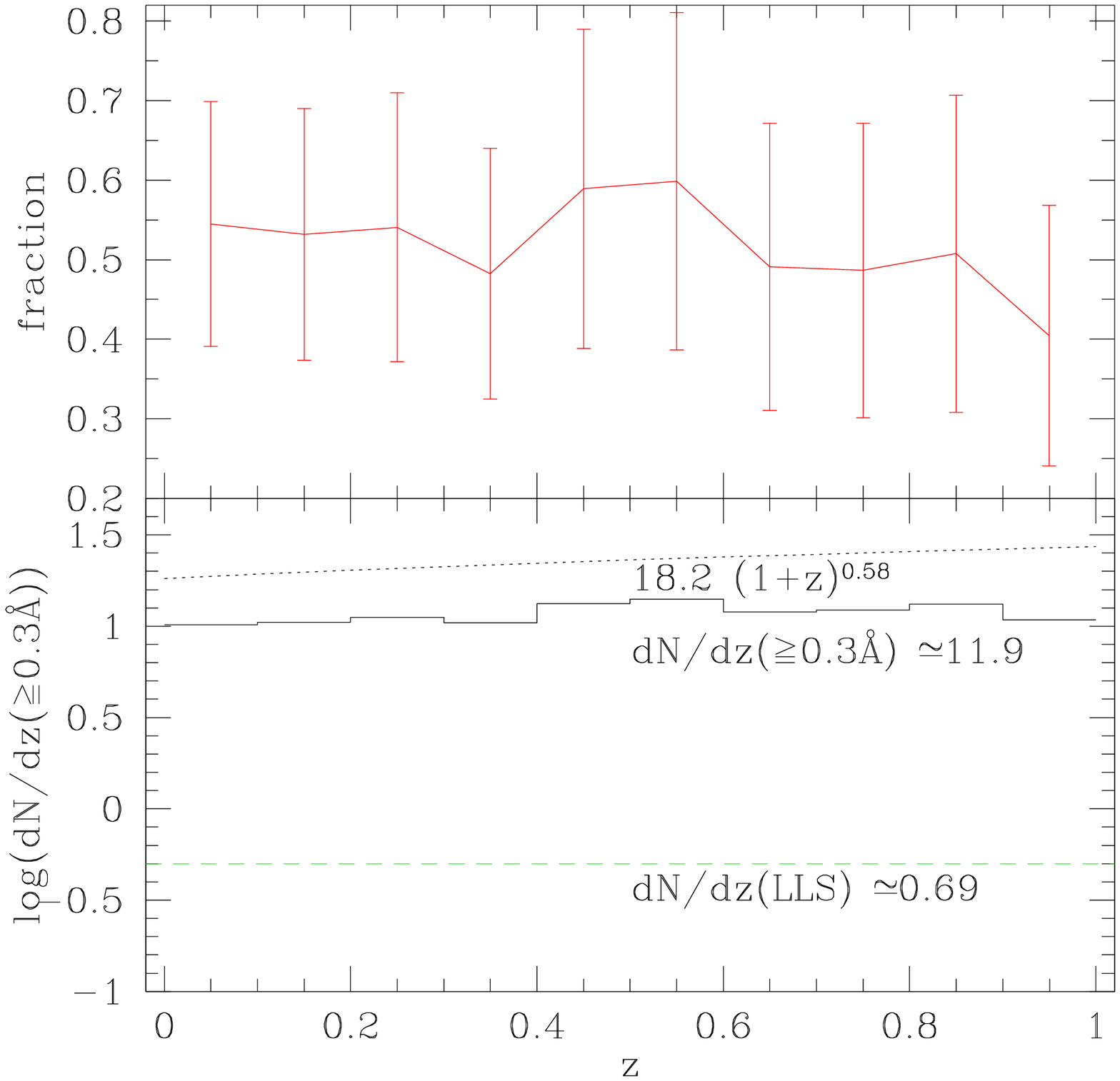,height=5.5cm,width=8.5cm}}
\caption{The distributions of equivalent line widths, projected distance, absolute magnitudes, circular velocity and redshifts, for model F3. In the lowest-right panel, the dotted line represents
the results of {\QALKP} (Bahcall et al. 1996), which is $\frac{dN}{dz}=(18.2\pm5.0)(1+z)^{0.58}$. The error bar is due to the uncertainty of the {\em HST} result. The number of lines per unit redshift
 for Lyman-limit systems is also shown as a dashed line.
}
\label{FigTest}
\end{minipage}
\end{figure*}

Our simulations provide additional information about the galaxy/absorber systems. 
For example, Fig.\ref{FigTest} presents the distribution histograms of the projected distances, equivalent line widths, luminosities, absolute magnitudes, circular velocities and redshifts of absorbing galaxies for the standard model F3. The results are summarized as follows:

\begin{enumerate}

\item The REW distribution has a peak near 0.3{\AA}, the majority of REWs are between 0.1{\AA} and 1.0{\AA}.

\item About 70 per cent of projected distances are between $20 h^{-1}\,\kpc$ and $200 h^{-1}\,\kpc$, only 10 per cent are at projected distances less then $20 h^{-1}\,\kpc$ and 20 per cent are at projected distances larger then $200 h^{-1}\,\kpc$.

\item About 70 per cent of the luminosities are between $0.1 L_{B*}$ and $1.0 L_{B*}$, which implies absorbing galaxies are relatively luminous galaxies in the model.

\item At least 80 per cent of the circular velocities are between $100 \km\,{\second}^{-1}$ to $300 \km\,{\second}^{-1}$, the fraction of absorbers with circular velocities less than $100 \km\,{\second}^{-1}$ is only about 15 per cent.

\item The number density for LLS is about 0.69 which is in good agreement with observations (Stengler-Larrea et al. 1995).

\item The distribution with redshift is flat and the average $n(z)$ for absorbers with $W_r \geq 0.3 {\rm \AA}$ is about 11.9 which can account for about 55 per cent of the sources observed by the {\em HST}.
\end{enumerate}

Note that in the lowest right panel of Fig.\ref{FigTest}, the predicted number density is almost independent of redshift.
This result allows us to average the number density over the whole redshift interval in Table~\ref{C}. 
In observation, the number density of strong \Lya line evolves slowly with redshift at low redshift ($z<1$).
It is valuable to investigate the correlations of REW versus projected distance, galaxy luminosity, galaxy/absorber redshift and study the fractions of absorbers produced by different morphological types of galaxies.

\subsubsection{Projected distance}

\begin{figure*}
\begin{minipage}{178mm}
\centerline{\psfig{figure=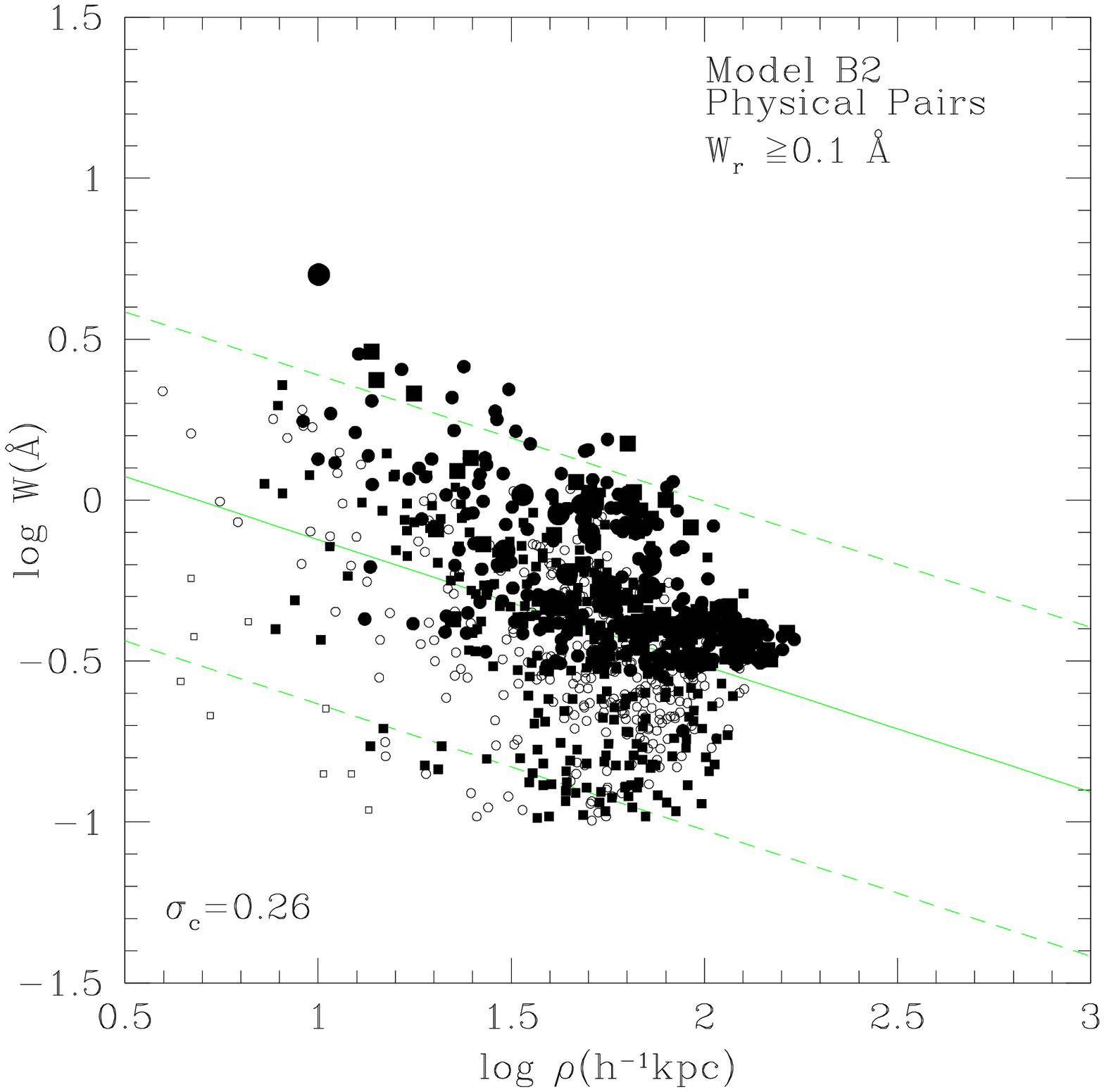,width=8.5cm} \psfig{figure=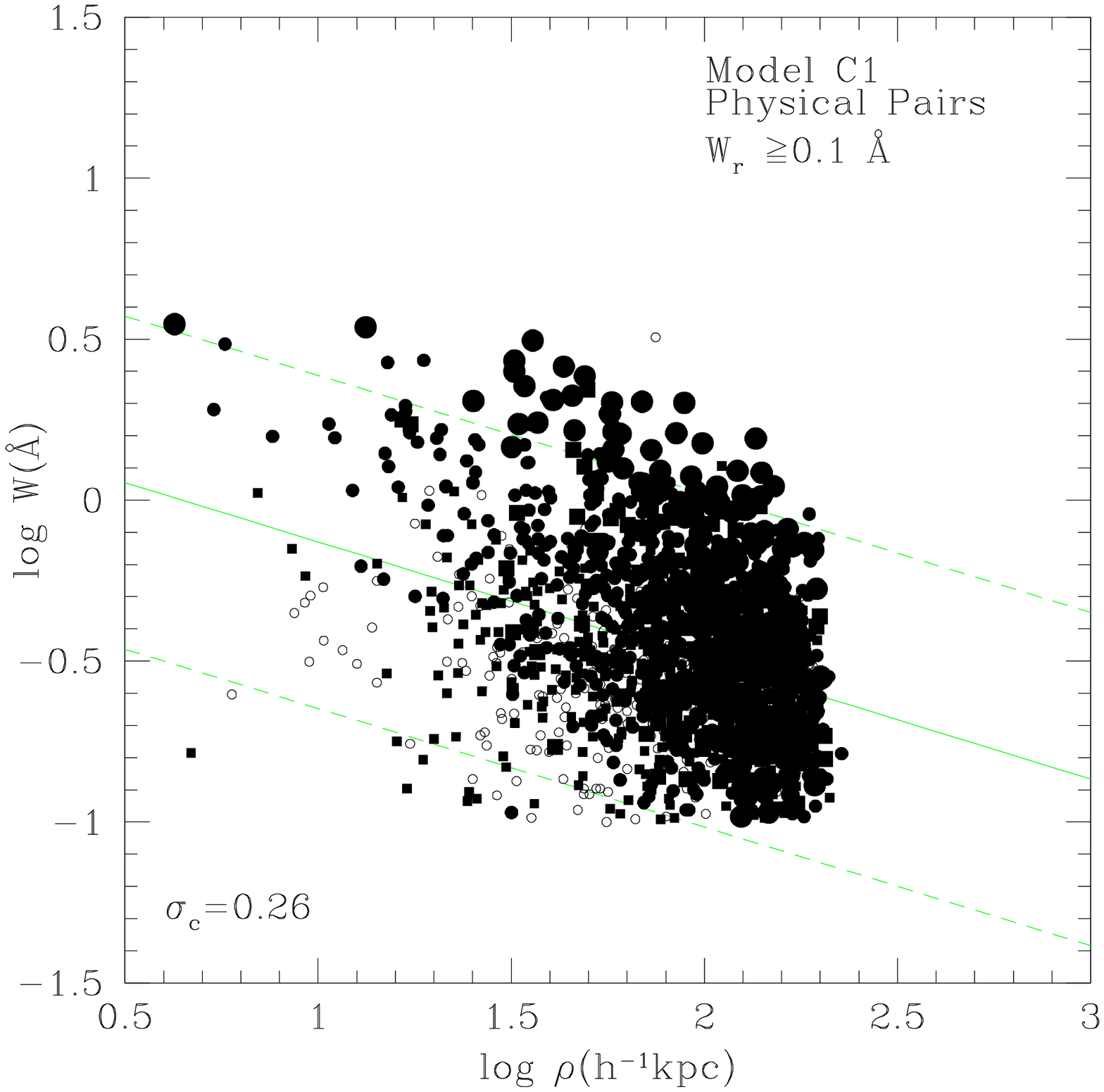,width=8.5cm}}
\centerline{\psfig{figure=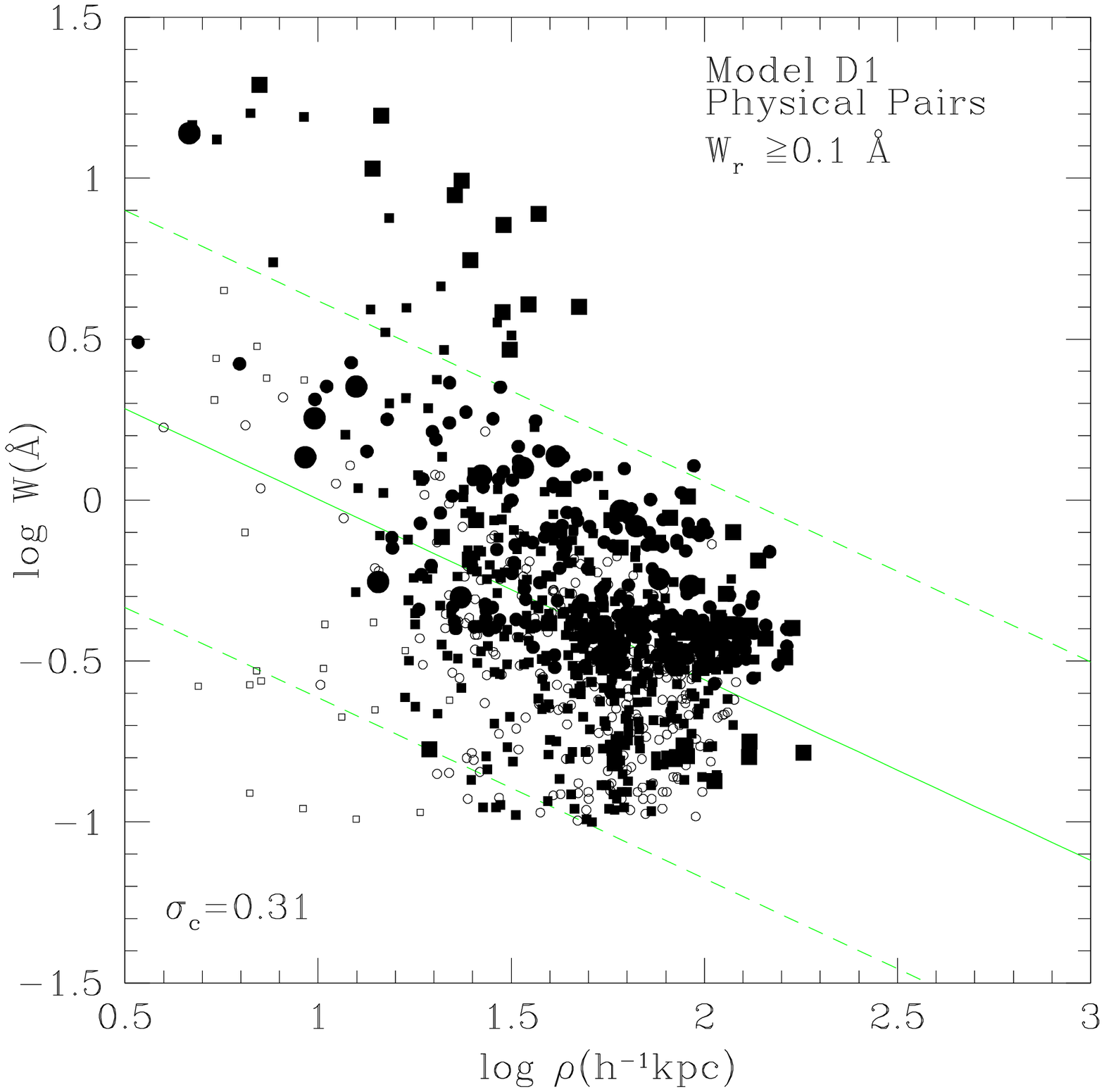,width=8.5cm} \psfig{figure=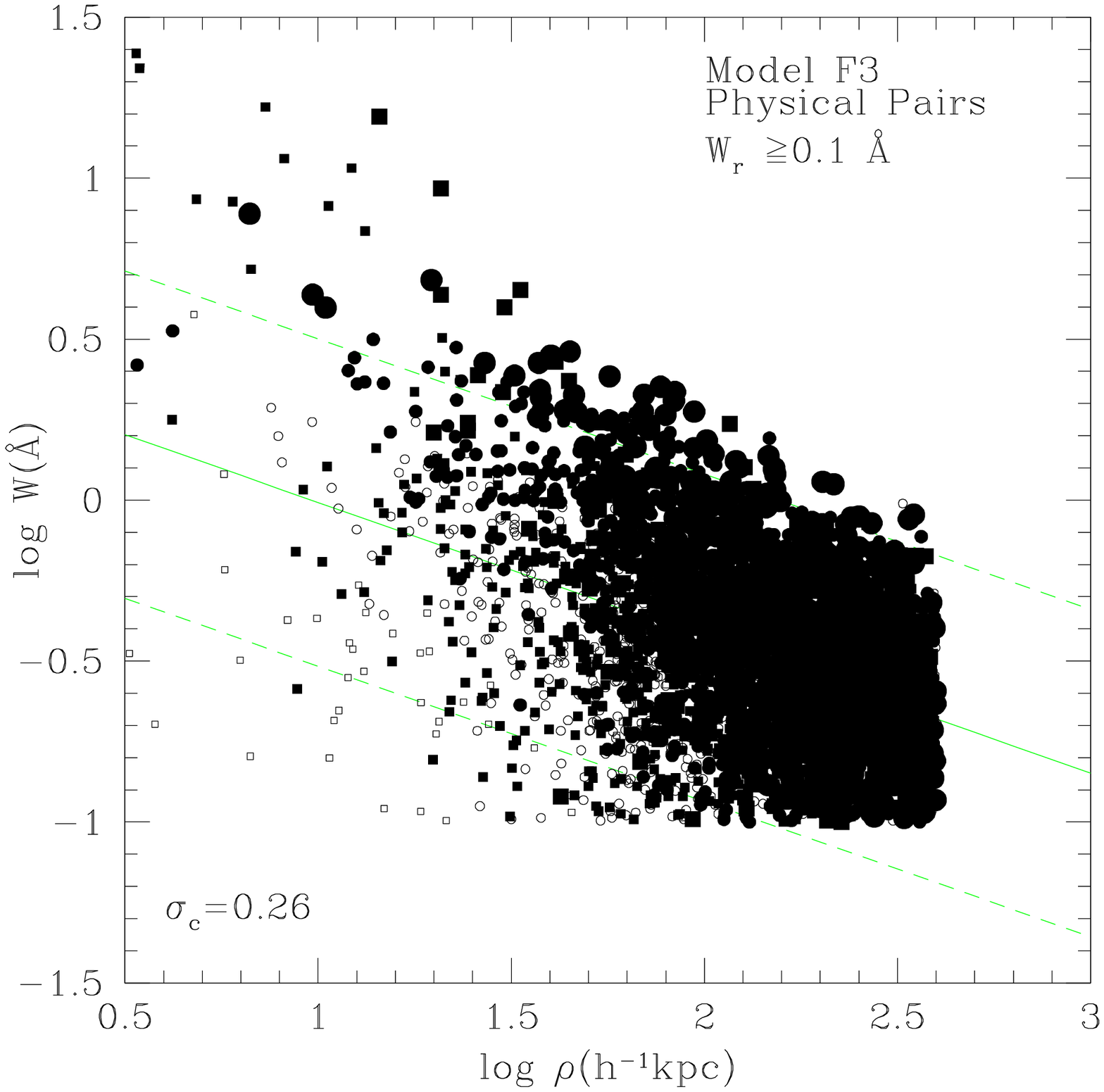,width=8.5cm}}
      \caption{Example Model B2, C1, D1, F3 ($\Lambda$CDM):
The solid lines in both panels are for the linear fit $\log W_r= {\rm C} -\alpha \log\rho$.
In each panel, about 95.4 per cent of the data points lie between the two dashed lines drawn with $2\sigma_c$ vertical shifts.
The squares and circles are for spiral galaxies and elliptical/S0 galaxies respectively.
Small open symbols represent galaxies of luminosity $L_B \leq 0.1 L_{B*}$,
small filled symbols represent galaxies of luminosity $0.1 L_{B*}< L_B \leq L_{B*}$,
large filled symbols represent galaxies of luminosity $L_B > L_{B*}$.
}
\label{FigB2}
\end{minipage}
\end{figure*}

We investigate the anti-correlation between REW, $W_r$ and projected distance of LOS to galaxy centre, $\rho$.
A power-law relation is adopted,
\beq
\log W_r =- \alpha \log \rho + C,
\label{eq-linfit}
\eeq
where $\alpha$ is the slope and $C$ is constant. The results are summarized in Table~\ref{C}.
In the table, we list results for typical models B2, C1, C3, D1, F1, F3 (the standard model), and F5.
In model Fs, we still get an apparent anti-correlation and a slope of $\sim$ -0.4.
Note that in all models the majority of absorption lines by halo clouds and/or satellites definitely decides the character of the anti-correlation (for example, see Fig.\ref{FigB2}).
The difference in the results found in $r_s$ and $r_p$ is reasonable because these three types of absorption components which do not necessarily have the same pattern of dependence. 
This difference has been noticed before (Le Brun et al. 1996).

We plot the distribution of $W_r$ and $\rho$ for some models. Panels in Fig.\ref{FigB2} are for the results of model B2, C1, D1, F3 respectively.

Small galaxies may have almost no cold gas because gas could be heated by supernova explosions or all the gas may have cooled to form stars, and there is no more gas to accrete. 
Thus these galaxies cannot contribute to absorption. In practice, if we exclude those with $V_{\rm cir} \leq 100\, \km\,{\second}^{-1}$ (as we can see in model F3, 
there are about 80 per cent of the absorbers with circular velocity of $100\, \km\,{\second}^{-1}$ or more), 
there will be a stronger anti-correlation and a steeper slope of the linear fit at a highly significant level for almost all models.
For model Fs, typically the Pearson co-efficient, $r_p$ is $\sim$ -0.5 and the slope is $\sim$ -0.5 at very high significance. This can be seen clearly from the results for `sample b' in Table~\ref{D}.

In summary, our predicted anti-correlation is significant but the dependence of $W_r$ on $\rho$ is not as strong as that of CLWB. 
However, as pointed out by Tripp et al. (1998), selection effects could artificially tighten this anti-correlation. We will discuss this problem in the next section.

\begin{table*}
\begin{minipage}{120mm}
\caption{The dependence of REW on projected distance}
\label{D}
\begin{tabular}{@{}lccccc}\hline\hline
{Model}& c & $\alpha$ & $r_s$ (SL$^c$)& $r_p (SL)$ & $\sigma_c$ \\
B2 sample a$^a$ & {$~.29\pm.05$}&{$.40\pm.03$}&-.36 (11.8)&-.42 (13.8)& 0.26 \\
B2 sample b$^b$ & {$~.66\pm.05$}&{$.60\pm.03$}&-.48 (15.4)&-.57 (19.8)& 0.23 \\ 
C1 sample a & { $.24\pm.04$}&{$.37\pm.02$}&-.33 (14.2)&-.37 (16.4)&  0.26\\
C3 sample a & { $.06\pm.03$}&{$.27\pm.01$}&-.28 (18.0)&-.34 (22.5)&  0.24\\ 
D1 sample a & { $.56\pm.06$}&{$.56\pm.03$}&-.35 (11.6)&-.47 (16.3)& 0.31 \\
D1 sample b & { $.97\pm.06$}&{$.77\pm.04$}&-.45 (14.8)&-.59 (21.1)& 0.28 \\
F1 sample a & { $.54\pm.04$}&{$.51\pm.02$}&-.42 (20.7)&-.46 (23.4)& 0.27 \\
F1 sample b & { $.95\pm.05$}&{$.70\pm.02$}&-.49 (22.1)&-.57 (29.1)& 0.26 \\
F3 sample a & { $.42\pm.03$}&{$.42\pm.01$}&-.38 (26.5)&-.48 (35.9)& 0.26 \\
F3 sample b & { $.65\pm.03$}&{$.52\pm.01$}&-.43 (28.9)&-.55 (40.0)& 0.25 \\ 
F5 sample a & { $.43\pm.02$}&{$.43\pm.01$}&-.44 (32.2)&-.50 (38.6)& 0.25  \\
F5 sample b & { $.69\pm.03$}&{$.53\pm.01$}&-.50 (36.5)&-.58 (45.0)& 0.24  \\ \hline
\end{tabular}

$^a$ In sample a, we include those galaxy-absorber pairs with $W_r \ge 0.1${\AA}.\\
$^b$ In sample b, we exclude those galaxy-absorber pairs with circular velocity of central galaxies $V_{\rm cir} < 100\, \km\, {\second}^{-1}$ and $W_r < 0.1${\AA}.\\
$^c$ The statistical significance level, which is equal to $r \sqrt{\frac{N-2}{1-r^2}}$, where $r$ is co-efficient and $N$ is number of points.
\end{minipage}
\end{table*}

\subsubsection{Galaxy luminosity}

As suggested by CLWB, there is a power-law relationship between $W$ and $\rho$ and $L_B$,
\beq
\log W_r = - \alpha \log \rho + \beta \log L_B + C.
\label{eq-linfit2}
\eeq
We apply the analysis to model F1, F3 and F5. The results are listed in Table \ref{E}. For example, for model F1, the analysis for sample b yields, $C=1.21$, $\alpha=0.79$, $\beta=0.19$.
We can determine the absorption radius of a galaxy $r \propto L_B^t$ ($t=\beta/\alpha$) with $t \sim 0.24-0.32$. 
The value of $t$ is comparable but smaller than 0.37 which was derived by CLWB. However it is similar to that derived from Mg{\small II} obsorbers (Bergeron \& Boiss\'e 1991; Bergeron et al. 1992; Le Brun et al. 1993; Steidel 1995).
Again, selection effects could lead to misleading conclusions (see discussion in \S~\ref{secImpact}). 

\begin{table*}
\begin{minipage}{120mm}
\caption{REW dependence on projected distance and luminosity}
\label{E}
\begin{tabular}{lcccccc}\hline\hline
{Model}& c & $\alpha$ & $\beta$ &  $r_s (SL)$& $r_p (SL)$ & $\sigma_c$ \\
F1 sample a & $~.89\pm.04$ &$~.63\pm.02$&$~.20\pm.01$ &-.53 (27.8)&-.59 (33.1)&.25\\ 
F1 sample b & $1.21\pm.05$ &$~.79\pm.02$&$~.19\pm.01$ &-.58 (29.7)&-.65 (36.4)&.24\\ 
F3 sample a & $~.62\pm.03$ &$~.48\pm.01$&$~.15\pm.01$ &-.46 (31.0)&-.57 (44.4)&.24\\ 
F3 sample b & $~.81\pm.03$ &$~.57\pm.01$&$~.14\pm.01$ &-.49 (34.4)&-.61 (46.5)&.23\\ 
F5 sample a & $~.66\pm.02$ &$~.50\pm.01$&$~.15\pm.01$ &-.51 (39.2)&-.58 (47.4)&.24\\ 
F5 sample b & $~.87\pm.03$ &$~.59\pm.01$&$~.14\pm.01$ &-.55 (41.9)&-.63 (51.6)&.23\\ \hline
\end{tabular}
\end{minipage}
\end{table*}

\subsubsection{Absorber redshift}

We also analyse the dependence of the line equivalent width on absorber redshift assuming
\beq
\log W_r = - \alpha \log \rho - \gamma \log (1+z) +C,
\label{eq-linfit3}
\eeq 
and
\beq
\log W_r = - \alpha \log \rho + \beta \log L_B - \gamma \log (1+z) +C.
\label{eq-linfit4}
\eeq
We apply the analysis to model F1, F3 and F5. The results are listed in Table \ref{F} and Table \ref{G}.
In summary, the relationship between REW and projected distance together with absorber redshift is marginally superior (with larger $|r_p|$ or $|r_s|$) to the relationship between REW and the projected distance 
but marginally inferior (with smaller $|r_p|$ or $|r_s|$) to the relationship between REW and projected distance accounting for $L_B$, 
and the anti-correlations between REW and projected distance accounting for $L_B$ together with $z$ is superior to the relationship between REW and projected distance accounting for $L_B$.
For the analysis of eq.(\ref{eq-linfit4}), we have $\alpha \sim 0.48-0.80$, $\beta \sim 0.14-0.21$, $\gamma \sim 0.49-0.61$. 
This means the absorption radius of a galaxy $r \propto L_{B}^t (1+z)^{-u}$ ($t=\beta/\alpha$, $u=\gamma/\alpha$) with $t \sim 0.24-0.32$ and $u \sim 0.76-1.02$.
Our result of dependence on absorber redshift is different with the result of CLWB. 
CLWB concluded that REW do not depend on absorber redshift. 
However as we will discuss below, selection effects should be considered in the imaging and spectroscopic survey and the total redshift interval of LOS in observation may be not large enough to determine the relations.

\begin{table*}
\begin{minipage}{120mm}
\caption{REW dependence on projected distance and redshift}
\label{F}
\begin{tabular}{lcccccc}\hline\hline
{Model}& c & $\alpha$ & $\gamma$ & $r_s (SL)$& $r_p (SL)$ & $\sigma_c$ \\
F1 sample a & $~.62\pm.04$ &$~.51\pm.02$&$.46\pm.07$ &-.44 (22.2)&-.48 (24.4)&.27\\ 
F1 sample b & $1.03\pm.05$ &$~.71\pm.02$&$.49\pm.07$ &-.51 (25.3)&-.58 (30.2)&.26\\ 
F3 sample a & $~.49\pm.03$ &$~.42\pm.01$&$.45\pm.05$ &-.41 (29.2)&-.50 (37.5)&.25\\ 
F3 sample b & $~.74\pm.03$ &$~.52\pm.01$&$.48\pm.05$ &-.46 (31.4)&-.57 (42.1)&.24\\ 
F5 sample a & $~.52\pm.02$ &$~.43\pm.01$&$.50\pm.04$ &-.47 (35.1)&-.52 (40.7)&.25\\ 
F5 sample b & $~.78\pm.03$ &$~.54\pm.01$&$.50\pm.05$ &-.53 (39.2)&-.60 (47.0)&.24\\ \hline
\end{tabular}
\end{minipage}
\end{table*}

\begin{table*}
\begin{minipage}{120mm}
\caption{REW dependence on projected distance, luminosity and redshift}
\label{G}
\begin{tabular}{lccccccc}\hline\hline
{Model}& c & $\alpha$ & $\beta$ & $\gamma$ & $r_s (SL)$& $r_p (SL)$ & $\sigma_c$ \\
F1 sample a & $1.01\pm.04$ &$~.65\pm.02$&$~.21\pm.01$ &$.56\pm0.06$&-.55 (29.6)&-.61 (34.7)&.25\\ 
F1 sample b & $1.34\pm.05$ &$~.80\pm.02$&$~.20\pm.01$ &$.61\pm0.07$&-.60 (31.7)&-.67 (38.3)&.23\\ 
F3 sample a & $~.71\pm.03$ &$~.48\pm.01$&$~.15\pm.01$ &$.49\pm0.04$&-.49 (36.3)&-.58 (46.3)&.24\\ 
F3 sample b & $~.90\pm.03$ &$~.57\pm.01$&$~.14\pm.01$ &$.51\pm0.05$&-.52 (37.0)&-.62 (48.5)&.23\\ 
F5 sample a & $~.75\pm.03$ &$~.50\pm.01$&$~.15\pm.01$ &$.51\pm0.04$&-.53 (41.9)&-.60 (49.6)&.23\\ 
F5 sample b & $~.96\pm.03$ &$~.59\pm.01$&$~.14\pm.01$ &$.51\pm0.04$&-.58 (44.6)&-.65 (53.8)&.22\\ \hline
\end{tabular}
\end{minipage}
\end{table*}

\subsubsection{Covering factor}
From eq. (\ref{eq-linfit4}), and the results in Table \ref{G}, we can estimate the average absorption radius and covering factor from redshift of 0 to 1. 
For example, for `sample a' (see notation in Table \ref{D} for its meaning) of model F1, absorbers with
REW larger than 0.3{\AA} follow the relation
\beq
\frac{r}{r_*} = \left(\frac{L_B}{L_{B*}}\right)^t (1+z)^{-u},
\eeq
where $r_*=0.23 h^{-1} \Mpc$ and $t=0.32$, $u=0.86$. Thus we can calculate the total number by integration
\[
N_{\rm total}=\int^{1}_{0}\pi {r_*}^2 \frac{c (1+z)^{2-2u}}{H(z)} C_l(z) dz 
\]
\beq
~~~~~~~ \times \,\phi^{*} \Gamma(1+2t-s,L_{B_{\rm min}}/L_{B*}),
\eeq
where $C_l(z)$ is the covering factor and $\Gamma$ is the incomplete gamma function. Note that for these absorbers with $W_r \geq 0.3$ {\AA} within $r_*$, it is not necessary that the covering factor is always larger than one,
that is, for a LOS with a large projected distance it is not always possible to find an absorber with $W_r \geq 0.3$ {\AA}.
If we choose $s=1.1$, $L_{B_{\rm min}}/L_{B*}=0.007$, $\phi^{*}=0.027 h^3 \Mpc^{-3}$ then we get $N_{\rm total}=20.4 \times F(z)$, where $F(z)=\int^{1}_{0} \frac{(1+z)^{0.28}}{\sqrt{\Lambda_0+\Omega_0(1+z)^3}} C_l(z) dz$.
Comparing with the total number listed in Table \ref{C}, $N_{\rm total}$ should be 7.3, the average value of $F(z)$ should be about 0.36. 
However the integration $\int^{1}_{0} \frac{(1+z)^{0.28}}{\sqrt{\Lambda_0+\Omega_0(1+z)^3}}dz$ is about 0.85. 
Thus the average covering factor within $230 h^{-1} \kpc$ should be $\sim$ 0.42 if $C_l(z)$ can be treated roughly as a constant. 
The effective gas absorption radius should be $\sqrt{0.42}\times r_* \sim 150 h^{-1} \kpc$.
Therefore the covering factor within $250 h^{-1} \kpc$ is $(150/250)^2 \sim 0.36$.

Our predicted covering factor is in good agreement with the LBTW paper. 
In that paper, almost every galaxy with $\rho < 70 h^{-1}\kpc$ gives rise to absorption, about five of 10 galaxies with $70 h^{-1}\kpc < \rho < 160 h^{-1}\kpc$ give rise to absorption, and 
just one of 9 galaxies with $\rho > 160 h^{-1}\kpc$ give rise to absorption. 
Thus the covering factor within $250 h^{-1}\kpc$ is about 0.31.
However our predicted effective absorption radius ($\sim 150 h^{-1}\kpc$) is a bit smaller than the $174 h^{-1} \kpc$ derived by CLWB. The reason of this difference could be the large covering factor used in CLWB.
Independently, Bowen et al. (1996) derive a covering factor of 0.50 within $\rho < 160 h^{-1} \kpc$  (three of six galaxies give rise to absorption).

\subsubsection{Galaxy morphological type}

CLWB conclude that galaxies which produce \Lya absorption systems span a wide range of morphological types from elliptical or S0 galaxies through late-type spiral galaxies. 
Consistently the models show every type of galaxy can produce \Lya absorption systems. 
The predicted fractions of absorbing galaxies in spiral, S0, elliptical galaxies are defined as f1, f1, f3 respectively and listed in Table \ref{C}. 
As we can see, the fractions vary from model to model. In model B and D, about half of the absorbers arise in spiral galaxies. 
In model C, this fraction is only about one third because E/S0 galaxies possess more satellites than spirals. 
In model F, about 35 per cent of the absorbers are produced by spirals.
It is, however, not clear, if there are many clouds inside haloes of ellipticals/S0 galaxies and how many absorbing systems can arise from tidal tails related with spirals. 
If we exclude absorption by halo clouds of ellipticals and S0 galaxies, for instance, comparing models B2 and F1, the fraction by spirals is estimated to be $(7.3\times 38 \%)/(7.3-3.7\times 54 \%)\simeq 52$ per cent. 
A crude estimate using Fig.2 of CLWB shows that this number is about 70 per cent, however the information on galaxy morphology in observations may be inadequate to tackle this problem.
We suggest that these numbers should be investigated further to see whether the gas absorption sections of different types of galaxies are the same or not.

\section{Selection effects: Mock Imaging And Spectroscopic Surveys}
\label{secMO}
In section 2 we have studied $(\frac{dN}{dz})$ and the overall properties of absorbers limited only by the lowest line width of 0.1{\AA}, but without considering selection criteria.
In carrying out comparisons with results of imaging and spectroscopic surveys, it is absolutely necessary to
construct absorber catalogues with observational selection criteria and investigate the possibility of mis-identification of absorbers (i.e., optically unseen absorbers mis-matched with a bright neighbour galaxy). 

\subsection{Selection criteria of absorber-galaxy pairs}

Our galaxy-absorber pair selection criteria are similar to those of CLWB. We only select absorbers with a rest frame line width
\beq
W_r \geq 0.1 {\rm \AA},
\eeq
and with the B band luminosity satisfying
\beq
m_B \leq 24.3.
\eeq
Similar to LBTW, we only select those galaxies within angular distances 
\footnote{See LBTW's Fig.2. for the variation of projected distance threshold with redshift for an angular distance threshold $\theta=1'3$.} 
to the QSOs satisfying 
\beq
\theta \leq 1'3.
\eeq
A small velocity separation between absorber and galaxy centre, $v=|c z_{\rm gal}-c z_{\rm abs}|< 500\, \km\, {\second}^{-1}$, is required to relate an absorber with a luminous galaxy (Lanzetta et al. 1997; CLWB; Bowen et al. 1996). 
This small velocity separation excludes almost all random galaxy-absorber pairs.


\subsection{Mis-identification}

LBTW argued that it is unlikely that the absorbing gas arises in dwarf companions of the luminous galaxies 
because no such dwarf galaxy was found in the LOS toward QSOs at redshifts $z <0.2$ down to a luminosity of $\approx 0.05 L_{*}$
\footnote{In our models, a spiral galaxy with $L_B \approx 0.05 L_{B*}$ may have $V_{vir} \approx 55 \km {\second}^{-1}$ and $m_B \approx 23.2$ at $z \approx 0.2$ in a $\Lambda$CDM cosmology, for example.}.
But van Gorkom et al. (1996) and Hoffman et al. (1998) have located faint dwarf galaxies at the redshifts of a few low-$z$ \Lya lines.
In addition, as we can see in \S\ref{sec.gsample}, the lower limit in the luminosity function can be as low as $\approx 0.007 L_{*}$. 
Satellites are even fainter with typical circular velocity around $30 \km {\second}^{-1}$.
Of course, at very low redshift, these faint satellites can be identified in imaging surveys with a large field of view. 
However, an angular threshold of $1'3$ at $z=0.1$ means in general a distance of $100 h^{-1} \kpc$ so that many of these satellites are excluded from pair samples because they typically have large distances to the central galaxy.

As suggested by Tripp, Lu, \& Savage (1998), the \Lya lines could be due to undetected faint dwarf galaxies that are clustered with the observed luminous galaxies. 
This kind of \Lya lines arise at small projected distances, but the host galaxies are too faint to be identified in galaxy imaging surveys, especially at modest to high redshifts.
These un-seen absorbers (optically uncatalogued galaxies) can cause
erroneous identifications, since one could make 
a mistake to relate the corresponding line with a nearby luminous galaxy at a larger
projected distance. We simulate nearby luminous galaxies around a central galaxy by the two-point correlation function of normal galaxies, which is 
\beq
\xi(r) =\left(\frac{5 h^{-1} \Mpc}{r}\right)^{\gamma},
\eeq
where $r$ is the separation of two galaxies and $\gamma \simeq 1.8$.
Thus the galaxy number in a volume with radius of $h^{-1} \Mpc$ is
\beq
N(r,z) = n_c (1+z)^3 \int^{1}_{0}4\pi r^2 [1+\xi(r,z)] dr,
\eeq
where $n_c$ is the co-moving galaxy density described in \S\ref{sec.gsample}, and $z$ is the redshift of the central galaxy.
The result of the Canada-France Redshift Survey shows a strong redshift evelution of the galaxy two-points correlation function which is
\beq
\xi(r,z)= \xi(r) (1+z)^{-(3+\epsilon)},
\eeq
where $\epsilon \sim 0-2$ at $z<1.3$ (Le F\`evre et al. 1996). Because $\epsilon$ is uncertain observationally, we use $\epsilon=1.5$ for simplicity (Shepherd et al. 1997). Our results are not sensitive to the choice, because the redshift range covered is small.
The luminosity distribution of these galaxies is consistent with the luminosity function and they are distributed around the central galaxy
following $\xi(r,z)$. The apparent magnitude of these galaxies is calculated from eq.~(\ref{mB}).
The distribution of the pairwise velocity is a Gaussian with dispersion $\sigma=400\, \km\, {\second}^{-1}$ (Efstathiou 1996; Mo, Jing \& B\"orner 1993; Jing, Mo \& B\"orner 1998).

We define a galaxy-absorber pair intimately linking the absorber with a bright galaxy (an absorbing galaxy with apparent magnitude brighter than the luminosity limit, i.e. $m_B \leq 24.3$) as a `physically associated pair' or a `physical pair' for simplicity. 
On the contrary a mis-matched galaxy-absorber pair is called `spurious pair'.
The method to find a mis-matched pair is as follows:
When there is a faint absorber (whose apparent magnitude is fainter than the luminosity limit), its neighbours will be simulated to see whether there is a nearby bright galaxy 
(brighter than the luminosity limit) with larger projected distance (however not larger than $400h^{-1} \kpc$). 
For a positive result, this bright neighbour will be chosen to pair with the absorption line arising from the faint absorber and the new projected distance will be chosen. 
For a negative result in the search of a bright neighbour, we classify the galaxy-absorber pair as a `missing pair'.
We define those bright `physical pairs' and `spurious pairs' as `bright pairs'. 
We also call a pair outside a certain angular distance threshold as a `missing pair'. For instance, at very low redshift some `bright pairs' have large angular separations to a QSO LOS.

\subsection{Effect on the $W_r - \rho - L_{B}$ relations}
\label{secImpact}
When comparing the $W_r - \rho - L_{B}$ relations for simulated galaxy/absorber pairs to the observations, the selection effects mentioned above may have impacts on the statistics.
For example, there could be some `spurious' galaxies at large impact parameters within the redshift window of the absorbers. 
In some of the surveys carrried out, the sky area surveyed is so large that there is always one galaxy (by chance) within the 500 $\km \second^{-1}$ of the redshift. 
This may cause mis-identification of the absorbing galaxy and add noise to the correlations.
On the other hand, faint absorbing galaxies without bright neighbours within the 500$\km \second^{-1}$ window will not be listed in the catalogues, which may reduce the noise and strengthen the correlations.
In observations, the correlations are the results of the balance of these two effects.
In principle, both effects may lead to misleading conclusions about the average galaxy/absorber distance.

We simulate 200 sight lines as in section 2 using methods described above. For model F1,  our result shows that, if all bright galaxy/absorber pairs are used , we get
$\log W_r=0.86 -0.59 \log \rho + 0.20 \log (L_{B}/L_{B*})$ 
and $\frac{r}{r_*}=\left(\frac{L_B}{L_{B*}}\right)^{0.34}$, where $r_*=220.7 h^{-1} \kpc$ ($W_r \geq 0.3${\AA}).
In contrast, for all physical pairs, we have 
$\log W_r=0.89 -0.63 \log \rho + 0.20 \log (L_{B}/L_{B*})$ 
and $\frac{r}{r_*}=\left(\frac{L_B}{L_{B*}}\right)^{0.32}$, where $r_*=174.8 h^{-1} \kpc$ ($W_r \geq 0.3${\AA}). 
As we can see, 
a larger $r_*$ is derived if spurious pairs at large distances are used.

\subsection{Results for mock observations of known QSO LOSs}
\label{secCLWB}
We simulate observations for 10 QSOs at the redshifts listed in the CLWB paper. 
The redshifts of these QSOs are 0.200, 0.264, 0.329, 0.371, 0.513, 0.534, 0.574, 0.616, 0.719, 0.927. The total redshift interval is about 5.
Because of the proximity effect, galaxies within $3000~\km~\second^{-1}$ of the quasar redshift are excluded.
With 100 mock observations, we get the statistical number of galaxy-absorber pairs and the statistical properties of galaxy-absorber pairs. 

The total number of `physical pairs' with $W_r \geq 0.3${\AA} for the ten QSOs is
\[
36.0\pm6.3, \,57.8\pm7.5, \,64.2\pm7.0
\]
for model F1, F3, F5 respectively.
After applying the selection criteria, the total number of `bright pairs' with $W_r \geq 0.3${\AA} is
\[
21.0\pm4.8, \,26.1\pm4.8, \,29.9\pm5.3
\]
for model F1, F3, F5 respectively. 
The predicted `bright pair' numbers are in good agreement with that of CLWB paper in which there are 26 galaxies giving rise to absorption.

\begin{figure*}
\begin{minipage}{178mm}
\centerline{\psfig{figure=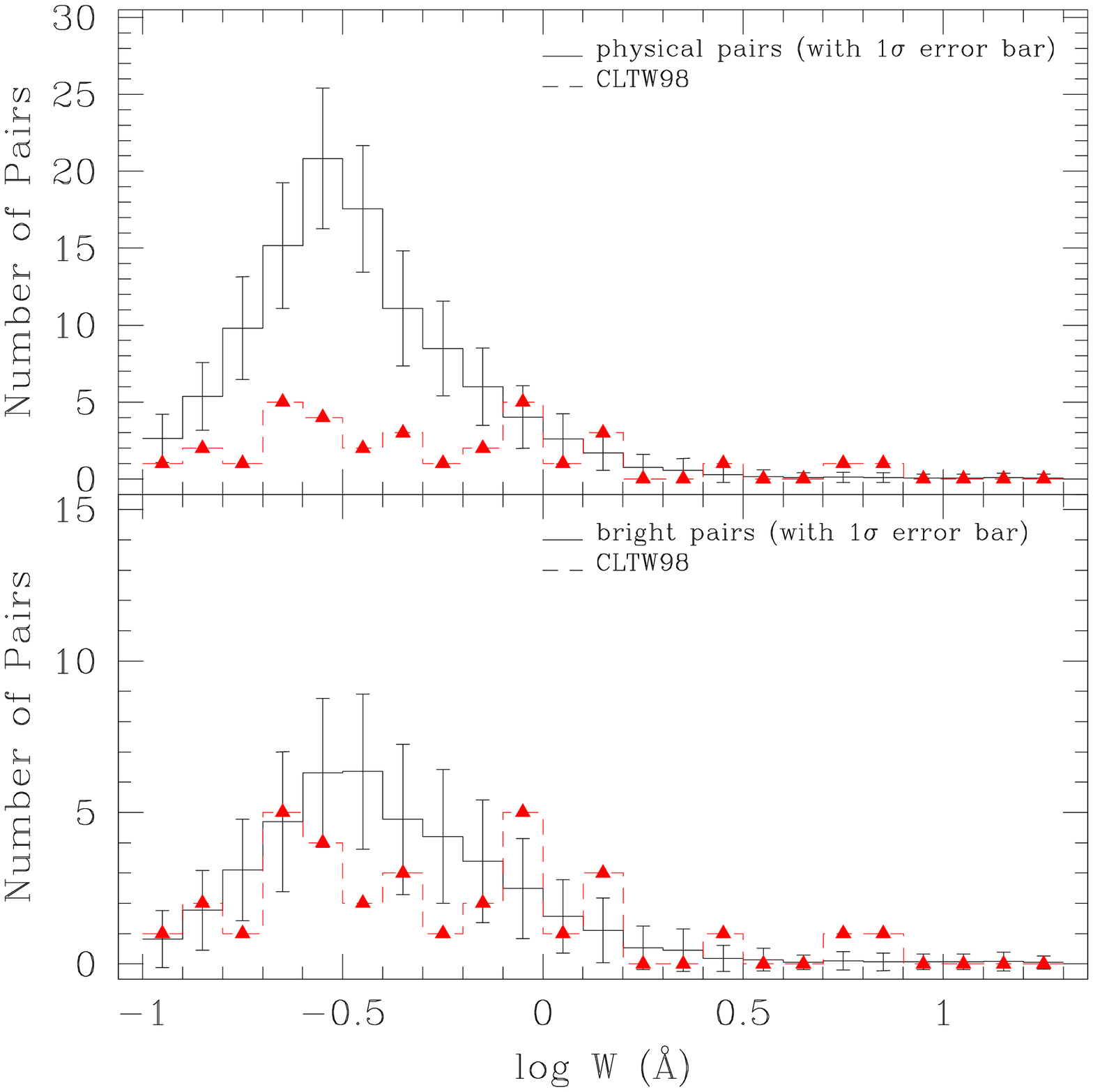,height=6.5cm,width=8.5cm}\psfig{figure=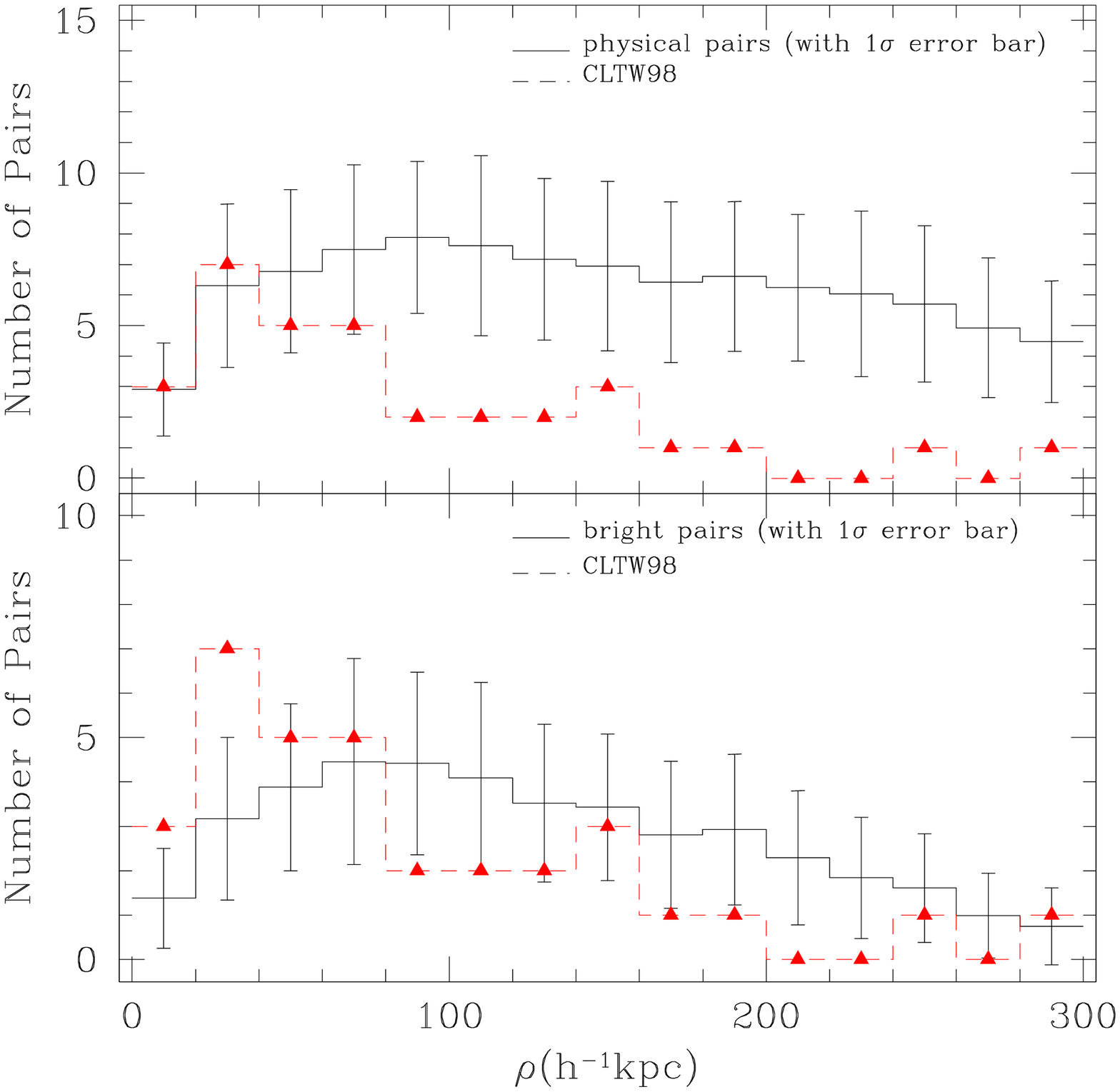,height=6.5cm,width=8.5cm}}
\centerline{\psfig{figure=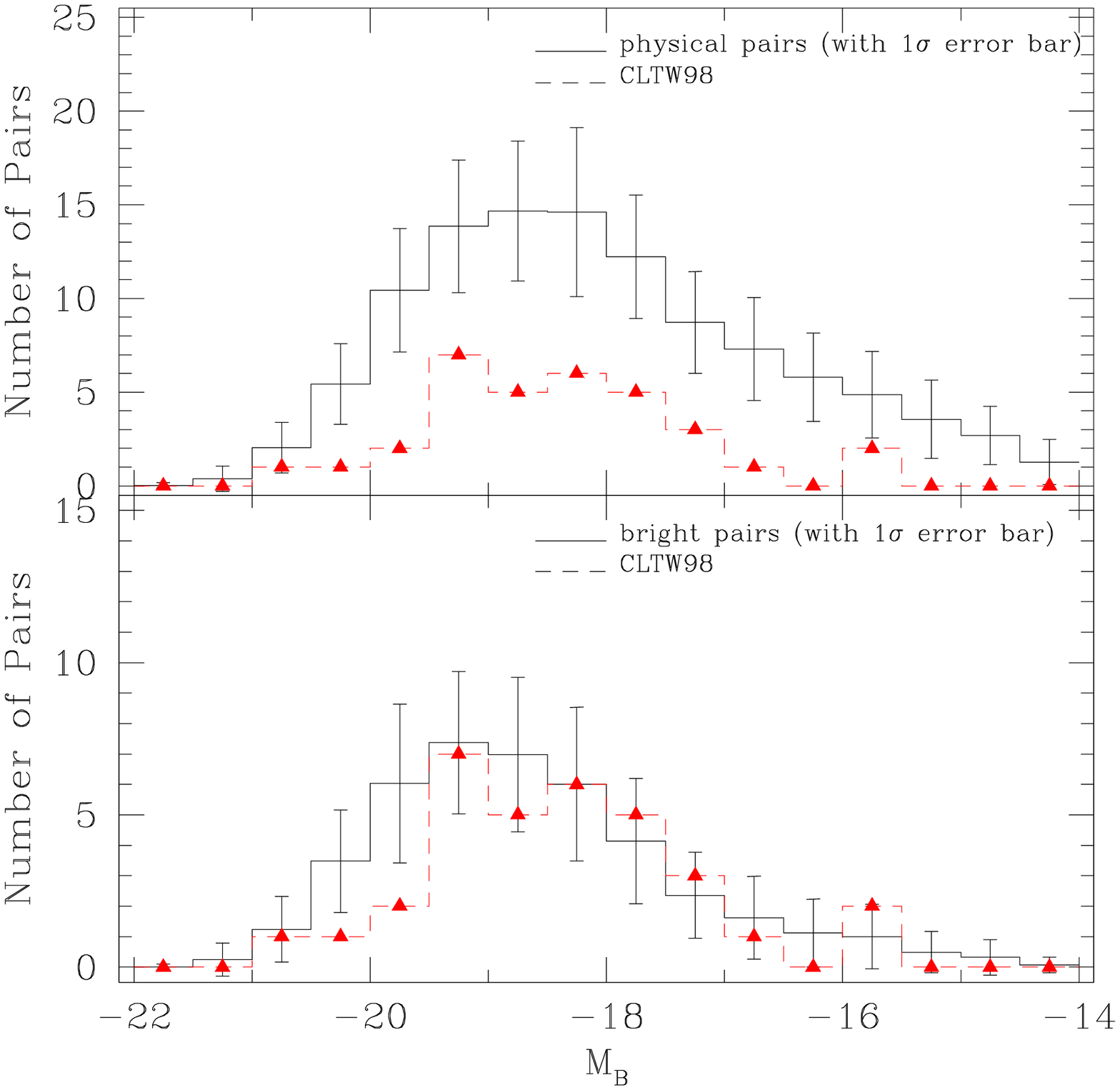,height=6.5cm,width=8.5cm}\psfig{figure=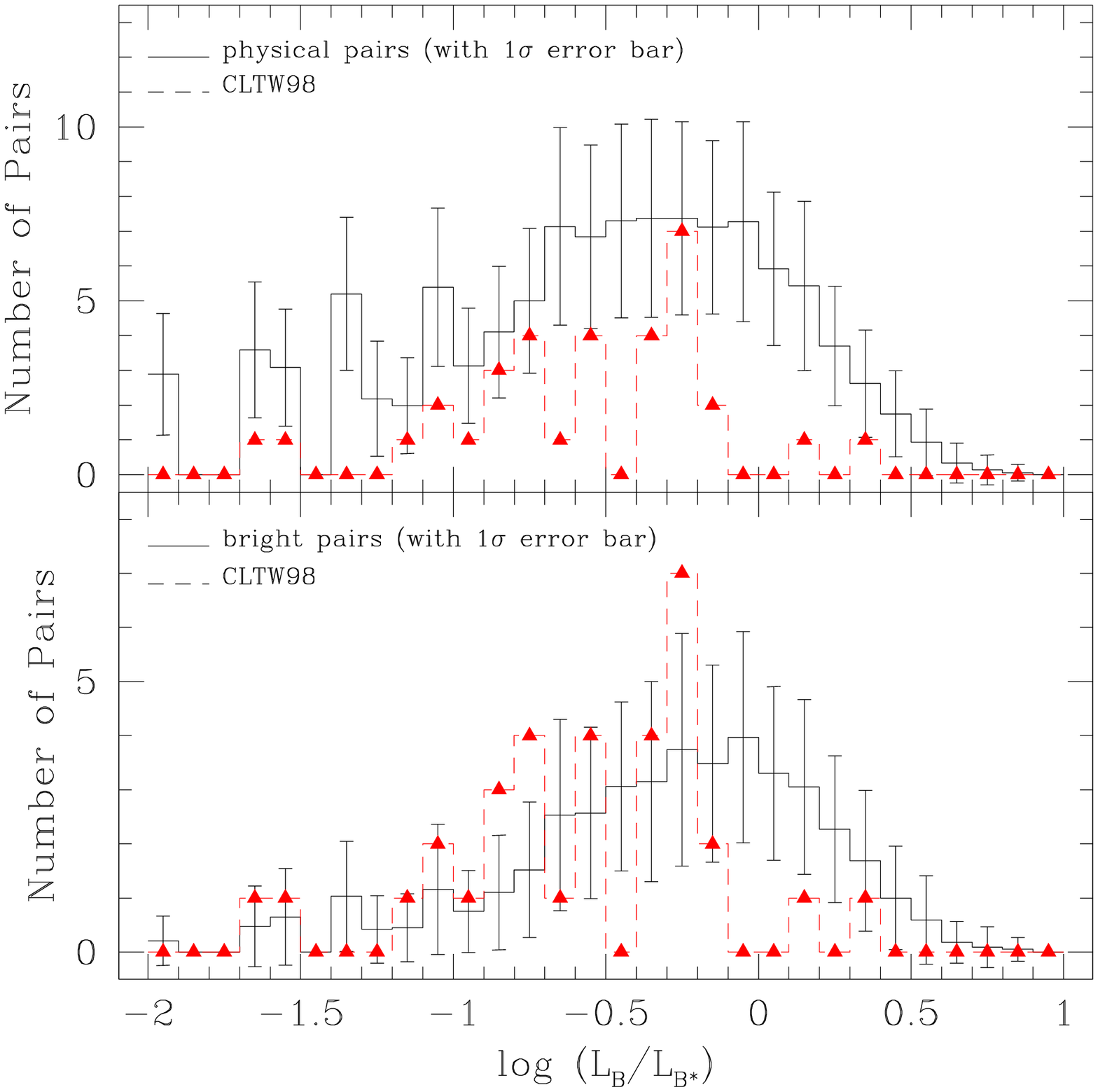,height=6.5cm,width=8.5cm}}
\centerline{\psfig{figure=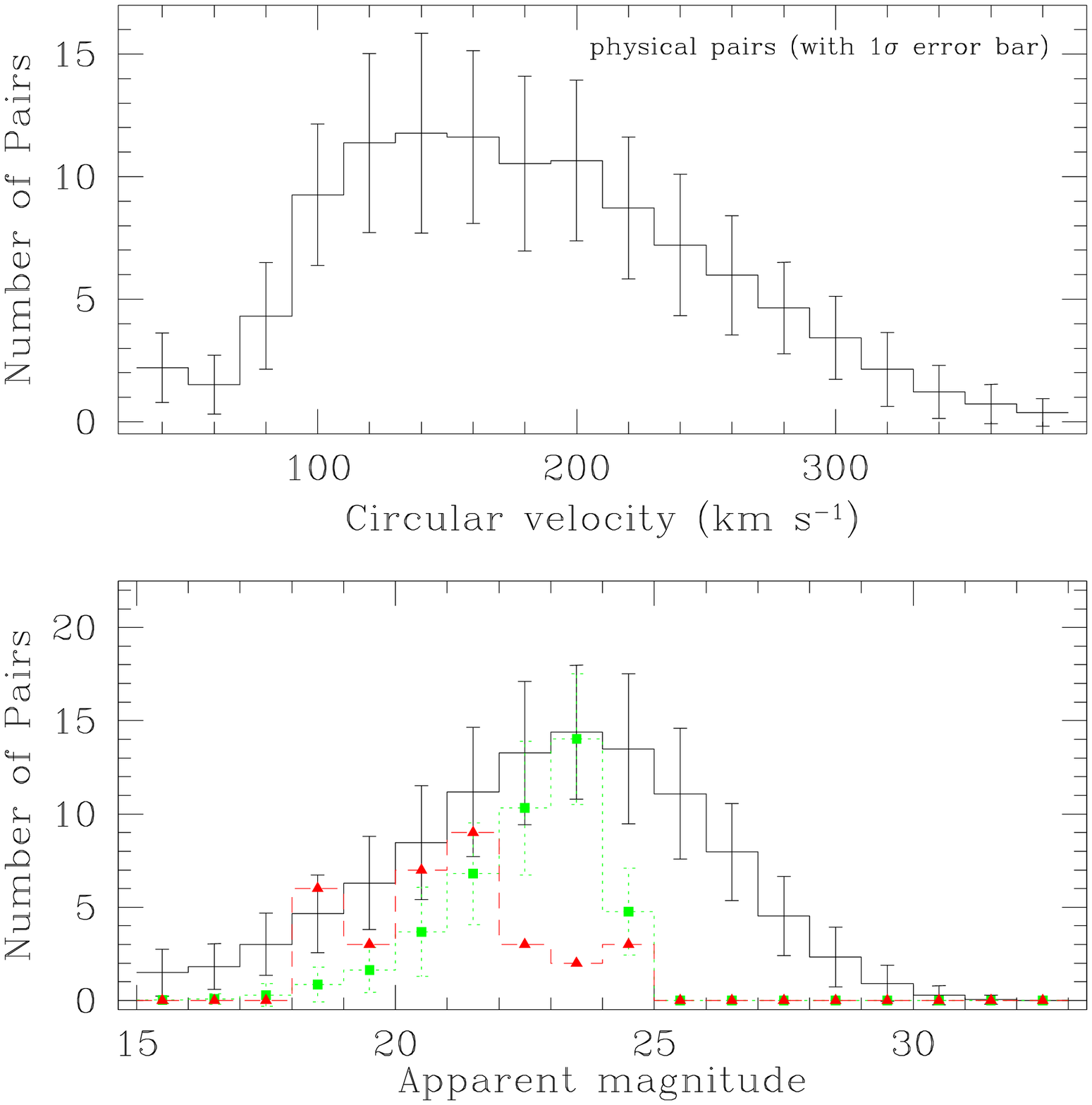,height=6.5cm,width=8.5cm}\psfig{figure=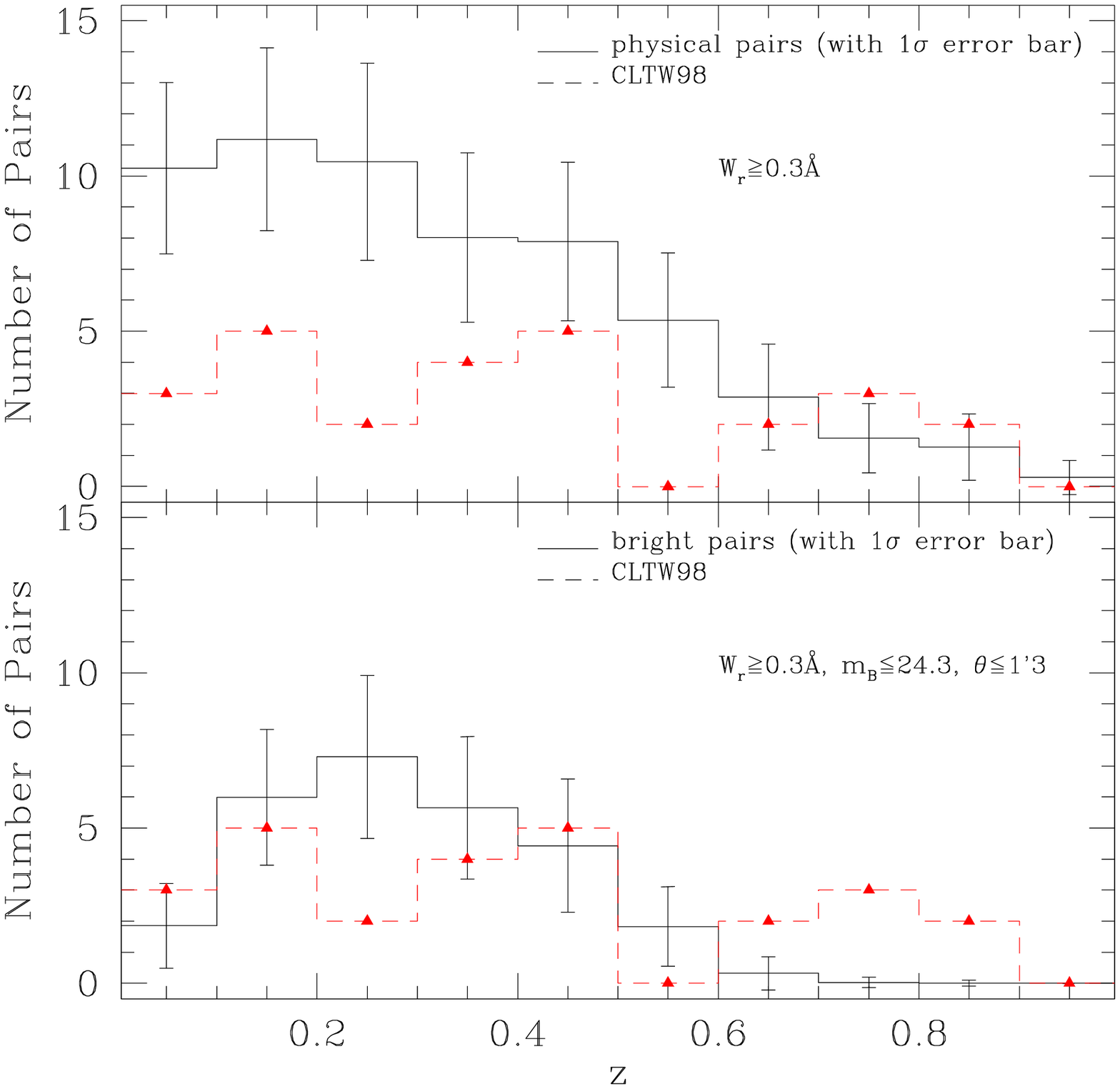,height=6.5cm,width=8.5cm}}
\caption{The distributions of equivalent line widths, projected distances, absolute magnitudes, circular velocities, apparent magnitudes, and redshifts for model F3.
We compare the results with CLWB's except for the circular velocity distribution. For `bright pairs', the distributions are in good agreement with those of CLWB.
However, the numbers of apparent magnitudes for `bright pairs' (dotted lines in the lowest left panel) in bins between 22.0 and 25.0 are higher than those of CLWB. This may mean that optical surveys could be incomplete in this apparent magnitude interval.
}
\label{FigCLWBF3}
\end{minipage}
\end{figure*}

If the model prediction is correct, we argue that the galaxy imaging survey at the faint end could be incomplete. 
This can be seen from the lower-left panel of Fig.\ref{FigCLWBF3}, in which the dotted line shows our predicted distribution of apparent magnitudes of `bright pairs' and 
the solid line represents the distribution for all `physical pairs', while the dashed line represents the distribution of apparent magnitudes of pairs from the CLWB paper.
As we can see, considerable numbers of the predicted $m_B$ of pairs are fainter than 25, and in the range of 23 to 25, our predicted `bright pairs' are more numerous than the observed ones.
In the lowest-right panel of Fig.\ref{FigCLWBF3}, the predicted number of `physical pairs' in redshift bins at $z<0.5$ is larger than that of CLWB, and comparable with that of CLWB at $z>0.5$, 
while for `bright pairs', the number is comparable with that of CLWB at $z<0.5$ but is lower than that of CLWB at $z>0.5$. 
However the observed number density at high redshift is uncertain because the redshift interval there is quite small.
Of course, if the difference is real, there are some implications. For example,
the luminosity function at $z > 0.5$ would be higher than at low redshift and galaxies could be brighter in the blue band because of intense star formation.
In conclusion, from the number of pairs and distribution of pair redshifts, the model predictions are consistent with current observations.

\begin{figure*}
\begin{minipage}{178mm}
\centerline{\psfig{figure=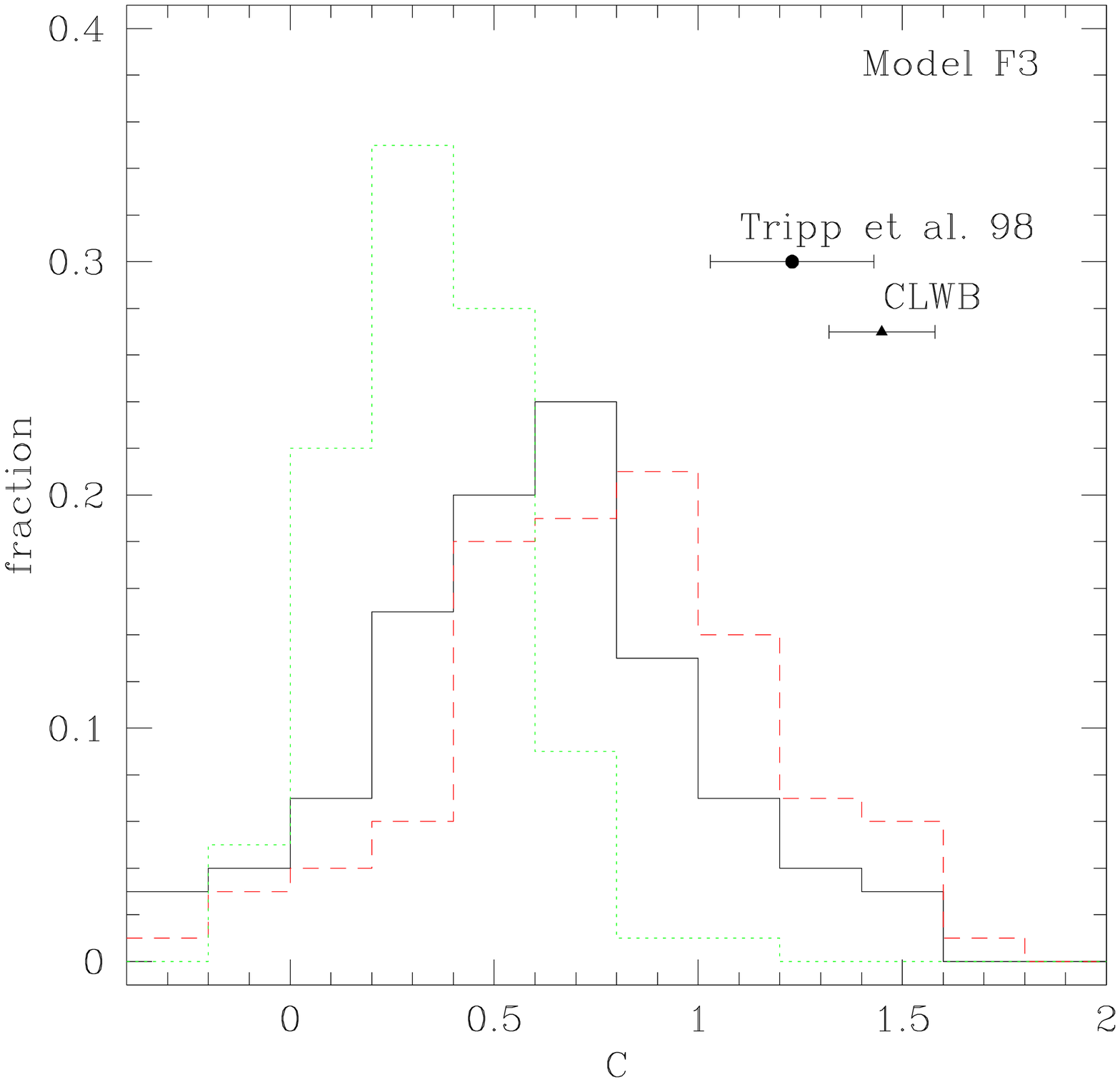,width=8.5cm}\psfig{figure=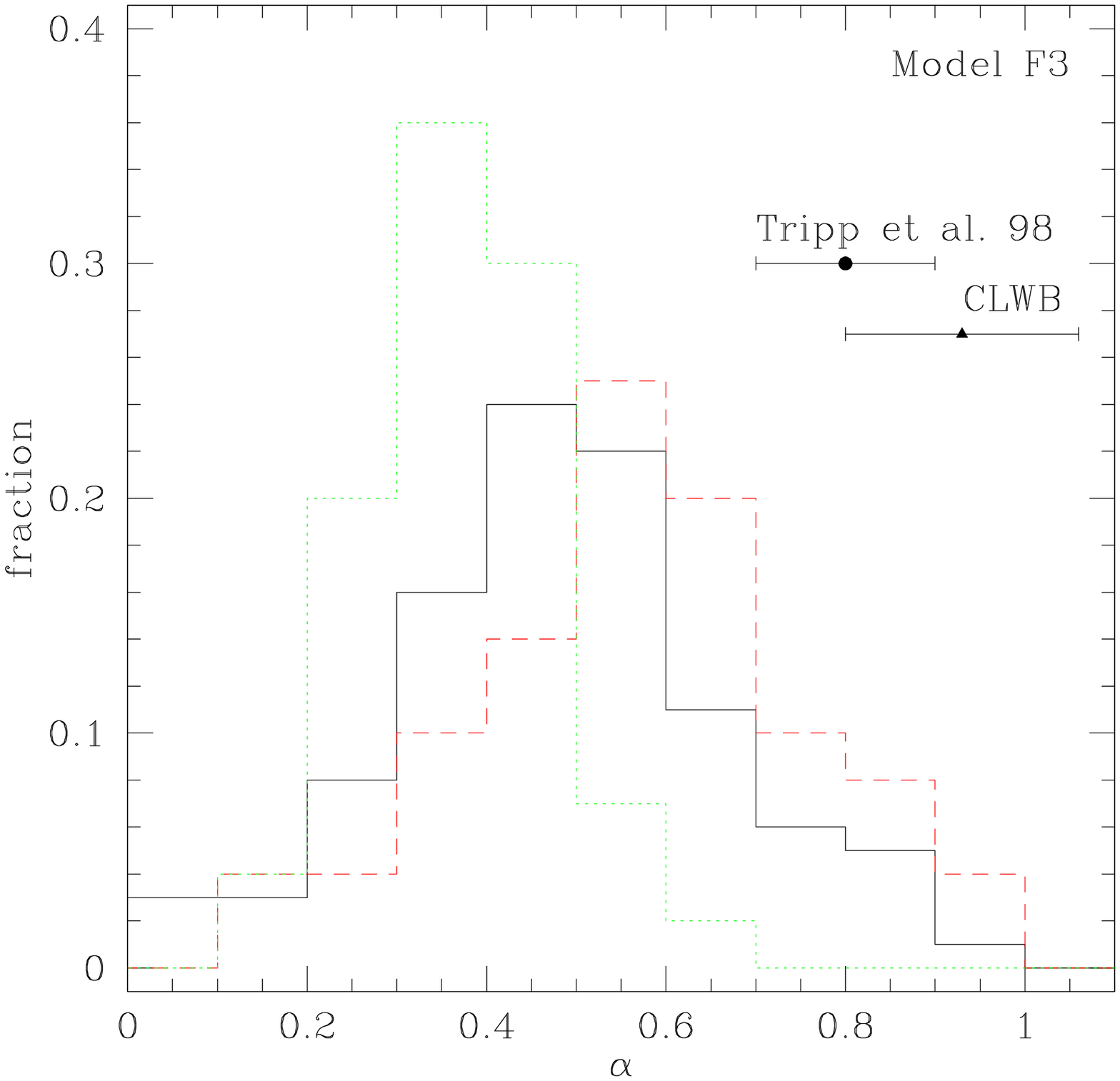,width=8.5cm}}
\centerline{\psfig{figure=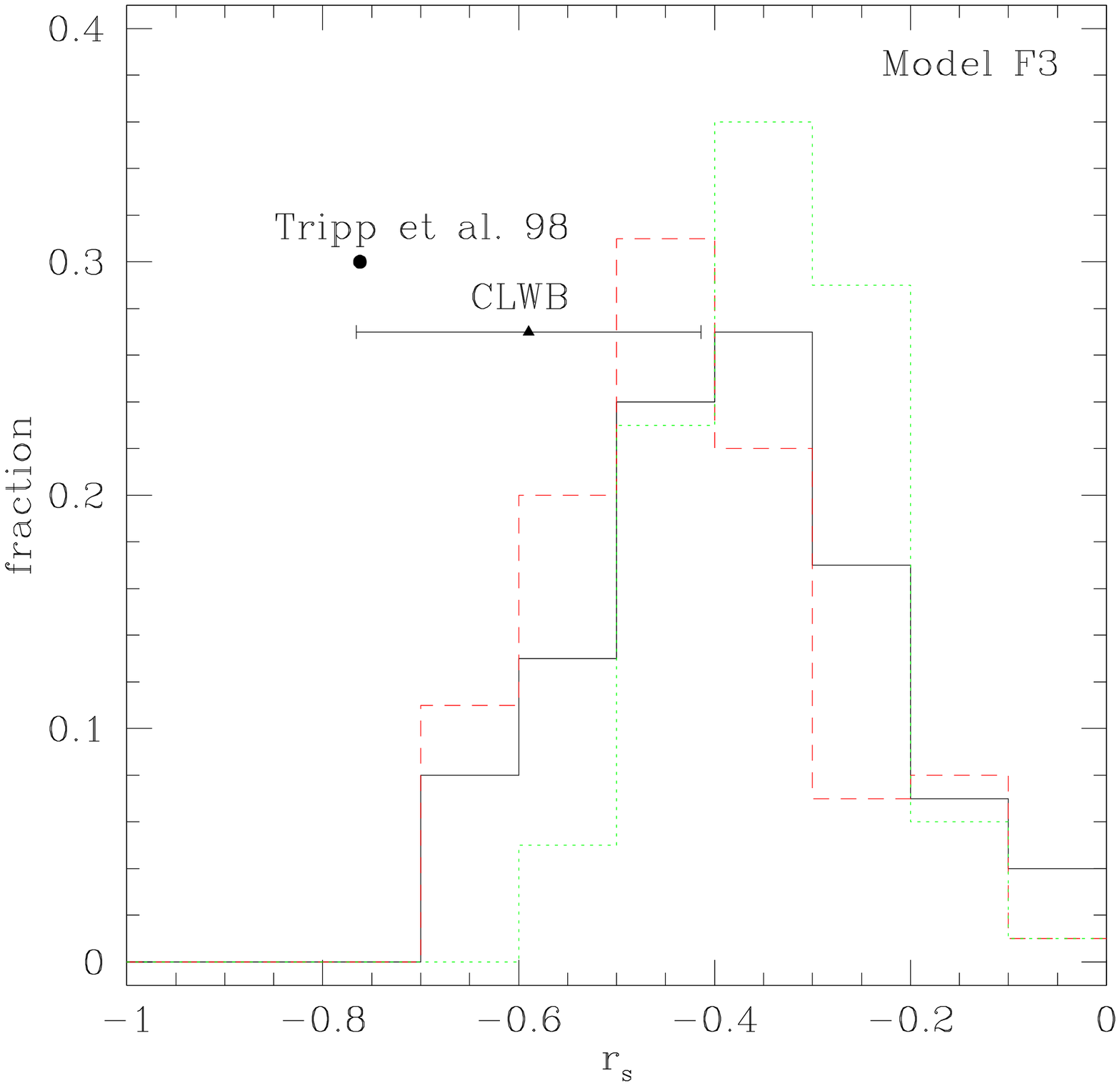,width=8.5cm}\psfig{figure=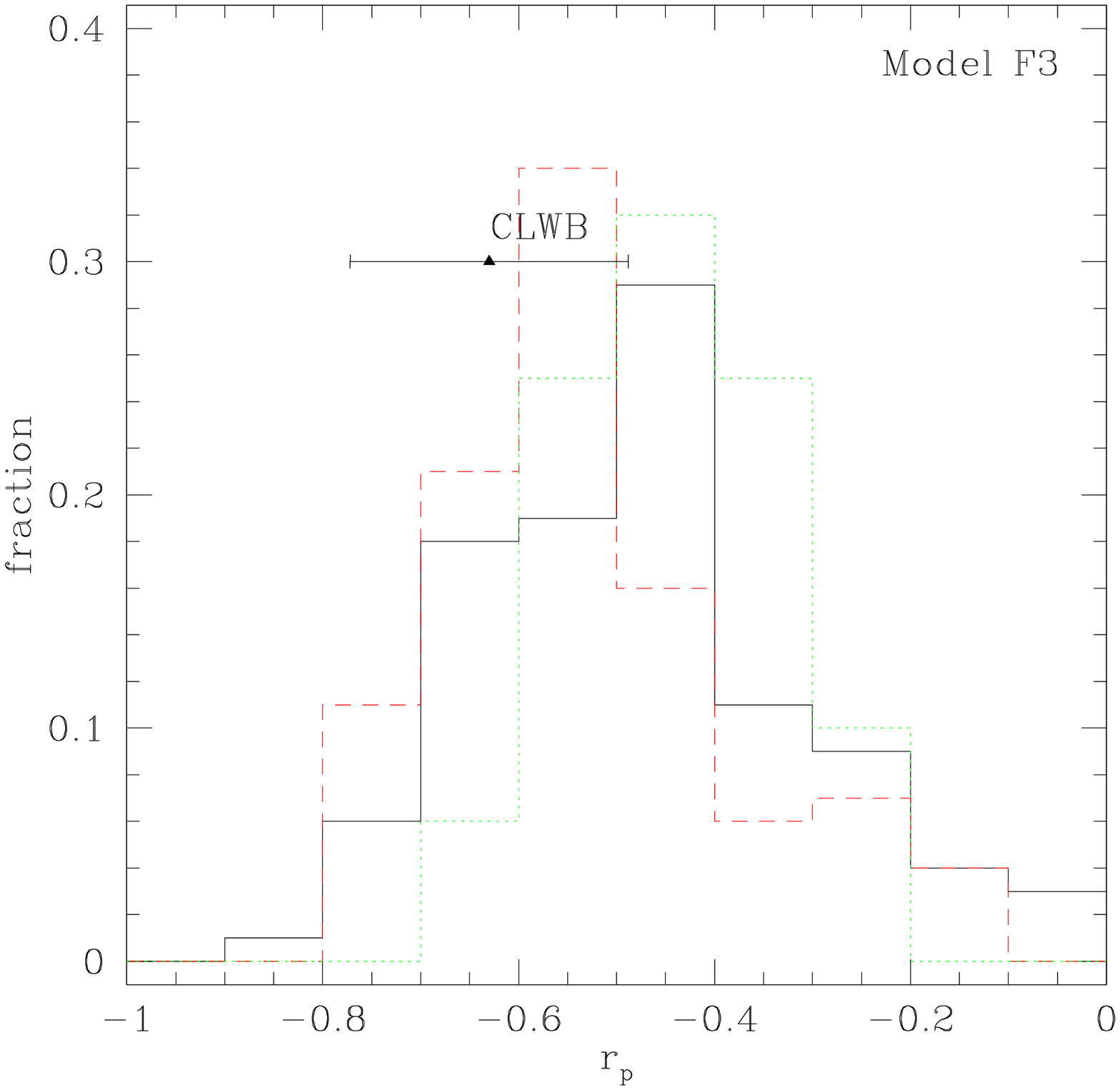,width=8.5cm}}
\caption{Statistical analysis of $\log W_r = - \alpha \log \rho + {\rm C}$ (for model F3). The upper left, upper right two panels are for distributions of constant C and slopes $\alpha$ for `brigh
t pairs' respectively.
The lower left and lower right panels are for distributions of the Spearman rank order co-efficient $r_s$ and the Pearson co-efficient $r_p$.
All the lines here are only for pairs with $W_r \geq 0.1$ {\AA}.
The dotted lines are for all `physical pairs'. The thick lines are for `bright pairs' with $\theta \leq 1'3$.
The dashed lines are for `bright pairs' with $\theta \leq 1'3$ and $V_{\rm cir} \geq 100 \km \second^{-1}$.
}
\label{FigHrp}
\end{minipage}
\end{figure*}

\begin{figure*}
\begin{minipage}{178mm}
\centerline{\psfig{figure=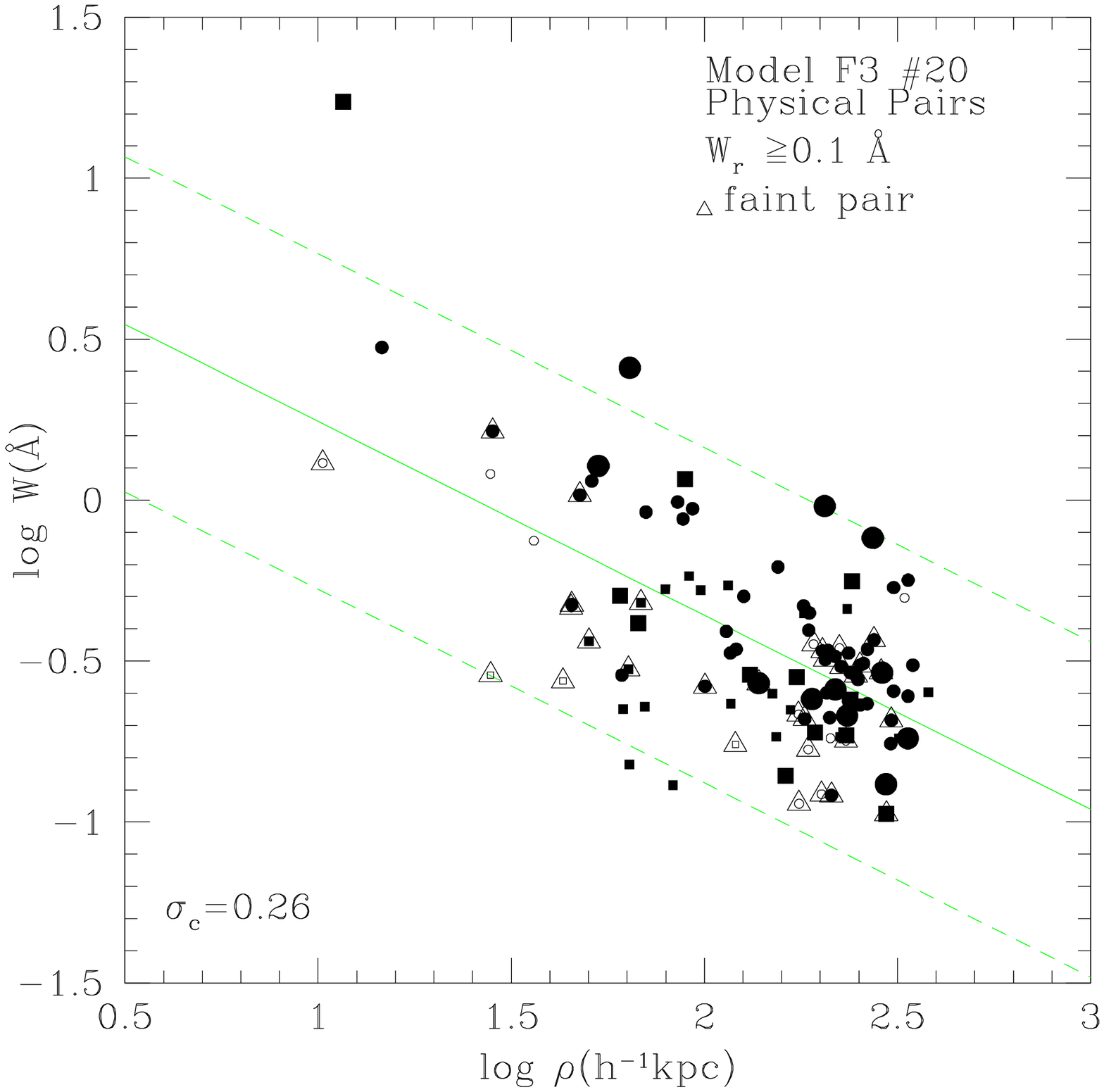,width=8.5cm}\psfig{figure=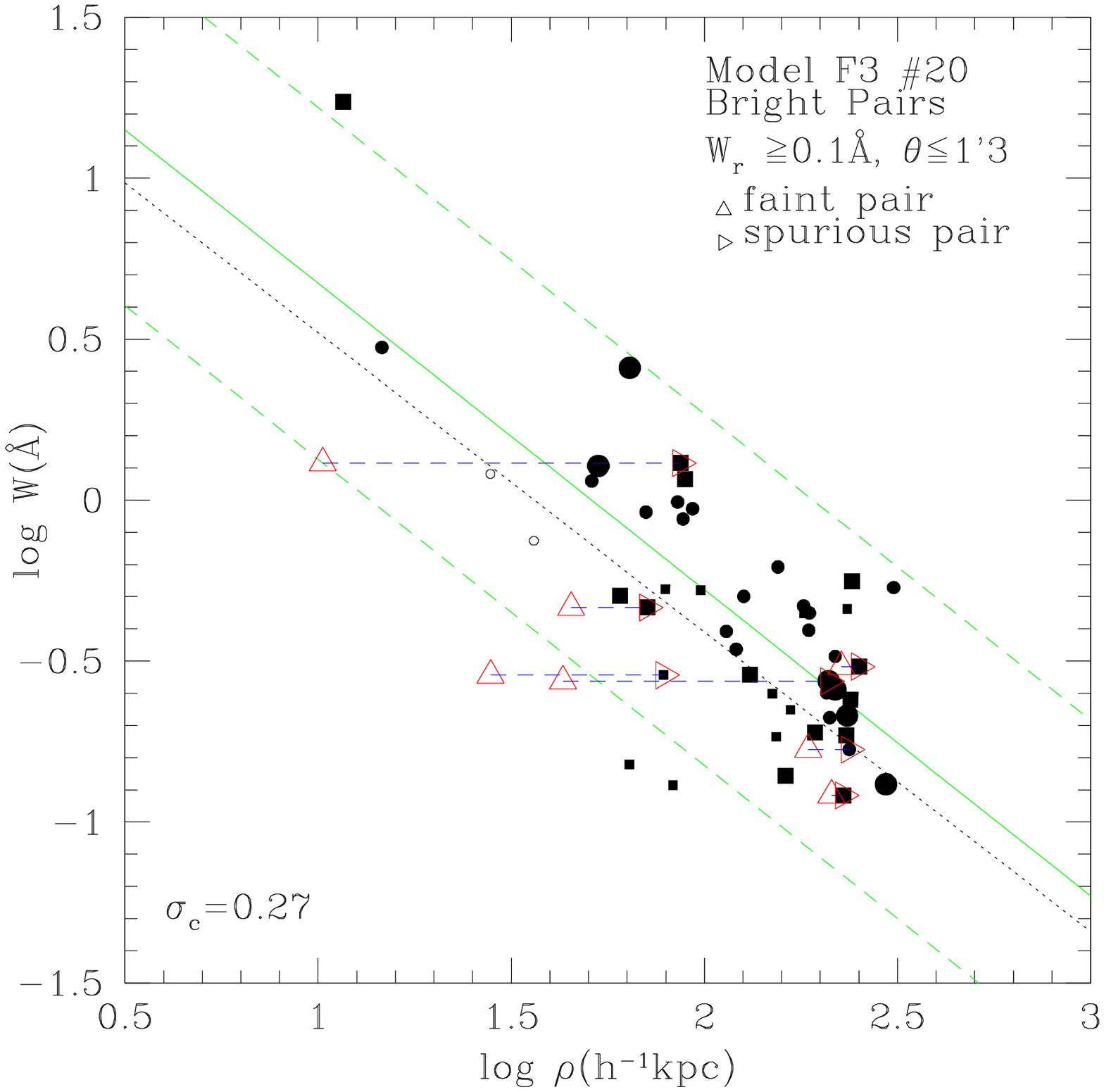,width=8.5cm}}
\caption{
The anti-correlation of $W_r$ versus $\rho$ for run No.20 of the mock data: The left, right panel are for `physical pairs' and `bright pairs' respectively.
The horizontal long-dashed-lines represent the possibility that a faint absorber (triangle at the left end of the line) may be mis-identified and paired with a nearby bright galaxy (triangle at th
e right end of the line).
The meanings of other symbols and lines are the same as in Fig.\ref{FigB2}. The dotted-line is the linear fit of CLWB.
}
\label{FigExample1}
\end{minipage}
\end{figure*}

\begin{figure*}
\begin{minipage}{178mm}
\centerline{\psfig{figure=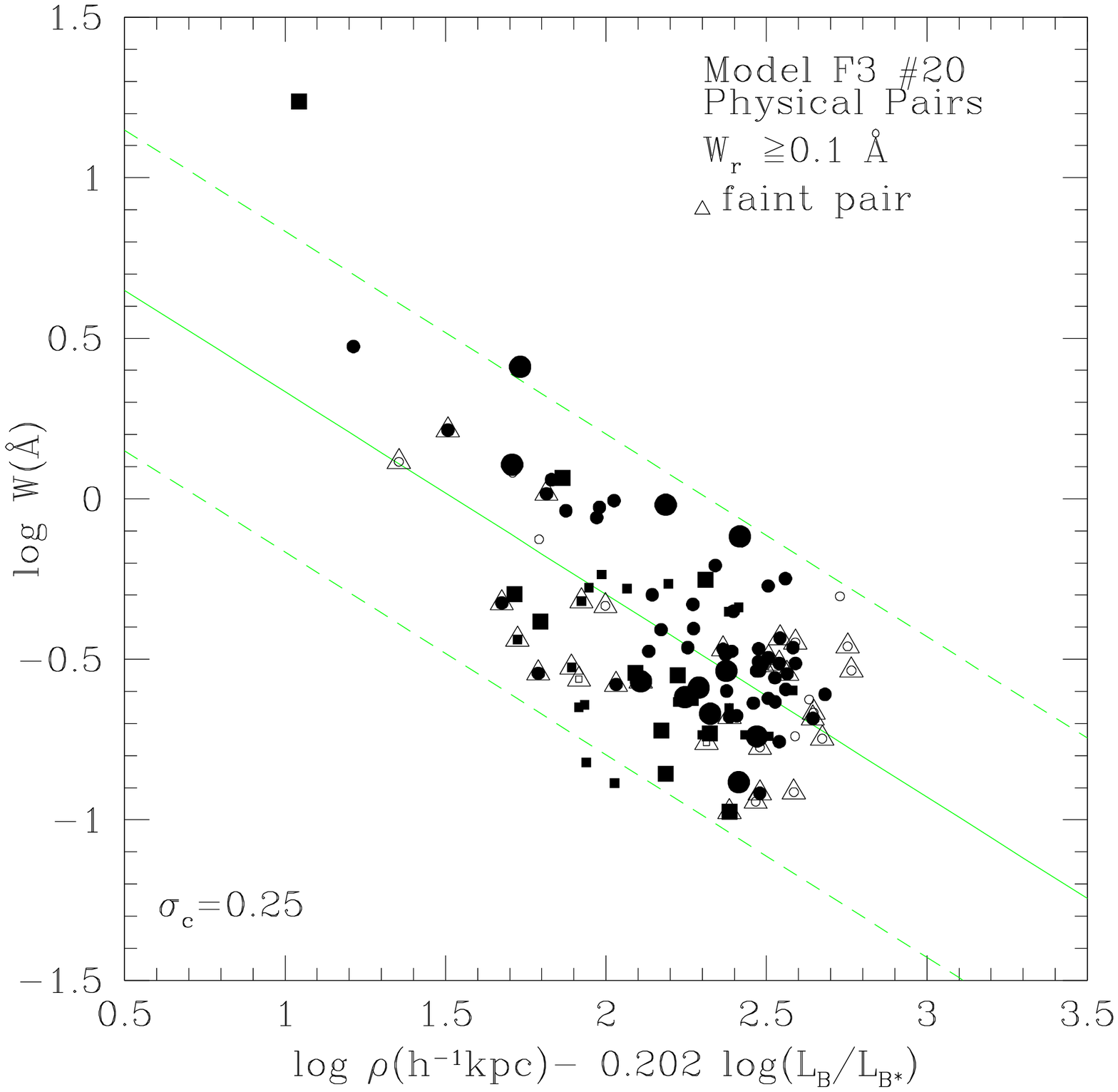,width=8.5cm}\psfig{figure=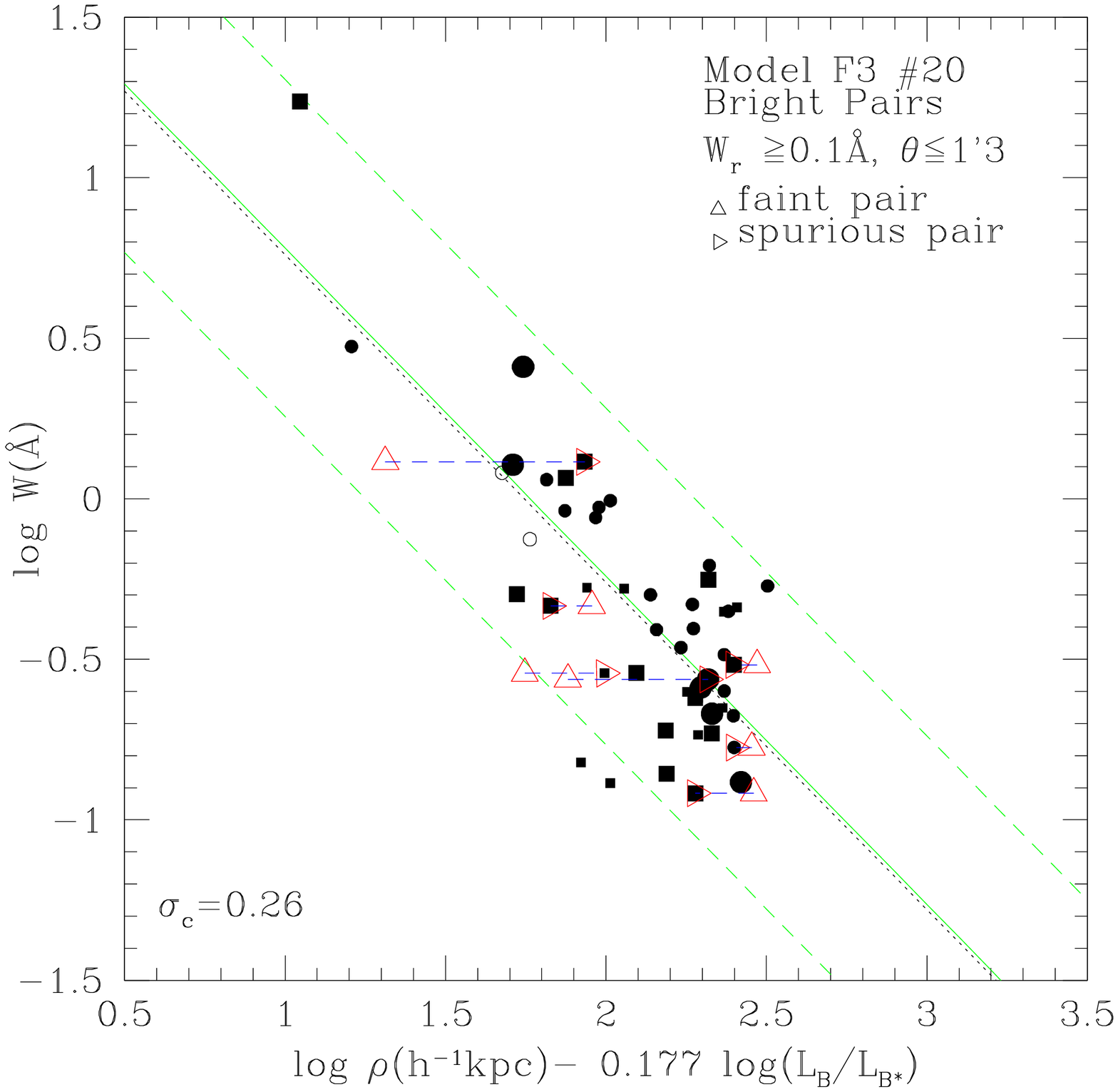,width=8.5cm}}
\caption{
As in Fig.\ref{FigExample1} but for the correlation of $W_r$ versus $\rho$ and $L_{B}$. Symbols are the same as in Fig.\ref{FigExample1}.
The dotted-line is the linear fit of CLWB.
}
\label{FigExample2}
\end{minipage}
\end{figure*}

In Fig.\ref{FigCLWBF3}, we also plot the distributions of $\rho$, $W_r$, $L_B/L_{B*}$, $M_B$ as well as $V_{\rm cir}$ and compare the results with those of CLWB except for circular velocity. 
We apply the Chi-Square test to see if our predicted distributions are similar to those of CLWB (the data are from their table 4).
The null hypothesis that the data sets are similar has probabilities of 97.0 per cent, 72.4 per cent, 65.7 per cent, 97.9 per cent, 31.1 per cent for $W_r$, $\rho$, $L_B/L_{B*}$, $M_B$, $z$ respectively.

We give statistical results of the anti-correlation of $W_r$ versus $\rho$ for all mock runs. 
Fig.\ref{FigHrp} contains the histograms of the Spearman rank-order coefficient $r_s$, the Pearson coefficiet $r_p$, and the zero point C, the slope $\alpha$ of the linear fit as in eq. (\ref{eq-linfit}). 
In the figure, the solid lines are for `bright pairs' satisfying the selection criteria, the dashed lines are for `bright pairs' with $V_{\rm cir}\geq 100 \km \second^{-1}$, and the dotted lines are for all `physical pairs'.
Statistically the distributions of the slope $\alpha$ shift from small to large values for `physical pairs', `bright pairs', `bright pairs' with $V_{\rm cir}\geq 100 \km \second^{-1}$ and so on. 
Similarly, there are also shifts of lines in other panels. 
These shifts mean that selection effects can statistically strengthen the anti-correlation of $W_r$ and $\rho$.
We have to point out that this strengthening only has a statistical meaning and does not necessarily occur for every specific run in our simulations, because in some cases selection effects may also weaken the anti-correlations.
Available results from observations (CLWB; Tripp et al. 1998) are also shown with $1 \sigma$ error bars. 
Obviously some simulation runs can give consistent results compared with the observations.
Note that the results of mock runs have considerable scatter, which may imply, as will be discussed later in \S\ref{secDis}, that in the models the same total redshift interval as in present observations is not adequate to predict the real correlation.

\begin{figure}
\centerline{\psfig{figure=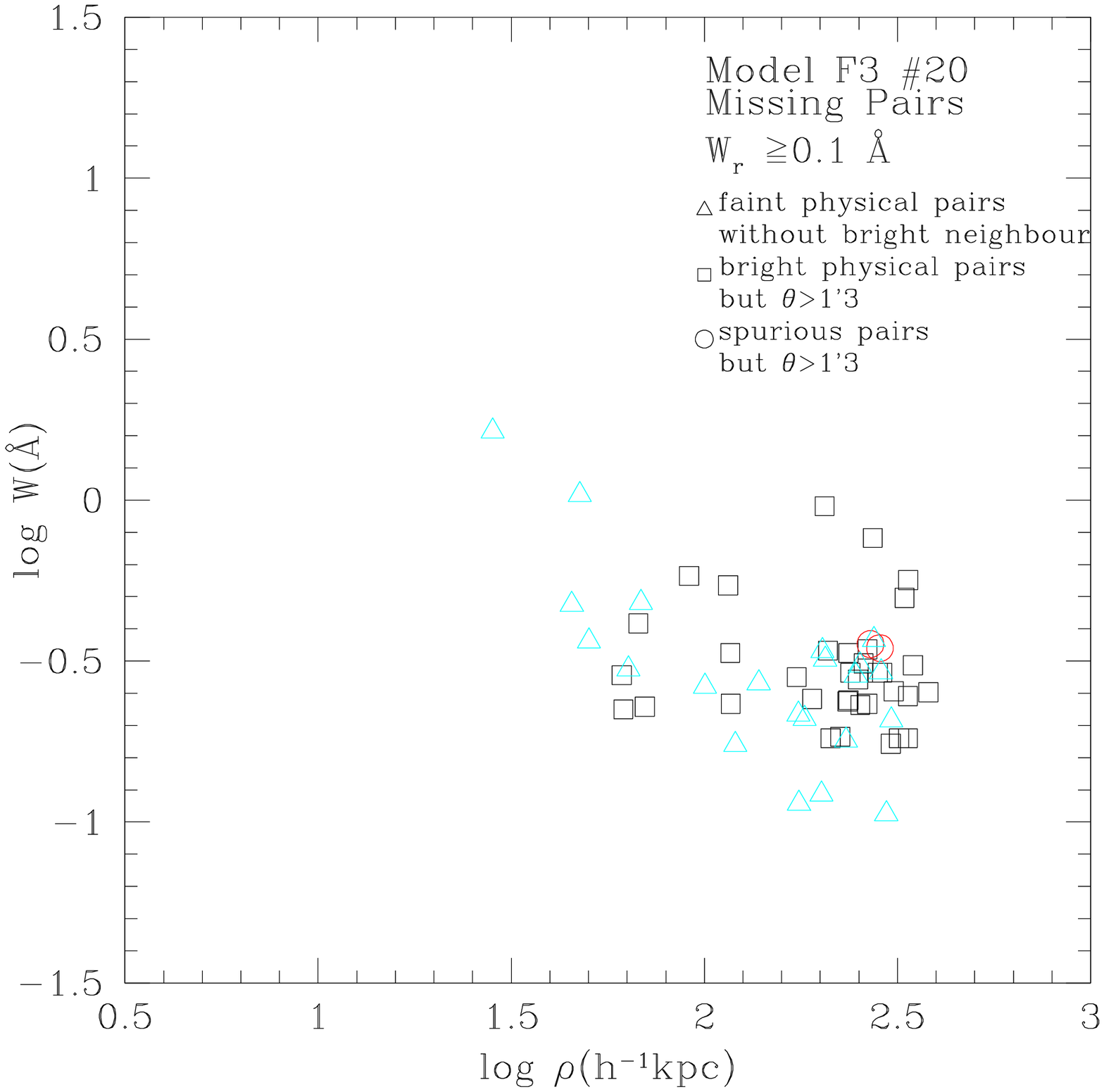,width=8.5cm}}
\caption{The missing pairs of run No.20 of the mock observations.
}
\label{FigExample3}
\end{figure}

As mentioned above, the outcome of each run could differ case by case. Some examples of mock observations are given in Fig.\ref{FigExample1}, Fig.\ref{FigExample2}, Fig.\ref{FigExample3}.

The results for the anti-correlation of REW versus projected distance are shown in Fig.\ref{FigExample1} (for run No.20).
All the real galaxy-absorber pairs (`physical pairs') are drawn in the left panel. For the $108$ `physical pairs' (of which $54$ pairs have $W_r \geq 0.3${\AA}), we get the Spearman rank-order coefficient, $r_s=-0.41$ (with significance level $4.6\sigma$) and the Pearson coefficient $r_p=-0.62$ (with significance level $8.0\sigma$) and best fit line $\log W_r=(0.85\pm0.16)-(0.60\pm0.07) \log\rho$.
The galaxy-absorber pairs after applying selection criteria are drawn in the right panel. For $49$ `bright pairs' (of which $30$ pairs have $W_r \geq 0.3${\AA}), we get $r_s=-0.58$ (with significance level $4.9\sigma$), $r_p=-0.75$ (with significance level $7.7\sigma$) and
$\log W_r=(1.63\pm0.26)-(0.95\pm0.12) \log\rho$.

In comparison, CLWB give
\[
\log W_r=(1.34\pm0.22)-(0.93\pm0.13) \log\rho
\]
and Tripp et al. (1998) give
\[
\log W_r=(1.32\pm0.20)-(0.80\pm0.10) \log\rho
\]
by adding more LOSs.
The results of `bright pairs' for this run are in good agreement with those of CLWB and Tripp et al. (1998). 
As we can see, selection effects do strengthen the anti-correlation in this specific run. 
However for some runs, selection effects do not strengthen the anti-correlation at all,
because they only have a statistical meaning in the simulations. 

We analyse the relation between REW and galaxy luminosity using eq. (\ref{eq-linfit2}) for the run. The results are shown in Fig.\ref{FigExample2}.
In the left panel, $C$, $\alpha$, $\beta$ are $0.96\pm0.16$, $0.63\pm0.07$, $0.13\pm0.04$ respectively.
In the right panel, they are $1.80\pm0.26$ of $1.02\pm0.12$, $0.18\pm0.09$ respectively.
Again, in comparison, CLWB give
\[
\log W_r=(1.78\pm0.20)-(1.02\pm0.12)\log\rho
\]
\[
~~~~~~~~~~~+(0.37\pm0.10)\log(L_B/L_{B*}).
\]
As we can see, the zero point $C$ and $\alpha$ of our prediction for `bright pairs' in run No.20 are in good agreement with those of CLTW. 
However our results for $\beta$ in this run are less than that of CLWB (still within the $2\sigma$ standard deviation), but in agreement with the value of $0.1-0.2$ suggested by Bowen et al. (1996).

There is a substantial number of missing pairs in our mock observations. Fig. \ref{FigExample3} gives the results for the run. As we can see, there are a number of bright `physical pairs' with angular distances to the LOS outside the threshold of $1'3$ and some faint `physical pairs' without bright neighbours located within $400 h^{-1} \kpc$. These pairs are defined as missing pairs and could not be listed in optical catalogues of absorbers. 
Note that there are also a few spurious pairs with angular distance outside the threshold.

\section{Discussion}
\label{secDis}

\begin{figure}
\centerline{
\psfig{figure=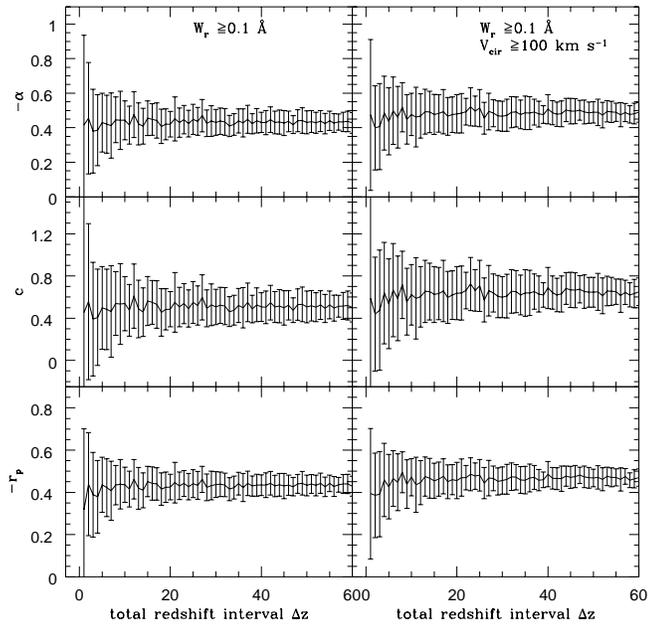,width=8.5cm}}
\caption{The statistical analysis of the anti-correlation between $W_r$ and $\rho$ (model F3)
for `bright pairs' with $m_b \leq 24.3$ and $\theta \leq 1'3$.
If the total redshift interval is 10,
$r_p$, C, $\alpha$ are $-0.50\pm0.11$, $0.55\pm0.26$, $0.46\pm0.11$ respectively and $-0.54\pm0.11$, $0.75\pm0.25$, $0.53\pm0.11$ respectively for the sample with $V_{cir} \geq 100 \km \second^{-1
}$.}
\label{FigHmri}
\end{figure}

As we remarked in the last section, if the total redshift interval is small, the results of correlations can have large statistical deviations. In order to investigate this, we randomly choose 40 groups of LOS with the same number of LOS in each group from the simulation. 
We calculate the correlation coeffecient, the slope and zero point of the linear fit for every group. 
At last the average values over all the groups as well as the standard deviations can be calculated. 
Then we increase the number of LOSs in each group. These values will change until the total redshift interval is large enough to get stable values with small deviations. We make plots of the dependence, from which we can determine how large a redshift interval is adequate for an observational sample.
The results for model F3 for `bright pairs' are presented in Fig.\ref{FigHmri}. Clearly for `bright pairs' a total redshift interval $\sim 10$ is necessary to get statistically accurate value of $r_p$, the Spearman coefficient, and $\alpha$, the slope of the linear fit in eq. (\ref{eq-linfit}), with standard deviation less than 0.1. If we want to determine the zero point (i.e. with standard deviation less than 0.2), the total redshift interval should be about 20 for model F3.
This implies that results of the anti-correlation of $W_r$ versus $\rho$ by different surveys may differ from each other if the total redshift interval in the survey is not large enough. 

In the models, only one third of the absorbers reside inside galactic haloes and two thirds of them are satelltes around central halos. 
This picture may reconcile different conclusions by various authors (Morris et al. 1993; LBTW; CLWB; Bowen, Blades, \& Pettini 1996; Le Brun et al. 1996; Tripp et al. 1998; Impey at al. 1999). 
Firstly, the models predict a large absorption radius and a reasonable covering factor. 
Secondly, the predicted absorbers are still closely related with galaxies. 
Furthermore, it is possible that there are still a number of satelltes around central haloes even at distance $>400 h^{-1}\kpc$. Thus if a catalogue of galaxy-absorber pairs includes those pairs with large projected distances, the anti-correlation of $W_r$ versus $\rho$ could be weaken.

As listed in Table~\ref{C}, our prefered models can explain up to $\sim$ 55 per cent 
(and even more if an absorption system with large velocity spread is counted as 2 lines as would be the case with the {\em HST} spectral resolution)
of the {\em HST} observed counterpart for lines wider than 0.3{\AA}.
Can our models predict more absorbers?
Actually the evolution of the galaxy luminosity functions with redshift can change our predicted absorber number density. 
Because the numbers of aborbers in our simulations are directly proportional to the galaxy number density,
our predicted number of absorbers can be higher at higher redshift if the galaxy number density there is higher. 
For example, if we chose the AUTOFIB luminosity function at $0.35 < z <0.75$ whose $\phi^{*}$ is about 1.5 times the $\phi^{*}$ at $ 0.02 < z < 0.15$, 
the predicted number density could be as high as 17.8, which may account for 80 per cent
of the observed counterparts. Furthermore, the higher galaxy density at higher redshift may also increase the absorber density at higher redshift in Fig.\ref{FigCLWBF3} and solve the possible discrepancy in high-redshift absorber number between our prediction and that of CLWB.

Caution should be taken with our results, not only because there could be some alternative \Lya absorption arising in other parts of the galaxies, but also because it is unclear under present circumstances whether there are many satellites in the vicinity of big central galaxies and whether they possess gas (see Klypin et al. 1999 for further discusion; see also Bullock, Kravtsov, \& Weinberg 2000; Charlton, Churchill, \& Rigby 2000).
After all we did not include all possible absorption components related with galaxies. 
For example, Morris \& van den Bergh (1994) suggested that a significant fraction of weak lines could arise from pressure-confined tidal debris in the enviroment of small groups and clusters of galaxies (cf. Mo 1994). Tidal debris can increase the total gas cross section so as to 
increase the absorber number density and the corresponding absorption line can have a large projected distance of $>100 h^{-1}\kpc$. 
However the absorption line number arising from tidal tails depends on the unknown gas cross section and the generally unknown lifetime of the gas in tidal tails. Thus it is not possible to detemine exactly what fraction of absorbers arises in tidal debris.
Another possibility is that some low surface brightness galaxies could possess
huge gaseous haloes or discs which can also give rise to absorption.
In addition, galactic wind is also another possible source (Wang 1995).
Of course, the absorption by the IGM still could play an important role. 

In observations, it is difficult to assign a galaxy to an absorber counterpart, 
because an imaging survey of galaxies is never quite complete down to the faint end. 
Also absorbing galaxies may be outside of the angular extent of the survey. 
More LOSs to QSOs with higher resolution of the UV spectroscopy and more complete imaging surveys are necessary to investigate the physical origin and environment of the absorbers.
At very low redshifts, it is possible to identify satellites with $V_{\rm cir} \geq 30 \km \second^{-1}$ in optical imaging surveys, and then we can examine whether these satellites can give rise to \Lya
absorption lines.
To discriminate models, it is also important to get more physical information about the absorbing components, such as size, temperature, metallicity, ionizing parameter, rather than only informations about $W_r$ and $\rho$.
Observations of multi-LOS (QSO pairs or lensed images) can provide useful tools to get more insight into absorbers (see Rauch 1998 and references therein).
Furthermore, the possible detection of line emission from extended gas in galactic haloes may also help to determine the properties of the gaseous haloes \cite{Cirkovic99}.

\section{Summary and Conclusions}
\label{secCl}
In this paper, we present results of Monte-Carlo simulations of \Lya absorption line systems at redshift $z<1$. 
To get constraints on the parameters, we simulate a set of models with different absorption components and various parameters. We compare the predicted absorption line densities for strong \Lya lines, Lyman-limit systems as well as damped \Lya systems with observations.
From these comparisons, some models can be excluded.
In summary: 
\begin{enumerate}

\item Models with a single absorption component (galaxy disc, or cold clouds in a galactic halo, or satellites) or models with disk and clouds in a galactic halo, cannot explain the observed number densites for strong \Lya lines, for Lyman-limit systems and for damped \Lya systems at low redshift. 

\item Models with all three components (galaxy disc, cold clouds in a galactic halo, and satellites) can explain up to 60 per cent of the observed number density for strong \Lya lines at low redshift. 
These models can also predict reasonable number densities of Lyman-limit systems and damped \Lya systems at low redshift.

\item The fraction of the line number density for strong \Lya lines due to satellites is $\sim 40$ per cent more than that due to clouds in galactic haloes (which is $\sim 20$ per cent) by about a factor of 2.
The exponential galaxy discs can only account for a small amount of strong \Lya lines.
If indeed there are large numbers of satellites surrounding big central galaxies and they possess gas, these satellites may play an important role for strong \Lya lines at low redshift. 

\item The predicted $(\frac{dN}{dz})$ for Lyman-limit systems due to cold clouds in galactic haloes is $\sim 0.4$, which can account for most of the observed Lyman-limit systems.  
The line number density of Lyman-limit systems due to satellites is only $\sim$ 0.1, which is four times smaller than that due to clouds in galactic haloes.

\end{enumerate}

The properties of the predicted absorbers, such as REW, projected distance, galaxy luminosity, circular velocity and absorber redshift have been analysed.
The predicted dependence of line width on projected distance is $W_r \propto \rho^{-\alpha}$ with $\alpha \sim 0.4-0.6$, rather than $\alpha \sim 0.8-0.9$ (cf. CLWB; Tripp et al. 1998).
This predicted anti-correlation is weaker than the observed one because we include all faint absorbers (with apparent magnitude fainter than magnitude limit in optical surveys) which have small impact parameters and we include absorbers with impact parameters larger than $200 h^{-1} \kpc$. 
Other correlations of REW versus luminosity and/or absorber redshift have also been investigated. In general, if we assume $W_r \propto \rho^{-\alpha} L_B^{\beta} (1+z)^{-\gamma}$, the analysis gives $\alpha \sim 0.5, \beta \sim 0.15, \gamma \sim 0.5$.
This means the average absorption radius of a galaxy $r \propto L_B^t (1+z)^{-u}$ with $t \sim 0.3$ and $u \sim 1.0$.
The average covering factor within $250 h^{-1} \kpc$ is estimated as $\sim 0.36$ which is in good agreement with previous results (LBTW).
The effective absorption radius is estimated to be $150 h^{-1}\kpc$, which is consistent with the observational result $\sim 170 h^{-1}\kpc$ derived by CLWB.

To compare with results of imaging and spectroscopic surveys, it is necessary to study selection effects.
Selection effects have impacts on the statistics of the galaxy/absorber properties. 
The present of `spurious' galaxies at larger impact parameters within the redshift window of the absorber and the `missing' of faint absorber at small impact parameter may lead to misleading conclusions that the average galaxy/absorber separation is very large. 
Our simulations show that this is indeed the case.
We construct mock observations with the same known QSO LOS as CLWB applying selection criteria which are similar.
By an adequate number of mock runs, the total number of galaxy-absorber pairs can be predicted and the correlations mentioned above can be analysed.
The predicted number of galaxy-absorber pairs with $W_r \geq 0.3${\AA} is $\sim 26\pm 5$, in good agreement with CLWB ($\sim 26$).
The analysis of the anti-correlation between $W_r$ and $\rho$ shows that
selection effects can statistically strengthen the anti-correlation.
Some results for mock runs can produce anti-correlations consistent with CLWB. We also predict some `missing galaxy-absorber pairs' which are excluded by the selection criteria.

We estimate the redshift interval adequate to predict accurate anti-correlations of $W_r$ versus $\rho$.
To get results with a small scatter, it is found that in the standard model a total redshift interval of $\sim 10$ is required. 
This redshift interval is twice that of the LOSs in CLWB. 
This may imply that the total redshift interval in present surveys is not large enough to reveal the real anti-correlation.

\section*{Acknowledgments}

The authors thank Shude Mao, Tom Theuns for helpful discussions, careful reading of the manuscript and useful comments. We thank the referee, Xavier Barcons for helpful suggestions.
WPL thanks Zhenglong Zou, Zugan Deng, Xiaoyang Xia for help.
WPL acknowledges the Max-Planck-Institut f\"ur Astrophysik for hospitality and,
gratefully acknowledges the predoctoral fellowship under the MPG-CAS exchange program. 
This work was also partly supported by SFB375.

{\appendix
\section{Self-similar solution for cooling flow}

From eq. (18),
we write,
\beq
p(x)=\frac{1}{(x+\bar{u})x^2}\frac{d}{dx}(x^2\bar{u})
\eeq
\beq
q(x)=\frac{-8}{(x+\bar{u})x^2 (1+x)^2},
\eeq
and get the general solution as
\beq
\bar{\rho_c}=e^{-\int p(x)dx}\left[\int q(x) e^{\int p(x)dx}dx +c\right].
\eeq
It is easy to solve this equation for the case that $\bar{u}$ does not depend
on $x$. Then
\[
\bar{\rho_c}=\left(\frac{x+b}{x}\right)^2 \times \left\{\frac{8}{(1+x)(b-1)^3}\right.
\]
\beq
\left. +\frac{16}{(b+x)(b-1)^3}
+\frac{4}{(b+x)^2(b-1)^2}+\frac{24 ln\frac{1+x}{b+x}}{(b-1)^4}+c\right\}
\eeq
Here we let $\bar{u} \equiv b$. For $x \rightarrow \infty$, $\rho_c(x) \rightarrow 0$, and have $c=0$.
}

\bsp
\label{lastpage}
\end{document}